\documentclass[noeprint, aps,prx,twocolumn,showpacs,amsmath,amssymb,floatfix,superscriptaddress]{revtex4-2}

\usepackage[latin9]{inputenc}
\usepackage{color}
\usepackage{float}
\usepackage{amsmath}
\usepackage{amssymb}
\usepackage{graphicx}
\usepackage[unicode=true,
 bookmarks=false,
 breaklinks=false,pdfborder={0 0 1},backref=false,colorlinks=false]
 {hyperref}
\hypersetup{
  colorlinks = true
  }

\providecommand{\tabularnewline}{\\}

\usepackage{amsfonts}
\usepackage{amsthm}
\usepackage{graphicx}
\usepackage{xcolor}

\newcommand{\ket}[1]{|{#1}\rangle}

\renewcommand\[{\begin{equation}}
\renewcommand\]{\end{equation}}

\theoremstyle{plain}

\usepackage[caption=false]{subfig}

\begin{document}

\title{Entanglement Renormalization Circuits for Chiral Topological Order}

\author{Su-Kuan Chu}

\affiliation{Joint Center for Quantum Information and Computer Science, NIST/University of Maryland, College Park, MD 20742, USA}
\affiliation{Joint Quantum Institute, NIST/University of Maryland, College Park, MD 20742, USA}
\affiliation{Kavli Institute for Theoretical Physics, University of California, Santa Barbara, CA 93106-4030, U.S.A.}

\author{Guanyu Zhu}
\affiliation{IBM Quantum, IBM T.J. Watson Research Center, Yorktown Heights, NY 10598}

\author{Alexey~V.~Gorshkov}
\affiliation{Joint Center for Quantum Information and Computer Science, NIST/University of Maryland, College Park, MD 20742, USA}
\affiliation{Joint Quantum Institute, NIST/University of Maryland, College Park, MD 20742, USA}

\begin{abstract}
Entanglement renormalization circuits are quantum circuits that can be used to prepare large-scale entangled states. For years, it has remained a mystery whether there exist scale-invariant entanglement renormalization circuits for chiral topological order. In this paper, we solve this problem by demonstrating entanglement renormalization circuits for a wide class of chiral topologically ordered states, including a state sharing the same topological properties as Laughlin's bosonic fractional quantum Hall state at filling fraction $1/4$ and eight states with Ising-like non-Abelian fusion rules. The key idea is to build entanglement renormalization circuits by interleaving the conventional multi-scale entanglement renormalization ansatz (MERA) circuit (made of spatially local gates) with quasi-local evolution. Given the miraculous power of this circuit to prepare a wide range of chiral topologically ordered states, we refer to these circuits as MERA with quasi-local evolution (MERAQLE).
\end{abstract}
\maketitle

\section{Introduction}

\label{sec:Introduction}

Quantum many-body systems at zero temperature manifest many phenomena that have no counterparts in the ordinary classical world. One distinctive feature that prevails only in the quantum realm is the notion of entanglement, which intuitively states that local degrees of freedom are organized in such a way that the whole system cannot be straightforwardly perceived as an assembly of uncorrelated individual pieces. However, complete understanding of the structure and nature of many-body entanglement remains an outstanding challenge for quantum physicists even today. Several proposals have been put forward in an attempt to capture its essential features, including tensor network states \cite{Verstraete2008,Cirac2019}, neural networks \cite{Deng2017}, and a wide variety of entanglement measures \cite{Walter2017}. One particularly useful definition of the entanglement structure of a many-body state is given by investigating the quantum circuits necessary to prepare the state, sometimes under certain restrictions, such as locality constraints or symmetries. One can start with a product state or some other easily prepared state and then use the circuit to generate the desired target state. This operational definition sheds light on the pragmatic aspect of entanglement.

\begin{figure}[b]
\begin{centering}
\includegraphics[width=\columnwidth]{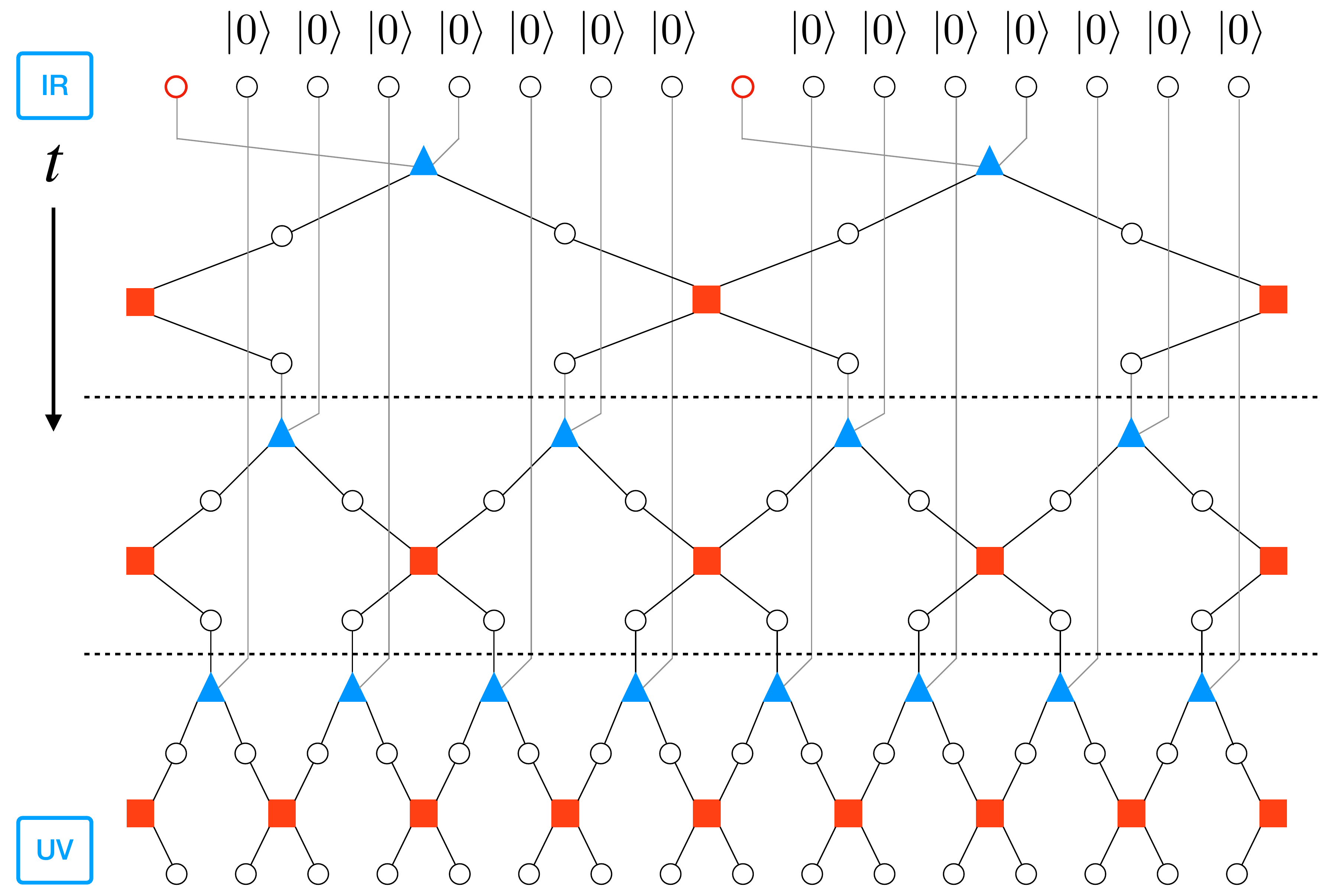}
\par\end{centering}
\centering{}\caption{A one-dimensional MERA circuit with time running downward. One starts at the top (the IR region) with a small system with sparse entanglement on a lattice with a large lattice constant (qubits represented by unfilled red circles without the $\ket{0}$ symbols above) and with many ancillary qubits (unfilled black circles) in state $\ket{0}$. We successively apply layers of entanglement renormalization consisting of isometries (blue triangles) and disentanglers (red squares) to progressively include and couple the ancillary qubits, creating a complex system with denser and more complicated entanglement structure in the UV region at the bottom. \label{fig:MERA1D}}
\end{figure}

Entanglement renormalization is a class of state-preparation quantum circuits marked by its repetitive operating procedures at varying length scales \cite{Vidal2007}, generating entanglement successively at different ranges. The earliest realization of this concept is the so-called multi-scale entanglement renormalization ansatz (MERA) \cite{Vidal2007,Vidal2008,Bridgeman2017}. A prototypical one-dimensional example of MERA composed of three steps (layers) of similar actions on a qubit system is depicted in Fig.~\ref{fig:MERA1D}.
In each step, two sets of quantum gates are applied on the system. While the isometry unitary operators (blue triangles) act on inputs in state $\ket{0}$ and in the state from the previous step, the disentangler unitary operators (red squares) act on the outputs of neighboring isometries. If we proceed with this protocol for a sufficiently large number of steps, we can create a complicated entangled state from an initial state that has almost all of the qubits unentangled, progressively introducing entanglement at various length scales. The hierarchical structure of the MERA circuit embodies the fact that entanglement can be present at different length scales. It is a convention to say that the initial time is at the infrared (IR) scale while the final time is at the ultraviolet (UV) scale. If we reverse the time arrow, going from UV to IR, the layers of the conjugated MERA circuit will progressively disentangle degrees of freedom in the order from smallest to largest scales. Ignoring the disentangled ancillary qubits, we are effectively arriving at lattices with larger and larger unit cells. This phenomenon is reminiscent of the coarse-graining procedure in the renormalization group in classical statistical mechanics, and hence the word renormalization is included in the name of the circuit. Since this circuit is an ansatz, there is no fundamental restriction on the gates, except for spatial locality of the gates at each length scale.
One can consider generalizations of this circuit to higher spatial dimensions \cite{Swingle2016, Aguado2008, Konig2009}, to fermions \cite{Corboz2009, Corboz2010}, to qudits \cite{Vidal2007, Vidal2008, Vidal2009, Gu2009, Konig2009, Swingle2016,Jha2022}, and to more types of unitaries \cite{Swingle2016}.
One can also consider gates acting on three or more qubits or even consider a fermionic system. If one is able to find an entanglement renormalization circuit for the target state, one is able to generate the final state from the initial state in time (i.e.\ circuit depth) logarithmic in the system size. Being a state with little or no entanglement, the initial state can be prepared by other means, such as adiabatic preparation \cite{Unanyan2002,Moller2008,LiDX2019,Semeghini2021,Giudici2022}, dissipative preparation \cite{Budich2015,Reiter2016}, or even a specially designed quantum process tailored for the structure of the state \cite{Muller2009,Haas2014}. Examples of states that entanglement renormalization circuits can prepare are the Greenberger-Horne-Zeilinger state (GHZ state) and the cluster state \cite{Bridgeman2017}, which are ground states of the transverse-field Ising model and the cluster state Hamiltonian, respectively. MERA is also capable of preparing gapless states in one dimension, such as the Ising model and the Potts model at the critical point \cite{Vidal2007,Evenbly2009}, which violate the area law logarithmically.

In Fig.\ \ref{fig:MERA1D}, despite the fact that we use the same red square symbol for all the disentanglers and the same blue triangle symbol for all the isometries, the unitaries can be different at different length scales and do not have to be translation-invariant. Nevertheless, in practice, if we study the ground state of a translation-invariant parent Hamiltonian, we can demand the unitaries to be translation-invariant. We can also require the MERA circuit to be scale-invariant, which means that the disentanglers and the isometries do not change from layer to layer, and see what kind of many-body quantum states keep the same local reduced density matrices after each step of the same entangling procedure. Such a scale-invariant circuit is appealing as it renders the preparation procedure of the corresponding quantum state simple conceptually and possibly in practice. A quantum state that can be prepared with a scale-invariant MERA circuit is termed a fixed-point wavefunction. 
A gapped quantum phase that has a zero-correlation-length wavefunction can serve as a fixed-point wavefunction of a scale-invariant MERA. After preparing the fixed-point wavefunction, one can reach any state in the same phase by adding an extra layer consisting of a finite-depth circuit with some locality constraints \cite{Chen2010LocalUnitary}. Models with known scale-invariant entanglement renormalization circuits are the toric code model \cite{Kitaev2003}, the quantum double model \cite{Kitaev2003, Aguado2008}, and, more generally, the Levin-Wen models \cite{Levin2005, Konig2009, Chen2010LocalUnitary}, as well as certain symmetry-protected topological phases with symmetry conditions imposed on the entanglement renormalization circuits \cite{Singh2013}.

The concept of entanglement renormalization has a wide range of interesting connections to other research areas. In particular, it is a unitary way to realize the concept of real space renormalization group without discarding any information. After each coarse-graining step, the information about the original wavefunction is encoded in the quantum gates, the present quantum state, and the ancillas in the $\ket{0}$ state. Therefore, it is drastically different from Kadanoff's real space renormalization group \cite{Kadanoff1966}, where the averaging operation to coarse-grain a system erases part of the information irretrievably. In addition, there have been some efforts to generalize the lattice version of entanglement renormalization  to devise a unitary approach to renormalizing quantum field theories, resulting in the continuous MERA (cMERA) \cite{Haegeman2013} and magic cMERA \cite{Zou2019}. 
Those formulations attempt to resolve the problem existing in traditional renormalization group approaches, where the integration out of high-momentum modes is an irreversible process.

From the perspective of experimental physics and quantum computing, a MERA circuit can serve as a practical quantum circuit to generate an initial quantum state in preparation for further quantum simulation or computation. One can implement the long-range gates in the IR with access to long-range interactions  \cite{Eldredge2017,Tran2021}. Depending on the specific experimental architecture, one may also realize these gates by first applying short-range gates to qubits and then
physically increasing the distance between them \cite{Browaeys2020, Bluvstein2022} before applying the next layer of short-range gates.
Therefore,
if long-range interactions are sufficiently strong or if qubits can be physically moved sufficiently quickly,
the MERA circuit can allow for the unitary preparation of a wide range of long-range entangled states in logarithmic time.

Even though entanglement renormalization is a powerful and beautiful concept for making sense of entanglement at different ranges, there is no guarantee that such structure exists for all states. In particular, there are phases of matter where a simple application of this concept does not work \cite{Swingle2016}, such as fracton phases in three dimensions \cite{Dua2020} or the Fermi sea in two dimensions \cite{Haegeman2018}. In those cases, one needs to use a generalized MERA formalism called the branching MERA \cite{Evenbly2014}, where entanglement is organized differently.

In two dimensions, it is hypothesized that log-depth  quantum circuits should be able to prepare all topological phases \cite{Chen2010LocalUnitary,Zeng2015,Swingle2016}. We know that the framework of scale-invariant MERA circuits is capable of preparing many quantum states belonging to the class of non-chiral topological orders (the toric code model, the quantum double model \cite{Aguado2008}, and the Levin-Wen models \cite{Konig2009} previously mentioned). 
However, it is still an open question whether we can employ scale-invariant MERA circuits to prepare chiral topological states, i.e., whether we can find a MERA circuit that has the desired chiral state as a fixed point of a single-layer application. (Here, we define chiral topologically ordered phases as quantum phases with nonzero thermal Hall conductivity \cite{Wen2017} .) One can prove no-go theorems under certain assumptions \cite{Barthel2010,Dubail2015,Li2019}. For example, Li and Mong \cite{Li2019} have shown that a free-fermionic system with a nonzero Chern number is incompatible with a scale-invariant MERA circuit with discrete strictly local quantum gates. One intuitive argument to understand the no-go theorems and the hardness of the problem is as follows. We first define the correlation length of a state to be the smallest $\ell>0$ such that the connected two-point correlation function $\left\langle {\cal O}(\mathbf{x})\,{\cal O}(\mathbf{y})\right\rangle_{\mathrm{c}} = \langle {\cal O}(\mathbf{x})\,{\cal O}(\mathbf{y})\rangle -\langle {\cal O}(\mathbf{x})\rangle \langle {\cal O}(\mathbf{y})\rangle$
of any local observable ${\cal O}(\mathbf{x})$ of finite support and unit operator norm $||{\cal O}(\mathbf{x})||=1$ can be bounded by an exponentially decaying function $C\exp(-\left|\mathbf{r}\right|/\ell)$ with $C>0$ ($C$ is  possibly dependent on the size of the support of $\cal O$), i.e., $\left|\left\langle {\cal O}(\mathbf{x})\,{\cal O}(\mathbf{y})\right\rangle_\mathrm{c} \right| \leq C\exp\left(-\left|\mathbf{x}-\mathbf{y}\right|/\ell\right)$. Suppose that we run a MERA circuit going from UV to IR. If all quantum gates in each layer of the MERA circuit are strictly local (i.e.~have an interaction range of finite radius), then in order to be scale-invariant under the coarse-graining operation, a fixed-point wavefunction must either have a zero correlation length or an infinite correlation length. The reason is that, after each step of the renormalization operation, the correlation length on the coarse-grained lattice $\ell'$ has to be the correlation length on the original lattice $\ell$ scaled down by a factor $b$ with $b>1$, i.e., $\ell'=\ell/b$ \footnote{The detailed argument is as follows. Suppose that we start with a state $\ket{\Psi}$ with its two-point connected correlation functions bounded by $C\exp(-\left|\mathbf{r}\right|/\ell)$ for any local observable, where $\ell$ is chosen to have the smallest possible value and where $C$ is possibly dependent on the size of the support of the observable. After a single-layer of the entanglement renormalization circuit $\mathcal{U}$, we arrive at a coarse-grained state $\ket{\Psi'}\equiv \mathcal{U}\,\ket{\Psi}$ and denote the connected two-point correlation function of the coarse-grained state with respect to the original lattice as $\left\langle {\cal O}(\mathbf{x})\,{\cal O}(\mathbf{y})\right\rangle_{\mathrm{c}}^{\prime}$. Since the circuit $\mathcal{U}$ is made up of strictly local gates, the operator ${\cal M}(\mathbf{y})\equiv \mathcal{U}^{\dagger}{\cal O}(\mathbf{y})\,\mathcal{U}$ is also a local observable with finite support and unit operator norm, which leads to the bound $\left\langle {\cal O}(\mathbf{x})\,{\cal O}(\mathbf{y})\right\rangle_{\mathrm{c}}^{\prime}=\left\langle {\cal M}(\mathbf{x})\,{\cal M}(\mathbf{y})\right\rangle_{\mathrm{c}}\leq C \exp(-\left|\mathbf{x}-\mathbf{y}\right|/\ell)$ with respect to the original lattice. In addition, as the expression ${\cal M}(\mathbf{x})$ explores (as we vary ${\cal O}(\mathbf{x})$) all possible local observables with finite support and unit operator norm, $\ell$ is actually the optimal length to bound the correlation functions of the coarse-grained state. Therefore, the correlation length of state $\ket{\Psi'}$ with respect to the original lattice is still $\ell$. However, because a certain fraction of the degrees of freedom are disentangled by a single layer of entanglement renormalization, we can define a new coarse-grained lattice with a lattice constant that is  $b>1$ times larger than the original one. We refer to this step as re-scaling \cite{Kardar2007}. Hence, with respect to the new coarse-grained lattice, the correlation length is $\ell'=\ell/b$.}. If a chiral wavefunction stays the same throughout all the coarse-graining operations, there are only two possibilities for its correlation length:~ $\ell=0$ and $\ell=\infty$. A system with an infinite correlation length means that some of its correlation functions cannot be bounded by any exponentially decaying function. For short-range Hamiltonians, this is generally a signature of gaplessness. As topologically ordered systems are defined as gapped phases, the case with an infinite correlation length is irrelevant to us. Hence, the only remaining question is: would it be possible to have a chiral topological state with a zero correlation length? Recall that, unlike the non-chiral states mentioned above, many well-known chiral topological states we know, such as many integer and fractional quantum Hall states, have nonzero correlation lengths. Additionally, it has been shown that a Chern insulator of free fermions (i.e.\ non-interacting integer quantum Hall state on a lattice) cannot have a zero correlation length \cite{Dubail2015,Bezrukavnikov2018}. Moreover, for an interacting chiral topological system with $U(1)$ symmetry and finite-dimensional on-site Hilbert spaces, a typical property of many known chiral topological phases, the Hamiltonian cannot be a sum of locally-commuting terms \cite{Kapustin2019}. As the condition of the correlation length being zero is usually a harbinger of the existence of a locally-commuting parent Hamiltonian \cite{Levin2005}, we expect that finding a representative wavefunction with zero correlation length for
a chiral phase should be a hard, if not impossible, task. With all the evidence mentioned above, it seems very unlikely that scale-invariant MERA circuits exist for chiral topological phases.

\begin{figure}[t]
\begin{centering}
\includegraphics[width=\columnwidth]{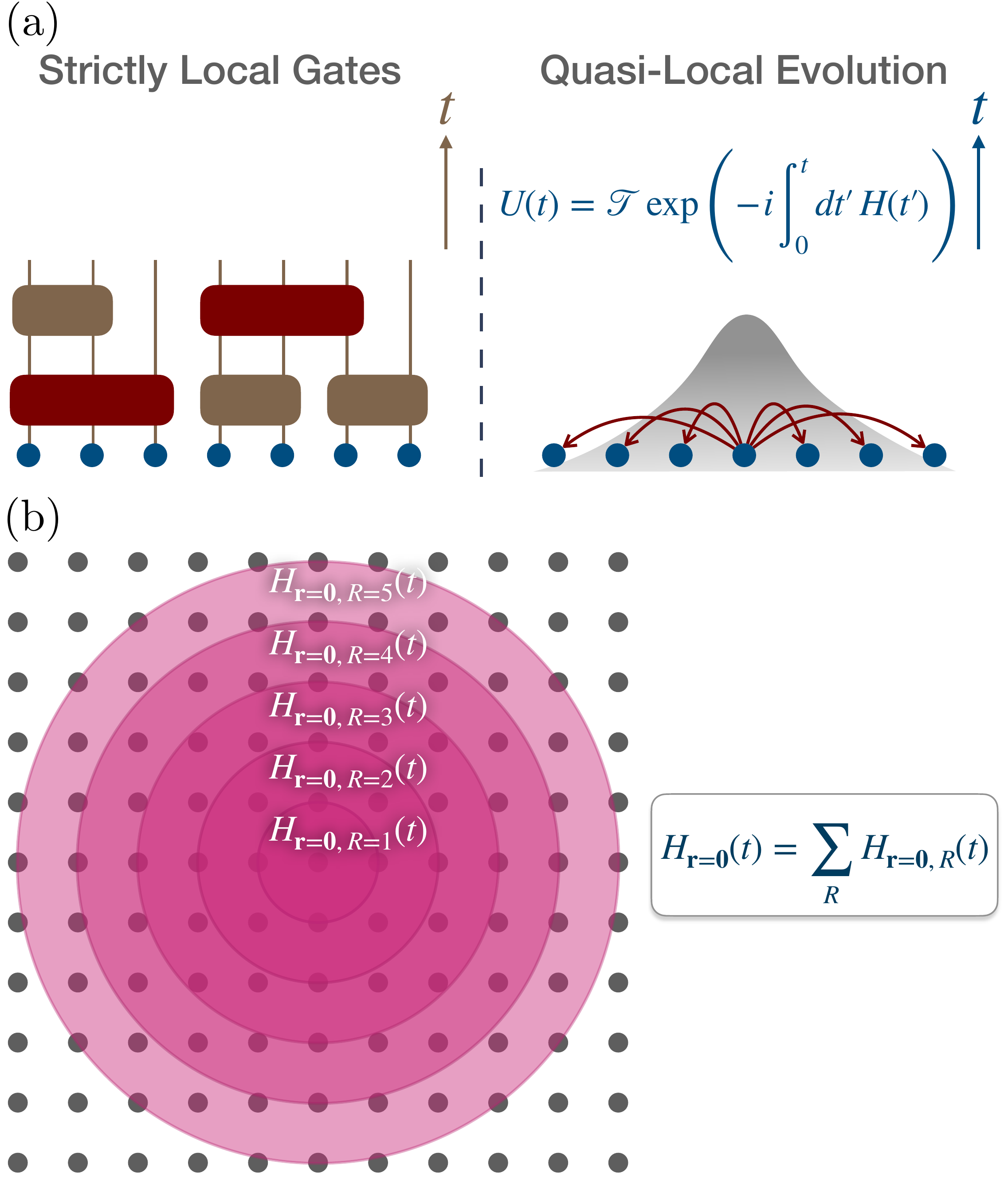}
\par\end{centering}
\centering{}\caption{(a) Left: A traditional local quantum circuit consisting of strictly local (i.e.\ with support bounded by a finite radius) discrete quantum gates. Right: A quantum circuit $U(t)$ based on quasi-local continuous time evolution under a time-dependent Hamiltonian $H(t)$ with interaction tails decaying faster than any 
power law function. To be precise, $H(t)$ is a sum of interaction terms $H(t)=\sum_{\mathbf{r}}H_{\mathbf{r}}(t)$, where each interaction term $H_{\mathbf{r}}(t)$ centered at position $\mathbf{r}$ has a decomposition in terms of Hermitian operators $H_{\mathbf{r},R}(t)$ supported on sites within disks of different radii $R\in \mathbb{N}$: $H_{\mathbf{r}}(t)=\sum_R H_{\mathbf{r},R}(t)$, and the Hermitian operators $H_{\mathbf{r},R}(t)$ with large $R$ can be uniformly bounded by $\left\Vert H_{\mathbf{r},R}(t) \right\Vert =\mathcal{O}(1/R^{\alpha})$ for any power $\alpha>0$.
(b) A schematic diagram of the decomposition of the interaction term $H_{\mathbf{r}=\mathbf{0}}(t)=\sum_R H_{\mathbf{r}=\mathbf{0},R}(t)$ for two-dimensional square-lattice systems.
\label{fig:strictlocalityandquasilocality}}
\end{figure}

Despite all the difficulties mentioned above, there are works pointing out how to overcome the issue mentioned above, at least for non-interacting fermions. The key insight is to relax the condition that quantum circuits for each layer of entanglement renormalization must be made up of strictly local and discrete quantum gates assumed in the conventional MERA framework. Instead, we allow the use of continuous time evolution under a time-dependent quasi-local Hamiltonian. By quasi-locality, we mean that the interactions are no longer restricted to be finite-range, but their strength should decay with distance faster than any power law. 
A comparison between a quantum circuit based on strictly local discrete quantum gates and one with quasi-local evolution is shown in Fig.~\ref{fig:strictlocalityandquasilocality}.
With quasi-local evolution, we can circumvent the no-go theorems and the intuitive argument of correlation length reduction stated above. A chiral state with a nonzero correlation length can now be a fixed-point wavefunction of a scale-invariant entanglement renormalization circuit. For example, in Ref.~\cite{Swingle2016}, an entanglement renormalization procedure is demonstrated for a lattice Chern insulator model, which has a nonzero finite correlation length in its ground state. The circuit is comprised of a series of subroutines, the so-called quasi-adiabatic evolutions, which are generated by quasi-local Hermitian operators derived from certain adiabatic evolutions of the Chern insulator model. In Ref.~\cite{Chu2019}, a different approach was presented. The authors used the formalism of cMERA to develop a scale-invariant entanglement renormalization circuit for a continuous Chern insulator model. The continuous MERA generalizes the discrete isometry unitaries and disentangler unitaries in the conventional MERA on a lattice to continuous unitary evolution generated by Hermitian operators acting on a continuum of spatial modes.
The Hermitian operators of the cMERA in Ref.~\cite{Chu2019} involve interactions with exponentially decaying tails. The evidence above suggests that quasi-locality may be an essential feature of scale-invariant entanglement renormalization circuits for gapped chiral topological states, which usually 
have nonzero finite correlation lengths. Despite the success in non-interacting chiral systems, designing scale-invariant entanglement renormalization circuits for interacting chiral topologically ordered systems based on this physics insight still remains a largely unexplored research area.

\begin{figure*}[t]
\begin{centering}
\includegraphics[scale=0.40]{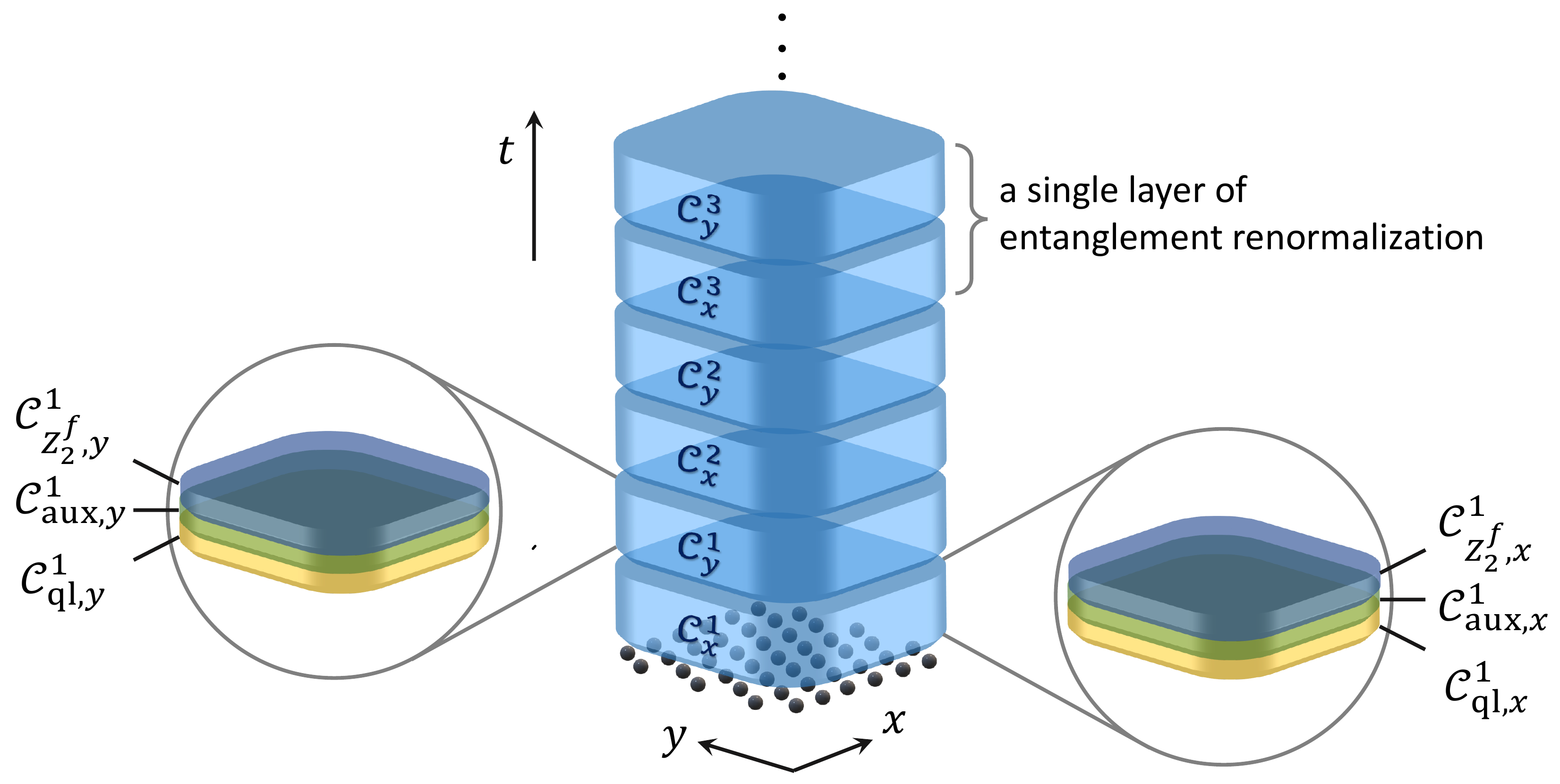}
\par\end{centering}
\centering{}\caption{
A schematic diagram of the scale-invariant MERAQLE framework presented in this paper. At the bottom is a two-dimensional many-body entangled state of spins (represented by black marbles) to be disentangled by the MERAQLE circuit. The full MERAQLE circuit is a tower consisting of layers labeled by the number $s\in\mathbb{N}$. The number $s$ labeling the layers is, roughly speaking, the logarithm of the length scale at which the entanglement of the system is being undone.
Each layer $s$ of the circuit consists of two subcircuits $\mathcal{C}_{x}^{s}$ and $\mathcal{C}_{y}^{s}$, which represent a single step of horizontal (i.e.~$x$ direction) entanglement renormalization and a single step of vertical (i.e.~$y$ direction) entanglement renormalization, respectively. The subcircuit $\mathcal{C}_{x}^{s}$ can be further decomposed into three circuit components: $\mathcal{C}_{\mathrm{ql},x}^{s}$ (a circuit based on quasi-local evolution in the right subfigure of Fig.~\ref{fig:strictlocalityandquasilocality}(a)), $\mathcal{C}_{Z_{2}^{f},x}^{s}$ (a layer of a conventional MERA circuit with strictly local gates designed to renormalize the toric code model, which is a pure $Z_{2}^{f}$ lattice gauge theory), and $\mathcal{C}_{\mathrm{aux},x}^{s}$ (an auxiliary circuit with strictly local gates designed to locally rearrange the fermionic modes). The precise definitions and meaning of the circuit components $\mathcal{C}_{\mathrm{ql},x}^{s}$, $\mathcal{C}_{Z_{2}^{f},x}^{s}$, and $\mathcal{C}_{\mathrm{aux},x}^{s}$ are discussed in detail in Sec.\ \ref{sec:MERAQLE}. The subcircuits $\mathcal{C}_{x}^{s}$ with different $s$ are essentially the same except acting on different length scales. 
A similar decomposition also holds for the subcircuit $\mathcal{C}_{y}^{s}$. \label{fig:layoutofMERAQLE}}
\end{figure*}

In this paper, we solve this problem by providing explicit circuits for several exactly solvable interacting chiral spin models, thus offering a glimpse of the entanglement structure of interacting chiral topological phases. The key idea---first briefly introduced in Ref.~\cite{Swingle2016} for a hybrid qubit-fermion system describing Ising topological order---is to marry 
the MERA circuits for interacting non-chiral topological states with quasi-local evolution that renormalizes non-interacting chiral topological states. We start by sketching the underlying logic behind our proposal. Consider several layers of non-interacting $p_{x}+ip_{y}$ topological superconductors on a lattice \cite{Bernevig2013,Bernevig2015}. Since the system has a $Z_2$ fermion parity symmetry, we can couple the system to a $Z_{2}^f$ lattice gauge field with $Z_2$ gauge variables and add the Gauss's law constraint to the coupled system. The superscript $f$ stands for ``fermion'' 
and helps us remember the role the gauge field plays in the fermion parity symmetry. We will refer to this procedure as gauging. The topological properties of the whole gauged system, including the original superconductors and the $Z_{2}^{f}$ gauge field, will fall into Kitaev's sixteenfold way classification \cite{Kitaev2006}. In the classification, the unitary modular structure of the quasiparticles suggests that it is possible to have bosons or spins (hard-core bosons) as the fundamental constituents of the theories rather than the original fermions and gauge field. In fact, inspired by the proposal in Refs.~\cite{Chen2018a,Chen2019Bosonization3D}, we are able to reformulate the gauged theory solely in terms of $S=1/2$ spins on a lattice.
To be more precise, this reformulation provides a duality between a theory with fermions coupled to a $Z_{2}^{f}$ gauge field on the one hand and an interacting spin theory on a lattice on the other. Because the original fermionic theories are chiral, the resulting spin theories are also chiral. Therefore, we will sometimes refer to the ground states of the resulting spin models as chiral spin liquids. The chiral spin liquids constructed this way include eight Abelian states and eight non-Abelian states. In particular, they include a state with the topological properties of Laughlin's bosonic fractional quantum Hall state at filling fraction $1/4$, a state with the 
Ising topological quantum field theory (Ising TQFT) fusion and braiding statistics, a state within the same universality class as the bosonic Moore-Read fractional quantum Hall state at filling fraction one (whose fusion rules are Ising-like), and six other states with Ising-like topological properties. Since the superconducting fermions in the original model are non-interacting and since the structure coming from the $Z_{2}^{f}$ gauge field is similar to the well-studied toric code \cite{Kitaev2003} (which can be interpreted as a $Z_2$ lattice gauge theory and as a non-chiral topologically ordered system), the spin models constructed this way are exactly solvable. Thanks to this nice property, we are able to analytically work out the corresponding (generalized) scale-invariant entanglement renormalization circuits. Intuitively speaking, we construct each layer of the entanglement renormalization circuit by incorporating the conventional MERA circuit for the interacting non-chiral toric code with quasi-local continuous time evolution that coarse-grains the 
non-interacting chiral $p_{x}+ip_{y}$ topological superconductors. In fact, under certain constraints on spins, Refs.~\cite{Chen2018a,Chen2019Bosonization3D} provide an additional duality between fermions and spins called bosonization. In this terminology, the quasi-local continuous time evolution of spins is simply the bosonization of a fermionic quasi-local continuous time evolution that coarse-grains layers of non-interacting $p_{x}+ip_{y}$ topological superconductors.
Even though the spin models have nonzero finite correlation lengths, due to the quasi-local structure of continuous time evolution, the resulting quantum circuits can evade the no-go arguments stated above. As shown schematically in Fig.~\ref{fig:layoutofMERAQLE},
our entanglement renormalization quantum circuits have strictly local quantum gates interleaved with quasi-local evolution. This figure sums up the core spirit of the entire manuscript. 
Note that, in this manuscript, we will consider unitary evolution that disentangles the state, with time going from UV to IR. One can get the entangling state-preparation unitary by performing Hermitian conjugation. Inspired by the miraculous power of combining the conventional MERA circuit with quasi-local evolution, we refer to this specific type of (generalized) entanglement renormalization circuit as MERA with quasi-local evolution (MERAQLE).

Since the notions mentioned above might not be familiar to all the readers, we will pedagogically review them in following sections. We will gradually introduce all the necessary concepts before presenting our results. The remainder of this paper is organized as follows. In Sec.\ \ref{sec:Toric-Code}, we review the toric code model \cite{Kitaev2003} and its MERA circuit. In Sec.~\ref{sec:-Topological-Superconductor}, we review the $p_{x}+ip_{y}$ topological superconductor model on a lattice \cite{Bernevig2013}, which is non-interacting and chiral, and its entanglement renormalization circuit, which uses the idea of quasi-local evolution. In Sec.~\ref{sec:Gauging-Fermion-Parity}, we first pedagogically review how to bosonize a fermionic theory. We describe in detail how to gauge the fermion parity symmetry of a fermionic theory and rephrase the fermionic modes and the gauge field purely in terms of spin degrees of freedom. Then, we use the bosonization technique to construct chiral spin liquid models belonging to Kitaev's sixteenfold way classification. Finally, in Sec.~\ref{sec:MERAQLE}, we present our main results. We use the entanglement renormalization circuits from Secs.~\ref{sec:Toric-Code} and \ref{sec:-Topological-Superconductor} to construct the MERAQLE circuits for all Kitaev's sixteenfold way chiral spin liquids. In Sec.~\ref{sec:conclusion}, we present conclusions and outlook. In Appendix~\ref{sec:appendix-quasiadiabatic}, we review the mathematical framework of quasi-adiabatic evolution, which forms the backbone of our quasi-local evolution for 
$p_x+ip_y$ superconductors. In Appendix~\ref{sec:appendix-compatibility}, we present some technical calculations related to the MERAQLE circuits and omitted from the main text.

\section{Toric Code}

\label{sec:Toric-Code}

Before we begin to discuss the framework of MERAQLE circuits, we first discuss an exact scale-invariant entanglement renormalization circuit for the simplest model with intrinsic topological order, i.e., the toric code \cite{Kitaev2003,Savary2016}, to make the reader more familiar with the notions of entanglement renormalization and fixed-point wavefunctions in two dimensions. We will first review the toric code Hamiltonian in Sec.~\ref{subsec:toriccodemodel} and then, in Sec.~\ref{subsec:MERAforToricCode}, present its entanglement renormalization circuit, which will be a simpler variant of the one initially proposed in Ref.~\cite{Aguado2008}. The toric code entanglement renormalization circuit belongs to the family of conventional MERA circuits with strictly local quantum gates. Despite its simplicity, the MERA circuit presented here will serve as an inspiration for the MERAQLE circuit constructed in Sec.~\ref{sec:MERAQLE}.

\subsection{Model}
\label{subsec:toriccodemodel}
\begin{figure}[t]
\begin{centering}
\includegraphics[scale=0.35]{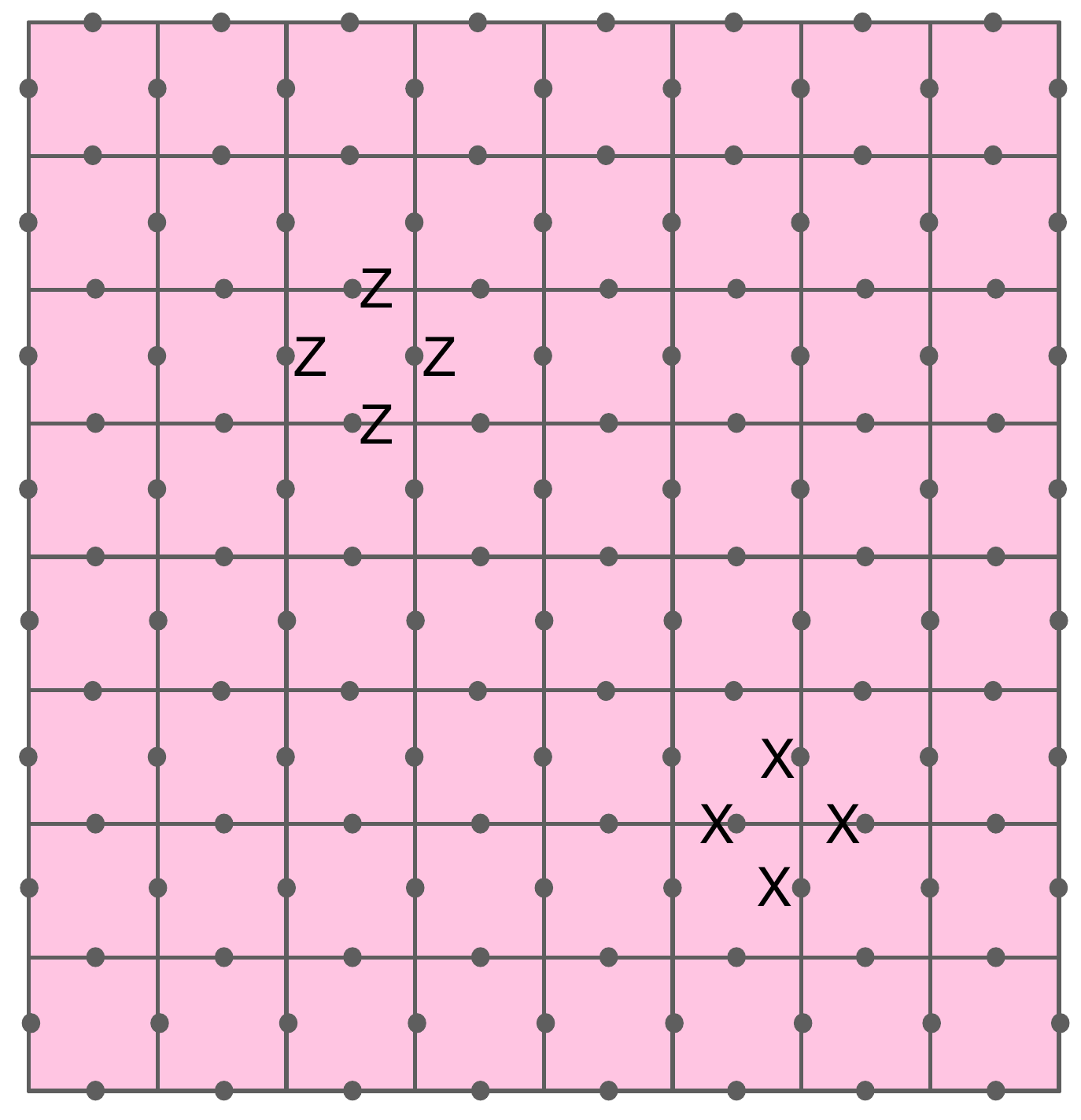}
\par\end{centering}
\centering{}\caption{The two sets of interaction terms for the toric code. The first set consists of  plaquette operators. A representative plaquette operator is illustrated as an operator made of four Pauli-$Z$ operators. The other set consists of  vertex operators. A representative vertex operator is illustrated as an operator made of four Pauli-$X$ operators. All translated versions of the illustrated plaquette and vertex operators are included in the Hamiltonian. For later convenience, we put $Z$ operator labels to the right of the qubits and $X$ operator labels to the left. We drop the subscript from $Z_{e}$ and $X_{e}$ since it is clear from the figure which qubits they act on. \label{fig:ToricCode-stabilizergenerators}}
\end{figure}

For the toric code model, we consider qubits living on the edges of a square lattice. The Hamiltonian is 
\begin{equation}
H_{\mathrm{TC}}=-\sum_{f}\prod_{e\in f}Z_{e}-\sum_{v}\prod_{e\in v}X_{e}\label{eq:ToricCode-Hamiltonian}
\end{equation}
with $X_{e}=\left(\begin{array}{cc}
0 & 1\\
1 & 0
\end{array}\right)$ and $Z_{e}=\left(\begin{array}{cc}
1 & 0\\
0 & -1
\end{array}\right)$ being Pauli matrices of the qubit on edge $e$, where the symbol $f$ labels all faces and the symbol $v$ labels all vertices on the square lattice. In this matrix representation, $\left(\begin{array}{c}
1\\
0
\end{array}\right)=\ket{0}$ and $\left(\begin{array}{c}
0\\
1
\end{array}\right)=\ket{1}$. The notation $e\in f$ means that the edge $e$ is one of the four edges of the face $f$, while the notation $e\in v$ means that the edge $e$ is incident on the vertex $v$. We will refer to $\prod_{e\in f}Z_{e}$ as a plaquette operator and $\prod_{e\in v}X_{e}$ as a vertex operator. The operators are shown in Fig.~\ref{fig:ToricCode-stabilizergenerators}.
This model is non-chiral and exactly solvable.
It can be considered to be a pure $Z_{2}$ lattice gauge theory \cite{Kitaev2003,Savary2016}. All the plaquette operators and all the vertex operators commute with one another, so the ground state can be chosen as simultaneous eigenstate of those operators with eigenvalue one. One can show that the correlation length of the toric code ground state is zero.

This model has four types of elementary excitations. Violating the first term while not violating the second term in Eq.~(\ref{eq:ToricCode-Hamiltonian}) implies the existence of
an $m$ particle that has bosonic self-braiding statistics. Violating the second term while not violating the first term in Eq.~(\ref{eq:ToricCode-Hamiltonian}) implies the existence of an $e$ particle that also has bosonic self-braiding statistics . The braiding of an $m$ particle and an $e$ particle results in a nontrivial $(-1)$ phase. The combination of $e$ and $m$ gives rise to a fermion $f$. 

For later convenience, we now introduce some quantum information terminology \cite{Nielsen2000,Bacon2022}. The Pauli group $\mathcal{P}_{n}$ on $n$ qubits is a non-Abelian group with group elements having the form of tensor products of Pauli matrices $i^{k}P_{1}\otimes P_{2}\otimes\cdots\otimes P_{n}$ with $k\in\left\{0,1,2,3\right\}$ and $P_{j}\in\left\{ I_{j},X_{j},Y_{j},Z_{j}\right\} $ being a Pauli matrix on the $j$-th qubit. The multiplication operation is defined using the matrix multiplication operation for each individual qubit. A stabilizer group, or stabilizer, $\mathcal{S}$ on $n$ qubits is an Abelian subgroup of $\mathcal{P}_{n}$ that does not contain the tensor product of identity matrices with a minus sign, $-I_{1}\otimes I_{2}\otimes\cdots\otimes I_{n}$, as its element. Sometimes, it is convenient to work with a set of generators that generate a stabilizer group so that any group element in the stabilizer group can be written as a product of the generators. Note that the choice of the set of generators is not unique. In addition, we say that a quantum state is stabilized by an operator if the state is a $+1$ eigenvector. We say that a state is stabilized by the stabilizer group $\mathcal{S}$ if the state is a $+1$ eigenvector of all the stabilizer generators. Therefore, in this terminology, the ground state of the toric code is stabilized by all plaquette operators and all vertex operators. The group generated by all the plaquette operators and all the vertex operators forms a stabilizer group, which stabilizes the ground state of the toric code model.

\subsection{Entanglement Renormalization Circuit}

\label{subsec:MERAforToricCode}

The full entanglement renormalization circuit for the toric code ground state has a hierarchical structure of multiple layers of smaller subcircuits, like the one-dimensional MERA in Fig.~\ref{fig:MERA1D}. Each layer represents the length scale at which the entanglement of the ground state is renormalized. Instead of having the circuit structure with isometries and disentanglers for each layer, like the toric code MERA initially proposed in Ref. \cite{Aguado2008}, we here have two kinds of subcircuits for each layer of entanglement renormalization. One is called a single step of horizontal entanglement renormalization, and the other is called a single step of vertical entanglement renormalization. The structure of the entanglement renormalization circuit is similar to the one shown in Fig.~\ref{fig:layoutofMERAQLE}.

\begin{figure}[t]
\begin{centering}
\includegraphics[width=\columnwidth]{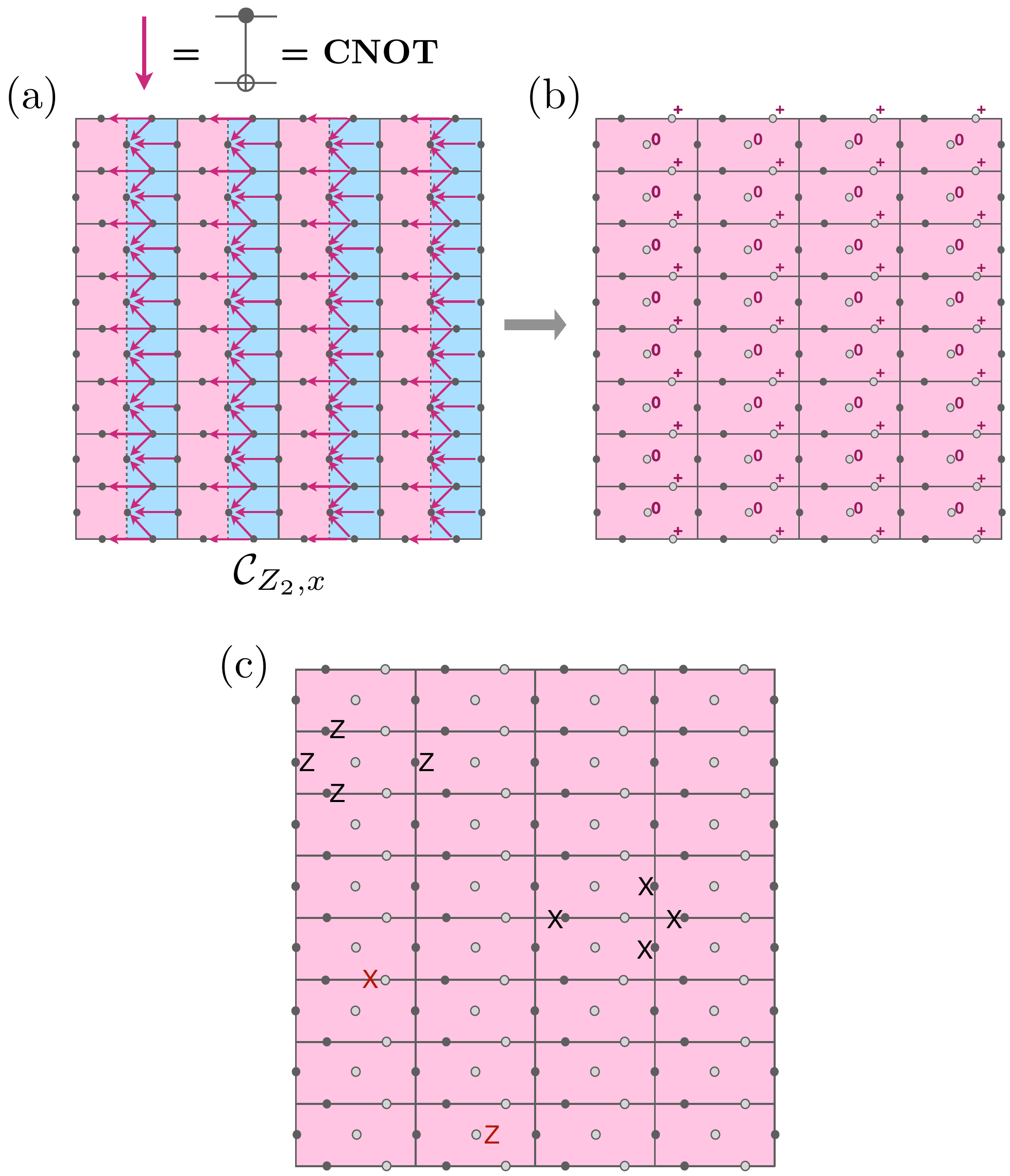}
\par\end{centering}
\centering{}\caption{(a) The circuit $\mathcal{C}_{Z_{2},x}$ for a single step of horizontal entanglement renormalization for the toric code. The filled circles represent qubits (spins) constituting the toric code model. (b,c) The state of the system after the circuit has been applied. Unfilled circles are the qubits (spins) that have been disentangled by the circuit into $\ket{0}$  and  $\ket{+}\equiv\frac{1}{\sqrt{2}}(\begin{array}{cc}
1 & 1\end{array})^{T}$ states, as indicated by the labels in (b). (c) shows the new stabilizer generators.  The red single-site $Z$ and $X$ generators stabilize the disentangled qubits, while the black 4-qubit generators stabilize the toric code defined on the new horizontally elongated square lattice.  The derivation of the new stabilizer generators is presented in Fig.~\ref{fig:MERA-ToricCode-hori-calulation}. \label{fig:MERA-ToricCode-hori}}
\end{figure}

\begin{figure}[t]
\begin{centering}
\includegraphics[width=\columnwidth]{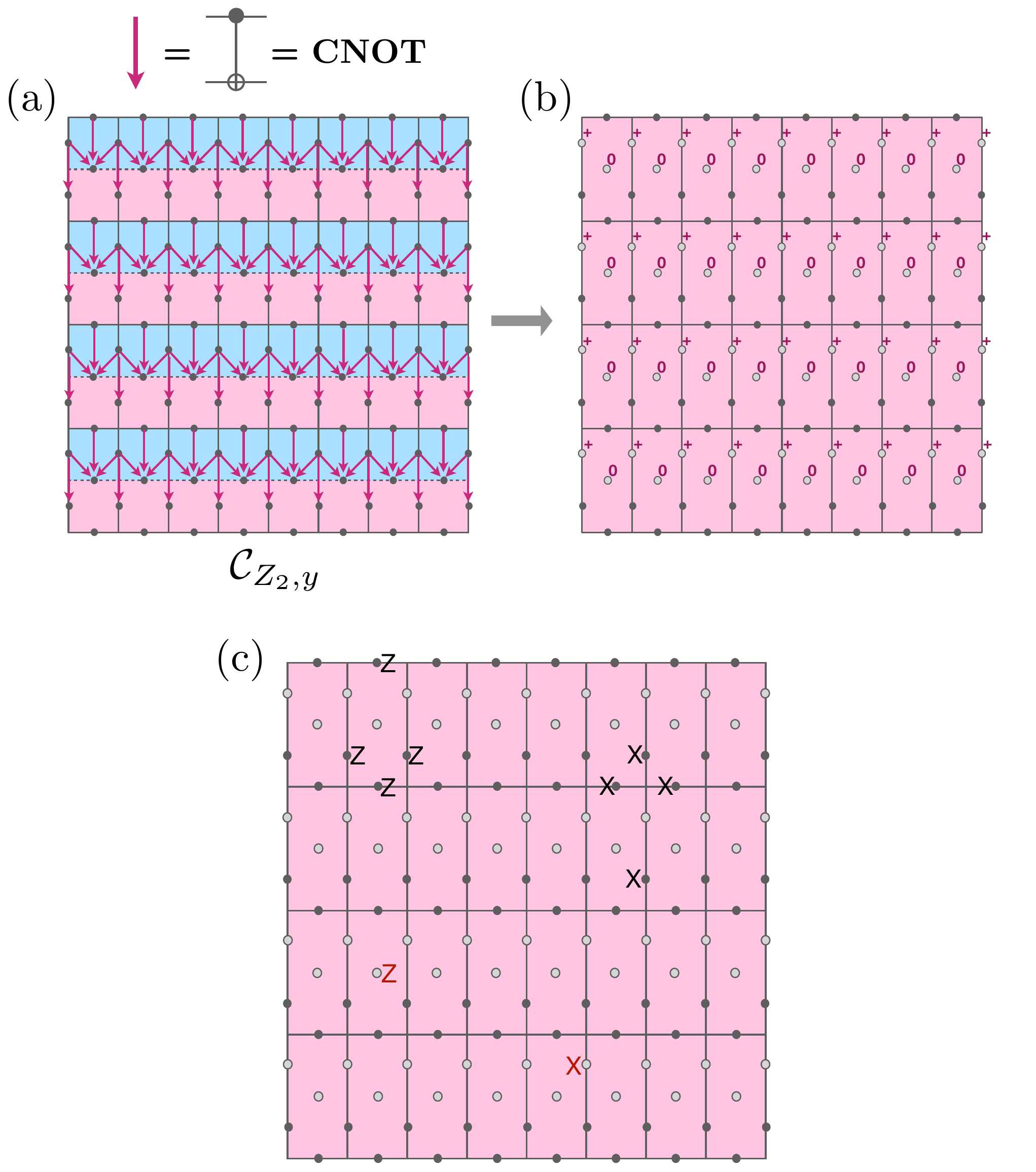}
\par\end{centering}
\centering{}\caption{(a) The circuit $\mathcal{C}_{Z_{2},y}$ for a single step of vertical entanglement renormalization for the toric code. The filled circles represent qubits (spins) constituting the toric code model. (b,c) The state of the system after the circuit has been applied. Unfilled circles are the qubits (spins) that have been disentangled by the circuit into $\ket{0}$ and $\ket{+}$ states, as indicated by the labels in (b). (c) shows the new stabilizer generators.  The red single-site $Z$ and $X$ generators stabilize the disentangled qubits, while the black 4-qubit generators stabilize the toric code defined on the new vertically elongated square lattice.  The derivation of the new stabilizer generators is presented in Fig.~\ref{fig:MERA-ToricCode-vert-calulation}. \label{fig:MERA-ToricCode-verti}}
\end{figure}

Those subcircuits for single steps of entanglement renormalization of the toric code consist of a series of controlled-NOT (CNOT) gates. A CNOT gate is a two-qubit gate defined by the following action: $\ket{00}\rightarrow\ket{00}$, $\ket{01}\rightarrow\ket{01}$, $\ket{10}\rightarrow\ket{11}$, and $\ket{11}\rightarrow\ket{10}$. The first qubit is called the control qubit, and the second qubit is called the target qubit. The horizontal entanglement renormalization subcircuit $\mathcal{\mathcal{C}}_{Z_{2},x}$ and the vertical entanglement renormalization subcircuit $\mathcal{\mathcal{C}}_{Z_{2},y}$ are shown in Fig.~\ref{fig:MERA-ToricCode-hori}(a) and Fig.~\ref{fig:MERA-ToricCode-verti}(a), respectively.
\begin{figure}[t]
\begin{centering}
\includegraphics[width=\columnwidth]{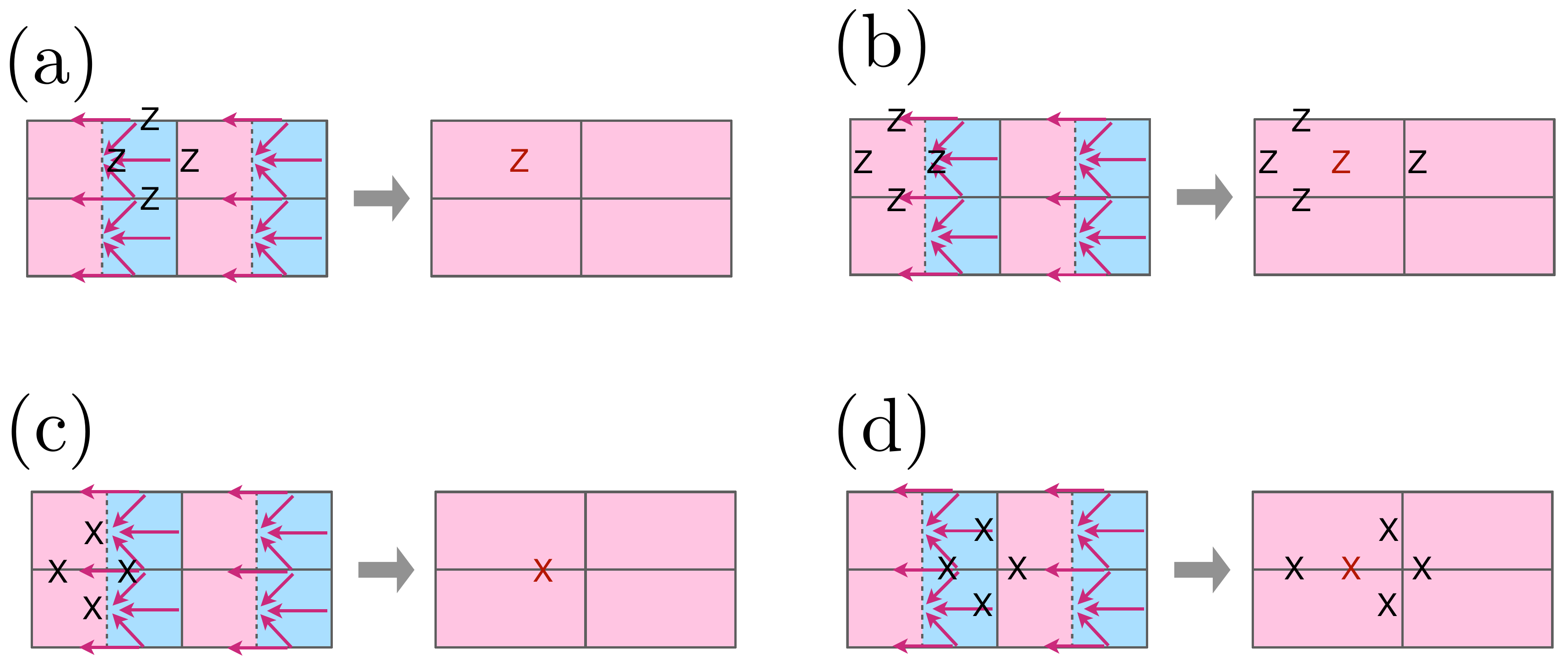}
\par\end{centering}
\centering{}\caption{Transformation of the stabilizer generators of the toric code model under conjugation by the horizontal entanglement renormalization subcircuit $\mathcal{C}_{Z_{2},x}$ in Fig.~\ref{fig:MERA-ToricCode-hori}(a). (a)(b) Transformation of the plaquette operators. (c)(d) Transformation of the vertex operators. The new stabilizer group generated by the operators on the right-hand sides of the subfigures is the same as the stabilizer group generated by the operators in Fig.~\ref{fig:MERA-ToricCode-hori}(c). The red Pauli operators in the subfigures are the red single-qubit stabilizer generators in Fig.~\ref{fig:MERA-ToricCode-hori}(c) acting on the disentangled qubits.  \label{fig:MERA-ToricCode-hori-calulation}}
\end{figure}
\begin{figure}[t]
\begin{centering}
\includegraphics[width=\columnwidth]{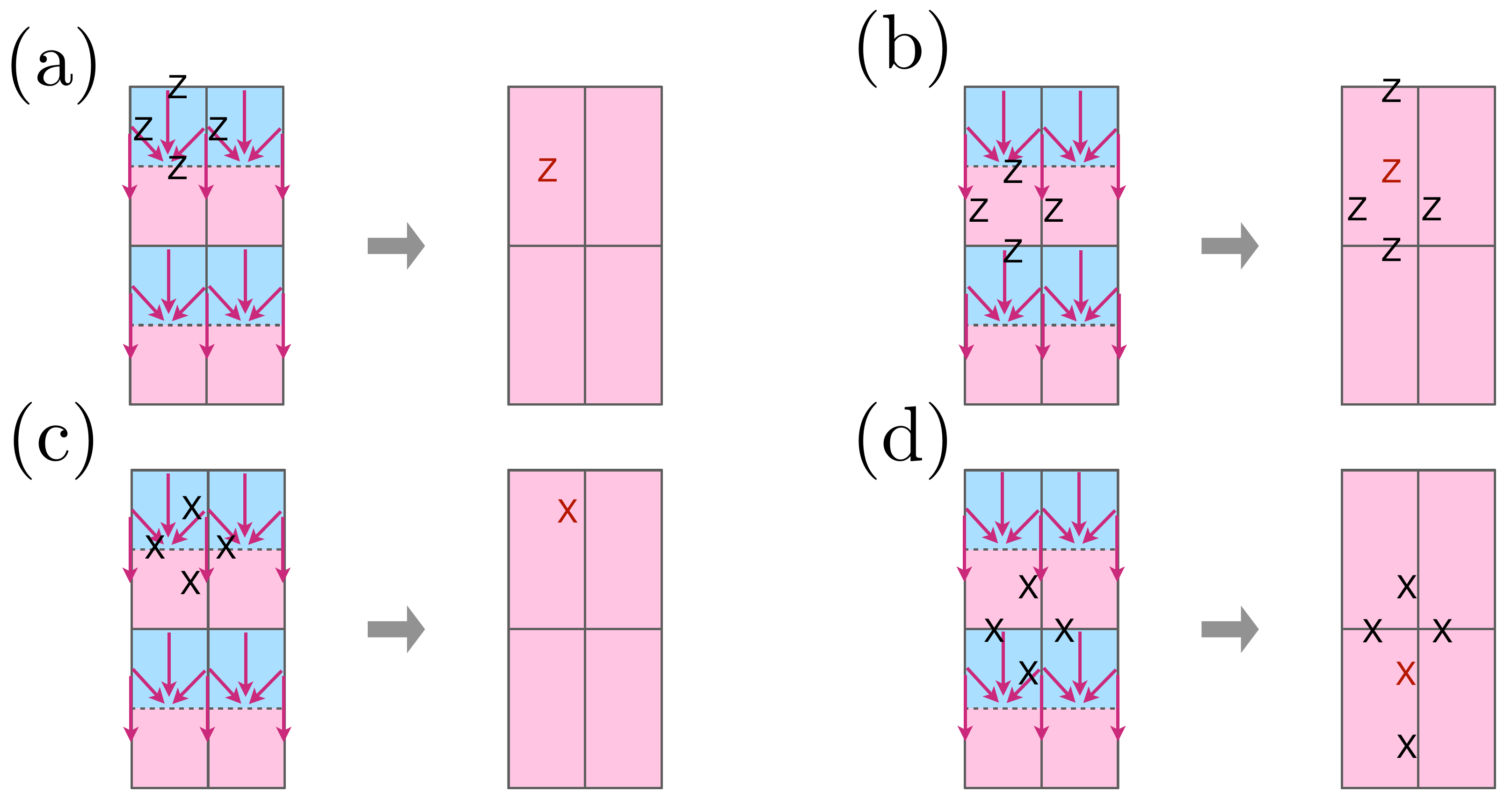}
\par\end{centering}
\centering{}\caption{Transformation of the stabilizer generators of the toric code model under conjugation by the vertical entanglement renormalization subcircuit $\mathcal{C}_{Z_{2},y}$ in Fig.~\ref{fig:MERA-ToricCode-verti}(a). (a)(b) Transformation of the plaquette operators. (c)(d) Transformation of the vertex operators. The new stabilizer group generated by the operators on the right-hand sides of the subfigures is the same as the stabilizer group generated by the operators in Fig.~\ref{fig:MERA-ToricCode-verti}(c). The red Pauli operators in the subfigures are the red single-qubit stabilizer generators in Fig.~\ref{fig:MERA-ToricCode-verti}(c) acting on the disentangled qubits.  
\label{fig:MERA-ToricCode-vert-calulation}}
\end{figure}
To represent a CNOT gate in the figures throughout the paper, we will use an arrow pointing from the control qubit to the target qubit. 
Note that all CNOT gates shown in Fig.~\ref{fig:MERA-ToricCode-hori}(a) commute; similarly, in Fig.~\ref{fig:MERA-ToricCode-verti}(a).

To understand how the ground state of $H_{\mathrm{TC}}$ transforms, it suffices to understand how the individual terms of $H_{\mathrm{TC}}$ change when conjugated by the subcircuits. In particular, if $\ket{\Psi}$ is a ground state of $H_{{\rm TC}}$ and hence an eigenvalue-one eigenvector of the $\prod_{e\in f}Z_{e}$ and $\prod_{e\in v}X_{e}$ operators, then $\mathcal{C}_{Z_{2},x}\ket{\Psi}$ must be an eigenvalue-one eigenvector of the $\mathcal{C}_{Z_{2},x}\,\prod_{e\in f}Z_{e}\,\mathcal{C}_{Z_{2},x}^{\dagger}$ and $\mathcal{C}_{Z_{2},x}\,\prod_{e\in v}X_{e}\,\mathcal{C}_{Z_{2},x}^{\dagger}$ operators. Since $\mathcal{C}_{Z_{2},x}$ and the constituent CNOT gates are in the Clifford group \cite{Nielsen2000,Bacon2022}, the normalizer of the Pauli group, the operators $\mathcal{C}_{Z_{2},x}\,\prod_{e\in f}Z_{e}\,\mathcal{C}_{Z_{2},x}^{\dagger}$ and $\mathcal{C}_{Z_{2},x}\,\prod_{e\in v}X_{e}\,\mathcal{C}_{Z_{2},x}^{\dagger}$ must be in the Pauli group and therefore generate a new stabilizer group. Instead of directly studying how the ground state is transformed by $\mathcal{C}_{Z_{2},x}$, we can investigate how the stabilizer group gets transformed under conjugation by $\mathcal{C}_{Z_{2},x}$. In quantum information, this approach is called the stabilizer formalism. A similar statement holds for the vertical entanglement renormalization subcircuit $\mathcal{C}_{Z_{2},y}$. 

To see the transformation of the stabilizer group, we first notice that, under conjugation, the CNOT gate transforms two-qubit operators as follows:
\begin{equation}
\begin{array}{cc}
1\otimes Z\leftrightarrow Z\otimes Z & Z\otimes I\leftrightarrow Z\otimes I\\
1\otimes X\leftrightarrow I\otimes X & X\otimes I\leftrightarrow X\otimes X
\end{array},\label{eq:stabilizertransformation}
\end{equation}
where the first qubit is the control qubit and the second qubit is the target qubit. With this insight, one can easily check that, as shown in Fig.~\ref{fig:MERA-ToricCode-hori}(b,c) and Fig.~\ref{fig:MERA-ToricCode-verti}(b,c), the stabilizer group is transformed under $\mathcal{C}_{Z_{2},x}$ and $\mathcal{C}_{Z_{2},y}$ into the stabilizer group of the toric code model defined on an (elongated) square lattice with larger unit cells, together with red single-qubit generators stabilizing disentangled qubits not associated with the new square lattice. Therefore, the ground state is transformed into the ground state of the toric code on the new lattice with some disentangled qubits. The disentangled ancillary qubits (which can be in either $\ket{0}$ or $\ket{+}\equiv\frac{1}{\sqrt{2}}(\begin{array}{cc}
1 & 1\end{array})^{T}$) are like the ancillary qubits in quantum state $\ket{0}$ in the one-dimensional example in Fig.~\ref{fig:MERA1D}. Notice that we will still refer to the qubits in state $\ket{+}$ as ancillary qubits since they only differ from $\ket{0}$ by single-qubit Hadamard gates
$\frac{1}{\sqrt{2}}\left(\begin{array}{cc}
1 & 1\\
1 & -1
\end{array}\right)$. 

Following the application of $\mathcal{C}_{Z_{2},y}\,\mathcal{C}_{Z_{2},x}$, we thus obtain a toric code ground state that is self-similar to the original toric-code ground state up to a scale transformations and up to the presence of disentangled qubits. This means that  the toric code ground state is indeed a fixed-point wavefunction under a single layer of entanglement renormalization $\mathcal{C}_{Z_{2},y}\,  \mathcal{C}_{Z_{2},x}$. If we iterate $\mathcal{C}_{Z_{2},y}\,  \mathcal{C}_{Z_{2},x}$ to further disentangle qubits at different length scales, we will obtain a scale-invariant entanglement renormalization circuit, which has a tower structure similar to the one  
shown in Fig.~\ref{fig:layoutofMERAQLE}. To further compare the circuit here with Fig.~\ref{fig:layoutofMERAQLE}, we introduce a number superscript $s\in \mathbb{N}$ to label the length scale (layer) the subcircuits are acting at, i.e., $\mathcal{C}_{Z_{2},x}\rightarrow \mathcal{C}^s_{Z_{2},x}$, $\mathcal{C}_{Z_{2},y}\rightarrow \mathcal{C}^s_{Z_{2},y}$. We can therefore say that the circuit components $\mathcal{C}^s_{Z_{2}^{f},x}$ and $\mathcal{C}^s_{Z_{2}^{f},y}$ in Fig.~\ref{fig:layoutofMERAQLE} for our purposes here are $\mathcal{C}^s_{Z_{2},x}$ and $\mathcal{C}^s_{Z_{2},y}$, respectively, while the other circuit components are trivial, i.e.,  $\mathcal{C}_{\mathrm{aux},\,x}^{s}=\mathcal{C}_{\mathrm{aux},\,y}^{s}=I$ and $\mathcal{C}_{\mathrm{ql},\,x}^{s}=\mathcal{C}_{\mathrm{ql},\,y}^{s}=I$.  
We categorize the whole scale-invariant entanglement renormalization circuit for the toric code as a conventional MERA circuit. It is a two-dimensional generalization of the one-dimensional MERA in Fig.~\ref{fig:MERA1D}, even though we do not specify which CNOT gates constitute the isometries and which CNOT gates constitute the disentanglers. The only reason why we call the circuit a MERA circuit is that it involves strictly local gates within each layer of the circuit. Since the toric code ground state has a zero correlation length, the fact that it serves as a fixed-point wavefunction of a conventional MERA circuit is consistent with the correlation length reduction argument presented in Sec.~\ref{sec:Introduction}.

\section{Lattice $p_{x}+ip_{y}$ Topological Superconductor}

\label{sec:-Topological-Superconductor}

Having introduced the entanglement renormalization circuit for the toric code model based on the conventional MERA framework in the previous section, in this section, we are going to discuss a different type of entanglement renormalization circuit in two dimensions. We will construct the scale-invariant entanglement renormalization circuit for a lattice $p_{x}+ip_{y}$ topological superconductor model, which is the most elementary non-interacting chiral topologically ordered system. We will review the model in Sec.~\ref{subsec:modelfortopological superconductor} and discuss the entanglement renormalization circuit for it in Sec.~\ref{subsec:QLEfortopological superconductor}. Following the construction of the entanglement renormalization circuit for a Chern insulator model in Ref.~\cite{Swingle2016}, the circuit will be based on the concept of adiabatic evolution. In Sec.~\ref{sec:MERAQLE}, we will  use the circuit constructed here together with the idea of conventional MERA circuits from the previous section to construct a wider class of entanglement renormalization circuits.  

\subsection{Model}

\label{subsec:modelfortopological superconductor}

\begin{figure}[t]
\begin{centering}
\includegraphics[width=\columnwidth]{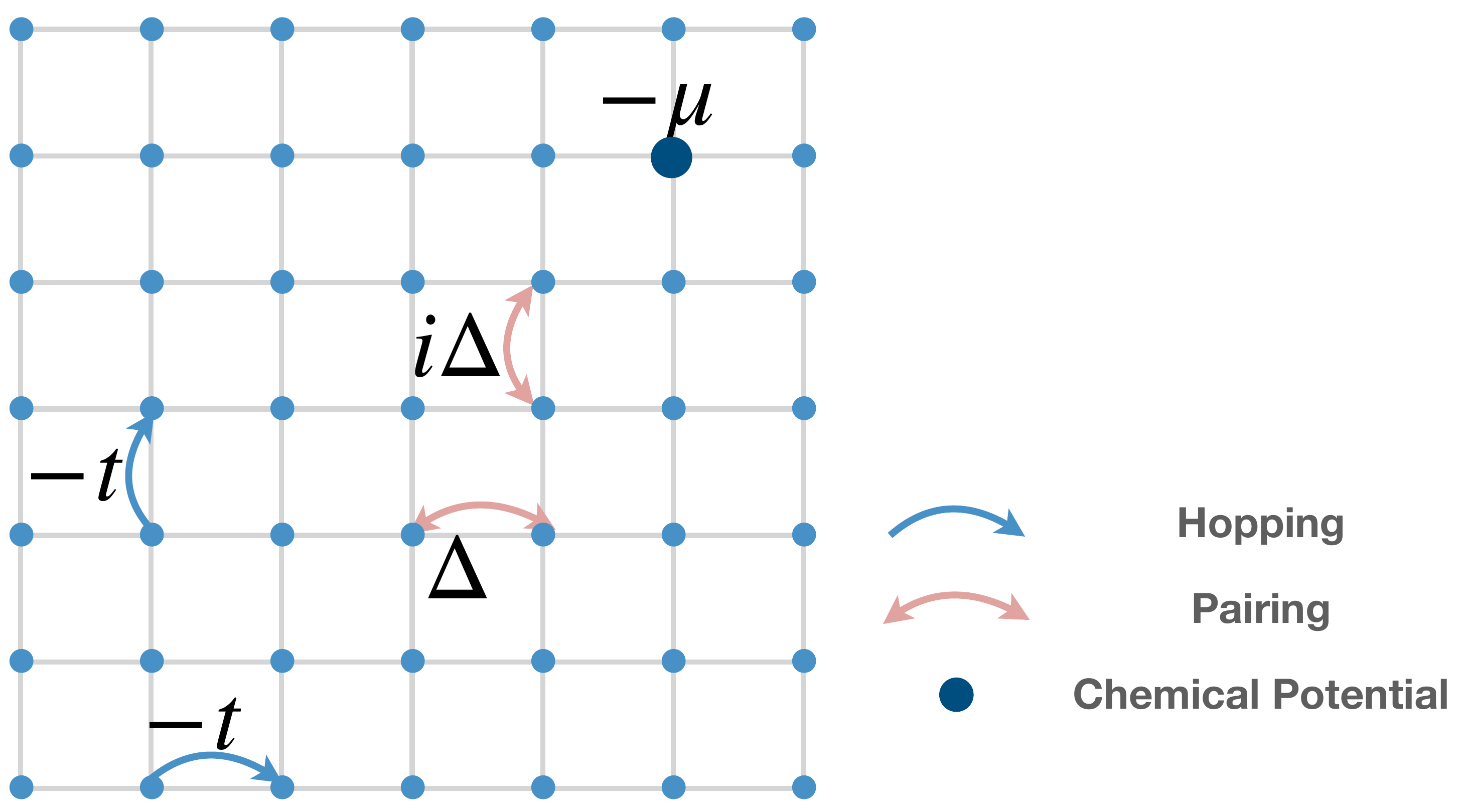}
\par\end{centering}
\centering{}\caption{An illustration of the Hamiltonian of the lattice $p_{x}+ip_{y}$ topological superconductor in Eq.~(\ref{eq:TopologicalSuperconductor-Hamiltonian}). We have nearest-neighbor hoppings with amplitude $-t$ and nearest-neighbor pairings with amplitudes $\Delta$ (for horizontal bonds) and $i \Delta$ (for vertical bonds). We also introduce a uniform chemical potential $\mu$ for each site. For hopping and pairing terms, an arrowhead pointing to a site represents a fermionic creation operator on that site, while a tail of an arrow for a hopping term represents a fermionic annihilation operator on that site. The Hermitian conjugates of the non-Hermitian hopping and pairing terms are not shown in the figure to avoid clutter, but are included in the Hamiltonian. \label{fig:topologcialsuperconductorHamiltonian}}
\end{figure}

We consider a two-dimensional $p_{x}+ip_{y}$ topological superconductor of spinless fermions on an infinite square lattice. The fermions live on the vertices. We use the vector $\mathbf{r}=(r_{x},r_{y})\in\mathbb{Z}^{2}$ to label lattice sites, where we have set the lattice spacing to one. We use $\hat{\mathbf{x}}=(1,0)$ and $\hat{\mathbf{y}}=(0,1)$ to represent horizontal and vertical unit vectors. In real space, the Hamiltonian is \cite{Bernevig2013}
\begin{align}
H_{p_{x}+ip_{y}}= & -t\sum_{\mathbf{r}}\left(c_{\mathbf{r}+\hat{\mathbf{x}}}^{\dagger}c_{\mathbf{r}}+c_{\mathbf{r}+\hat{\mathbf{y}}}^{\dagger}c_{\mathbf{r}}+{\rm h.c.}\right)-\mu\sum_{\mathbf{r}}c_{\mathbf{r}}^{\dagger}c_{\mathbf{r}}\nonumber \\
 & +\sum_{\mathbf{r}}\left(\Delta\,c_{\mathbf{r}+\hat{\mathbf{x}}}^{\dagger}c_{\mathbf{r}}^{\dagger}+i\Delta\,c_{\mathbf{r}+\hat{\mathbf{y}}}^{\dagger}c_{\mathbf{r}}^{\dagger}+{\rm h.c.}\right),\label{eq:TopologicalSuperconductor-Hamiltonian}
\end{align}
where $t$ and $\Delta$ are real positive numbers. We will refer to the parameter $\mu$ as chemical potential even though there is no charge conservation here. This parameter satisfies $-4t<\mu<0$ 
\footnote{We will treat this quadratic Hamiltonian as an exact expression for the superconducting model and not as a mean-field Hamiltonian. Therefore, we will not deal with the
gap equation in the following, and $U(1)$ charge-conservation symmetry will be explicitly broken.}. The Hamiltonian $H_{p_{x}+ip_{y}}$ is illustrated in Fig.~\ref{fig:topologcialsuperconductorHamiltonian}.

To analyze the spectrum, we perform a Fourier transformation to momentum space $\mathbf{k}=\left(k_{x},k_{y}\right)\in\left[-\pi,\pi\right)\times\left[-\pi,\pi\right)$
to obtain \cite{Bernevig2013}
\begin{align}
H_{p_{x}+ip_{y}} & =\frac{1}{2}\sum_{\mathbf{k}}(\begin{array}{cc}
c_{\mathbf{k}}^{\dagger} & c_{-\mathbf{k}}\end{array})
M_k
\left(\begin{array}{c}
c_{\mathbf{k}}\\
c_{-\mathbf{k}}^{\dagger}
\end{array}\right),\label{eq:TopologicalSuperconductor-spinorform}
\end{align}
where
\begin{align}
M_k \!=\! \left(\begin{array}{cc}
-2t (\cos k_{x}\!+\!\cos k_{y})\!-\!\mu & -i2\Delta\left(\sin k_{x}\!+\!i\sin k_{y}\right)\\
i2\Delta\left(\sin k_{x}\!-\!i\sin k_{y}\right) & 2t\cos k_{x}\!+\!2t\cos k_{y}\!+\!\mu
\end{array}\right).
\end{align}
Here and in future derivations we will be omitting the constant term. In the continuum limit, where $\mathbf{k}$ is close to $(k_{x},k_{y})=(0,0)$, we have $\sin k_{x}+i\sin k_{y}\rightarrow k_{x}+ik_{y}$. This confirms the fact that the lattice model is indeed a lattice regularization of the continuum $p_{x}+ip_{y}$ superconductor. This model can be solved by the standard Bogoliubov transformation and is gapped and topologically nontrivial with a nonzero spectral Chern number \cite{Bernevig2013}. If, instead of an infinite lattice, we had a lattice with a boundary, this model would have had a chiral propagating Majorana edge mode on the boundary  \cite{Bernevig2013}. The chiral central charge is $c=1/2$.
Hence, this model has chiral topological order. However, unlike the toric code model in Sec.~\ref{sec:Toric-Code}, this model does not have intrinsic topological order  in the sense that this model does not have anyonic quasiparticles. One can show that the ground state has a nonzero finite correlation length. For further details regarding the  $p_{x}+ip_{y}$ superconductor, we refer the reader to Refs.~\cite{Read2000,Nayak2008,Alicea2012,Bernevig2013}.

\subsection{Entanglement Renormalization Circuit}

\label{subsec:QLEfortopological superconductor}

We now construct the entanglement renormalization circuit for the lattice $p_{x}+ip_{y}$ topological superconductor. Similar to Sec.~\ref{subsec:MERAforToricCode}, a single layer of the entanglement renormalization procedure 
will consist of two different kinds of subcircuits. One is for a single step of horizontal entanglement renormalization; the other is for a single step of vertical entanglement renormalization.

\begin{figure}[t]
\begin{centering}
\includegraphics[scale=0.44]{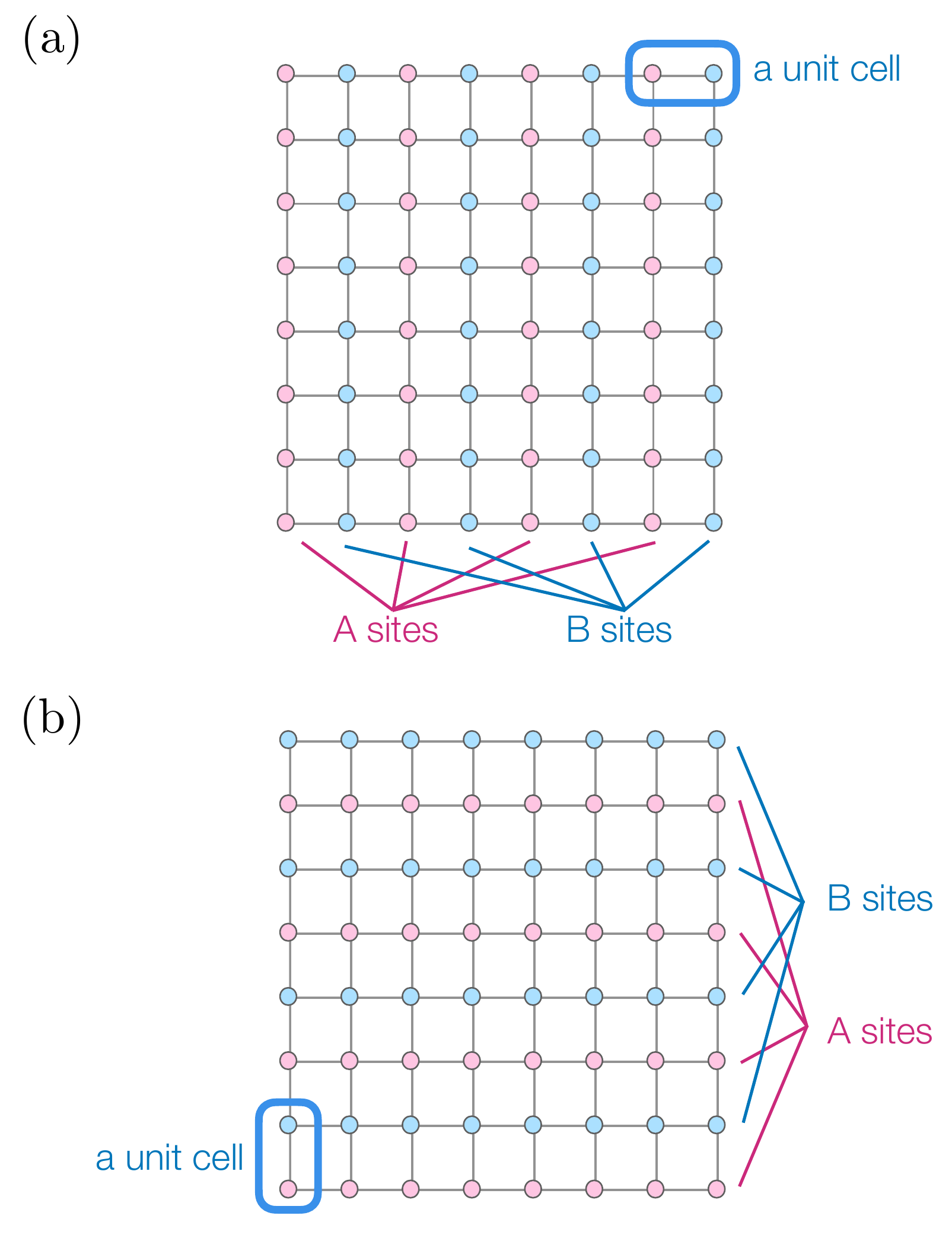}
\par\end{centering}
\centering{}\caption{
We introduce an $AB$  sublattice structure for the entanglement renormalization of the $p_{x}+ip_{y}$ topological superconductor (a) in the  horizontal direction (i.e.~$x$) and (b) in the  the vertical direction (i.e.~$y$). \label{fig:p+iprenor}}
\end{figure}

We first demonstrate how to perform a single step of horizontal entanglement renormalization of the ground state of the lattice $p_{x}+ip_{y}$ topological superconductor. This construction is a variant of the entanglement renormalization circuit for a Chern insulator model in Ref.~\cite{Swingle2016}. We introduce an $AB$ sublattice structure to the superconductor model, as shown in Fig.~\ref{fig:p+iprenor}(a).
Each unit cell has a pink $A$ site on the left and a blue $B$ site on the right. Our goal is to design a renormalization procedure that produces a superconducting state only on the pink $A$ sites, while disentangling the blue $B$ sites from the pink $A$ sites and from each other. Up to a scale transformation, the new superconductor state should be the same as the original superconductor state. The blue $B$ sites will be disentangled by ensuring they are empty. To achieve this goal for this non-interacting fermionic model, instead of using discrete strictly local gates like the ones in the Sec.\ \ref{sec:Toric-Code}, we will find an adiabatic path between the initial Hamiltonian (with every site participating in the superconducting state) and the final Hamiltonian, in which only pink $A$ sites participate in the superconducting state while blue $B$ sites are kept empty with on-site potential terms \cite{Swingle2016}. We require that the Hamiltonian gap along the entire adiabatic path between the initial and final Hamiltonians remains open in the thermodynamic limit.

In this framework, we can rewrite the initial Hamiltonian  $H_{p_{x}+ip_{y}}$ in Eq.~(\ref{eq:TopologicalSuperconductor-Hamiltonian}) using the notation induced by the $AB$ sublattice structure:
\begin{widetext}
\begin{align}
H_{p_{x}+ip_{y},\,x,\,{\rm initial}}= & -t\sum_{\mathbf{r}}\bigg(c_{B\mathbf{r}}^{\dagger}c_{A\mathbf{r}}+c_{A\mathbf{r}+\hat{\mathbf{x}}}^{\dagger}c_{B\mathbf{r}}+c_{A\mathbf{r}+\hat{\mathbf{y}}}^{\dagger}c_{A\mathbf{r}}+c_{B\mathbf{r}+\hat{\mathbf{y}}}^{\dagger}c_{B\mathbf{r}}+{\rm h.c.}\bigg)-\mu\sum_{\mathbf{r}}\left(c_{A\mathbf{r}}^{\dagger}c_{A\mathbf{r}}+c_{B\mathbf{r}}^{\dagger}c_{B\mathbf{r}}\right)\nonumber \\
 & +\sum_{\mathbf{r}}\bigg(\Delta\,c_{A\mathbf{r}+\hat{\mathbf{x}}}^{\dagger}c_{B\mathbf{r}}^{\dagger}+\Delta\,c_{B\mathbf{r}}^{\dagger}c_{A\mathbf{r}}^{\dagger}+i\Delta\,c_{A\mathbf{r}+\hat{\mathbf{y}}}^{\dagger}c_{A\mathbf{r}}^{\dagger}+i\Delta\,c_{B\mathbf{r}+\hat{\mathbf{y}}}^{\dagger}c_{B\mathbf{r}}^{\dagger}+{\rm h.c.}\bigg).\label{eq:p+ip-initial}
\end{align}
The $x$ subscript reminds us that we are performing horizontal entanglement renormalization here. In momentum space, the Hamiltonian becomes
\begin{align}
H_{p_{x}+ip_{y},\,x,\,{\rm initial}}=\frac{1}{2}\sum_{\mathbf{k}} & \left(\begin{array}{c}
c_{A\mathbf{k}}^{\dagger} \\ c_{B\mathbf{k}}^{\dagger} \\ c_{A\mathbf{-k}} \\ c_{B\mathbf{-k}}\end{array}\right)^T
\left(\begin{array}{cccc}
-2t\cos k_{y}-\mu & -t-te^{-ik_{x}} & 2\Delta\sin k_{y} & -\Delta+\Delta e^{-ik_{x}}\\
-t-te^{ik_{x}} & -2t\cos k_{y}-\mu & \Delta-\Delta e^{ik_{x}} & 2\Delta\sin k_{y}\\
2\Delta\sin k_{y} & \Delta-\Delta e^{-ik_{x}} & 2t\cos k_{y}+\mu & t+te^{-ik_{x}}\\
-\Delta+\Delta e^{ik_{x}} & 2\Delta\sin k_{y} & t+te^{ik_{x}} & 2t\cos k_{y}+\mu
\end{array}\right)\left(\begin{array}{c}
c_{A\mathbf{k}}\\
c_{B\mathbf{k}}\\
c_{A\mathbf{-k}}^{\dagger}\\
c_{B\mathbf{-k}}^{\dagger}
\end{array}\right).
\end{align}

We choose our final Hamiltonian to be
\begin{align}
H_{p_{x}+ip_{y},\,x,\,{\rm final}}= & -t\sum_{\mathbf{r}}\left(c_{A\mathbf{r}+\hat{\mathbf{x}}}^{\dagger}c_{A\mathbf{r}}+c_{A\mathbf{r}+\hat{\mathbf{y}}}^{\dagger}c_{A\mathbf{r}}+{\rm h.c.}\right)-\sum_{\mathbf{r}}\left(\mu\,c_{A\mathbf{r}}^{\dagger}c_{A\mathbf{r}}+\mu'\,c_{B\mathbf{r}}^{\dagger}c_{B\mathbf{r}}\right)\label{eq:p+ip-final}\\
 & +\sum_{\mathbf{r}}\left(\Delta\,c_{A\mathbf{r}+\hat{\mathbf{x}}}^{\dagger}c_{A\mathbf{r}}^{\dagger}+i\Delta\,c_{A\mathbf{r}+\hat{\mathbf{y}}}^{\dagger}c_{A\mathbf{r}}^{\dagger}+{\rm h.c.}\right).
\end{align}
Therefore, the Hamiltonian for the pink $A$ sites has the same form as $H_{p_{x}+ip_{y}}$ in Eq.~(\ref{eq:TopologicalSuperconductor-Hamiltonian}). The ground state for the pink $A$ sites will thus still be the original $p_{x}+ip_{y}$ topological superconductor up to a horizontal lattice rescaling. On the other hand, the blue $B$ sites in the final Hamiltonian are not coupled to pink $A$ sites or each other. We choose the chemical potential $\mu'$ for the blue $B$ sites to be negative so that they are empty (and therefore disentangled) in the ground state of $H_{p_{x}+ip_{y},\,x,\,\mathrm{final}}$. These properties make $H_{p_{x}+ip_{y},\,x,\,\mathrm{final}}$ a proper parent Hamiltonian for a horizontally entanglement renormalized $p_{x}+ip_{y}$ topological superconducting state.

In momentum space, the final Hamiltonian becomes
\begin{align}
H_{p_{x}+ip_{y},\,x,\,{\rm final}}= & \frac{1}{2}\sum_{\mathbf{k}}
\left(\begin{array}{c} c_{A\mathbf{k}}^{\dagger} \\ c_{B\mathbf{k}}^{\dagger} \\ c_{A\mathbf{-k}} \\ c_{B\mathbf{-k}}\end{array}\right)^T
\left(\begin{array}{cccc}-2t\cos k_{x}-2t\cos k_{y}-\mu & 0 & -i2\Delta\left(\sin k_{x}+i\sin k_{y}\right) & 0\\
0 & -\mu' & 0 & 0\\
i2\Delta\left(\sin k_{x}-i\sin k_{y}\right) & 0 & 2t\cos k_{x}+2t\cos k_{y}+\mu & 0\\
0 & 0 & 0 & \mu'
\end{array}\right)\left(\begin{array}{c}
c_{A\mathbf{k}}\\
c_{B\mathbf{k}}\\
c_{A\mathbf{-k}}^{\dagger}\\
c_{B\mathbf{-k}}^{\dagger}
\end{array}\right).
\end{align}
\end{widetext}

\begin{figure*}[t]
\begin{centering}
\includegraphics[scale=0.52]{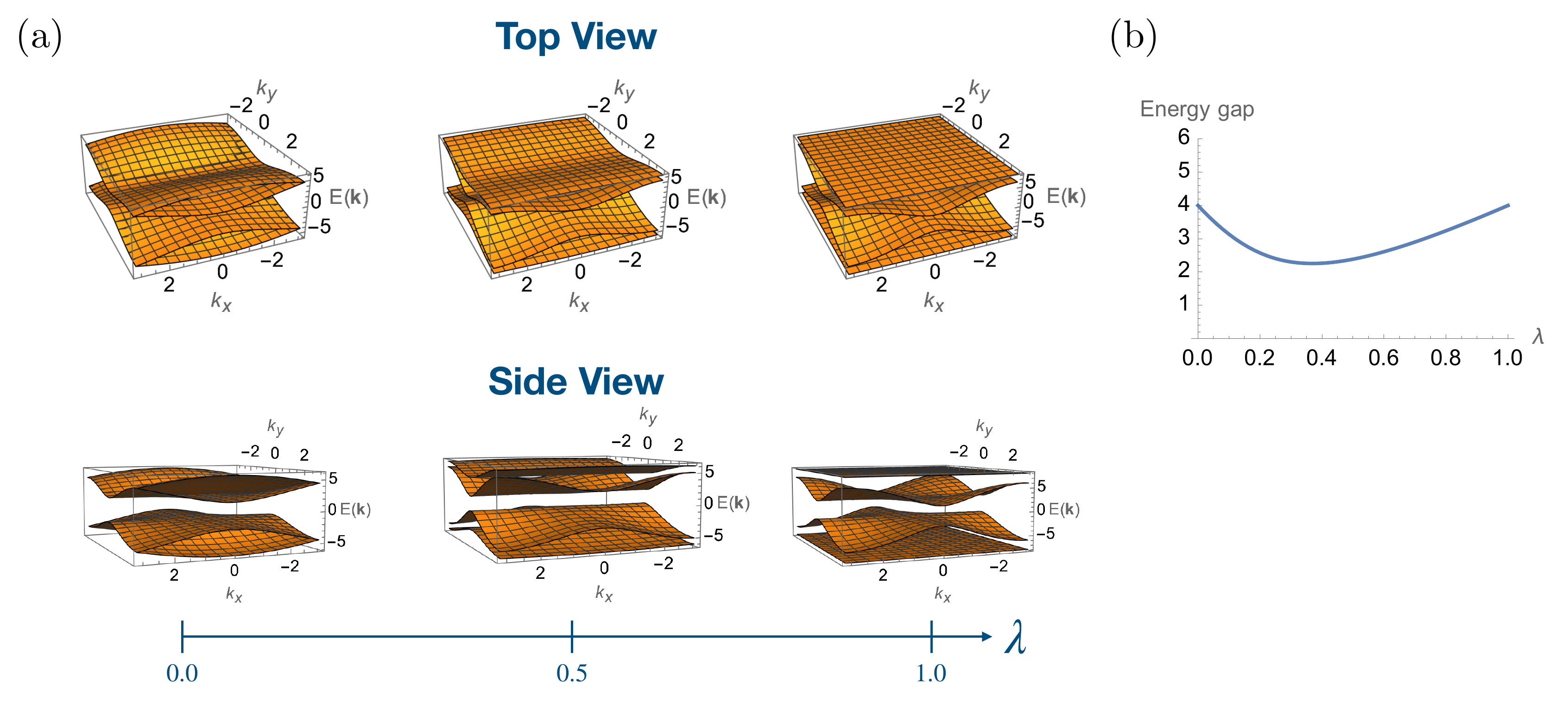}
\par\end{centering}
\centering{}\caption{(a) The Bogoliubov quasiparticle bands of the interpolating Hamiltonian $H_{p_{x}+ip_{y},\,x}(\lambda)$ in Eq.~(\ref{eq:p+ipadiabaticHamiltonian}) with $t=1.0$, $\mu=-2.0$, $\mu'=-8.0$, and $\Delta=1.0$. Along the entire path, the many-body ground state has the lower two (i.e.~negative-energy) quasiparticle bands filled and the upper two (i.e.~positive-energy) bands empty. Along the entire path, it takes a finite amount of energy to create a quasiparticle in the upper bands or remove a quasiparticle from the lower bands. This demonstrates that the many-body system is gapped throughout the adiabatic evolution for horizontal entanglement renormalization. There is no hopping or pairing for blue $B$ sites at the end of the adiabatic process, so we get two flat bands at $\lambda=1$. (b) We restrict the Hamiltonian to the same fermion parity superselection sector as the ground state and, within this restricted Hilbert space, plot the spectral gap above the many-body ground state. \label{fig:Topologicalsuperconductor-adiabaticrenorm}}
\end{figure*}

For a general system, a gapped path 
between two Hamiltonians can be hard to find. In this case, however, with a proper choice of parameters ($t=1.0$, $\mu=-2.0$, $\mu'=-8.0$, $\Delta=1.0$), a gapped path from $H_{p_{x}+ip_{y},\,x,\,\mathrm{initial}}$ to $H_{p_{x}+ip_{y},\,x,\,\mathrm{final}}$ can be found using the following simple linear interpolation:
\begin{equation}
H_{p_{x}+ip_{y},\,x}(\lambda)=(1-\lambda)H_{p_{x}+ip_{y},\,x,\,\mathrm{initial}}+\lambda\,H_{p_{x}+ip_{y},\,x,\,\mathrm{final}},\label{eq:p+ipadiabaticHamiltonian}
\end{equation}
with $\lambda\in[0,1]$. We can use the standard Bogoliubov transformation to analyze the spectrum of this Hamiltonian. As we show in Fig.~\ref{fig:Topologicalsuperconductor-adiabaticrenorm},
the system is gapped throughout the whole process. While the simple linear interpolation as in Eq.~(\ref{eq:p+ipadiabaticHamiltonian}) may not yield a gapped path for some other choices of the parameters ($t$, $\mu$, $\mu'$, $\Delta$), we can always find a gapped path by first adiabatically tuning the parameters to the case studied above, then using linear interpolation in Eq.~(\ref{eq:p+ipadiabaticHamiltonian}), and then adiabatically tuning the parameters back to the original desired set of parameters.

The renormalization along the $y$-direction is similar, and the corresponding $AB$ sublattice structure is depicted in Fig.~\ref{fig:p+iprenor}(b). Once again, the pink $A$ sites are for the remaining active fermions representing the renormalized lattice $p_{x}+ip_{y}$ topological superconductor, while the blue $B$ sites are to be emptied and thus disentangled at the end of the renormalization step. We will again use a simple linear interpolation like in Eq.~(\ref{eq:p+ipadiabaticHamiltonian}) between the initial Hamiltonian and the final renormalized Hamiltonian. In fact, if we start with the horizontal renormalization Hamiltonian $H_{p_{x}+ip_{y},\,x}(\lambda)$ and map $k_{y}\rightarrow k_{x}$, $k_{x}\rightarrow-k_{y}$, and $c_{A,B,\,\mathbf{r}}\rightarrow e^{i\pi/4}c_{A,B,\,\mathbf{r}}$, we will obtain the desired vertical renormalization Hamiltonian $H_{p_{x}+ip_{y},\,y}(\lambda)$.

Now that we have a gapped path, we could consider traditional adiabatic evolution along this path, but perfect state preparation fidelity would require perfect adiabaticity and therefore an infinite amount of time.  Instead of doing this, we will use  quasi-adiabatic evolution  \cite{Osborne2007,Hastings2010,Hastings2010b,Bachmann2012}. For any given adiabatic path of gapped Hamiltonians $H(\lambda)$ with time $\lambda \in [0,1]$, the quasi-adiabatic evolution is a unit-time evolution
\begin{equation}
{\cal U}_{{\rm qa}}={\cal T}\exp\left(i\int_{0}^{1}d\lambda\,{\cal D}(\lambda)\right)\label{eq:quasiadiabaticevolution}
\end{equation}
generated by a time-dependent Hamiltonian \footnote{Following the convention in Refs.~\cite{Osborne2007,Hastings2010,Hastings2010b}, the quasi-adiabatic evolution in Eq.~(\ref{eq:quasiadiabaticevolution}) has the opposite sign in the exponent as compared to the traditional time evolution operator. For convenience, we will still call ${\cal D}(\lambda)$ a Hamiltonian, while keeping in mind that, in order to realize ${\cal U}_{{\rm qa}}$ experimentally, we need to engineer the Hamiltonian $-{\cal D}(\lambda)$.} (see Appendix \ref{sec:appendix-quasiadiabatic} for details)
\begin{equation}
\mathcal{D}(\lambda)=-i\int_{-\infty}^{\infty}dt\,F(E_{\mathrm{gap}}t)\,e^{iH(\lambda)t}\partial_{\lambda}H(\lambda)e^{-iH(\lambda)t},\label{eq:quasi-adabatic-continuation-operator}
\end{equation}
where ${\cal T}$ denotes $\lambda$-time-ordering and where $F(x)$ is an odd function decaying subexponentially, i.e., there exist $x$-independent constants $C_\alpha$ such that $\left|F(x)\right|\leq C_{\alpha}\exp\left(-\left|x\right|^{\alpha}\right)$ for any $\alpha\in(0,1)$.
The evolution ${\cal U}_{\mathrm{qa}}$ uses unit time to take the ground state of the initial Hamiltonian $H(\lambda=0)$ to that of the final Hamiltonian $H(\lambda=1)$. The parameter $E_{\mathrm{gap}}$ of the quasi-adiabatic evolution is chosen to be the minimum energy gap between the ground state and the first excited state of $H(\lambda)$ along the entire path $\lambda\in[0,1]$. Later, we will refer to the Hamiltonian ${\cal D}(\lambda)$ as the quasi-adiabatic continuation operator. Since the quasi-adiabatic continuation operator ${\cal D}(\lambda)$ is derived from the Hamiltonian $H(\lambda)$, ${\cal D}(\lambda)$ possesses many properties similar to those of $H(\lambda)$. In particular, if $H(\lambda)$ is translationally invariant and has an on-site symmetry, the operator ${\cal D}(\lambda)$ will also be invariant under translations and the on-site symmetry. Using Lieb-Robinson bounds \cite{Lieb1972} and the fact that $F(x)$ decays subexponentially, one can also show that, if $H(\lambda)$ is a strictly local Hamiltonian (as is the case in this section), then the quasi-adiabatic continuation operator ${\cal D}(\lambda)$ is a sum of interaction terms ${\cal D}(\lambda)=\sum_{\mathbf{r}}{\cal D}_{\mathbf{r}}(\lambda)$, where each interaction term ${\cal D}_{\mathbf{r}}(\lambda)$ can be further decomposed into ${\cal D}_{\mathbf{r}}(\lambda)=\sum_{R}{\cal D}_{\mathbf{r},\,R}(\lambda)$ with the Hermitian operator ${\cal D}_{\mathbf{r},\,R}(\lambda)$ supported on sites within a distance $R$ from site $\mathbf{r}$ and satisfying $\left\Vert {\cal D}_{\mathbf{r},\,R}(\lambda) \right\Vert \leq{\cal O}([E_\mathrm{gap}]^{-\alpha}R^{-\alpha+1})$
for any integer $\alpha>1$ \cite{Osborne2007,Hastings2010b}. 
Therefore, the strength of the operator ${\cal D}_{\mathbf{r},\,R}(\lambda)$ decays with $R$ superpolynomially, i.e.~faster than any inverse power law. We call such a Hamiltonian quasi-local (see Fig.~\ref{fig:strictlocalityandquasilocality}), in contrast to the strict locality of $H(\lambda)$. We also see that a smaller minimum energy gap $E_\mathrm{gap}$ results in a slower decay of the bound on $\left\Vert {\cal D}_{\mathbf{r},\,R}(\lambda) \right\Vert $ as a function of $R$. We present further details regarding quasi-adiabatic evolution in  Appendix~\ref{sec:appendix-quasiadiabatic}.

Therefore, for a single layer of entanglement renormalization, we first apply the horizontal entanglement renormalization subcircuit ${\cal C}_{p_{x}+ip_{y},\,x}^{s}$ constructed by inserting the interpolating Hamiltonian  Eq.~(\ref{eq:p+ipadiabaticHamiltonian}) into  Eqs.~(\ref{eq:quasiadiabaticevolution},\ref{eq:quasi-adabatic-continuation-operator}). The resulting quasi-adiabatic subcircuit ${\cal C}_{p_{x}+ip_{y},\,x}^{s}$ renormalizes every other site horizontally. Then, using a similar construction, we apply the vertical entanglement renormalization subcircuit ${\cal C}_{p_{x}+ip_{y},\,y}^{s}$ to renormalize every other site vertically. We can successively apply the same horizontal and vertical entanglement renormalization subcircuits to get a quantum circuit that renormalizes the degrees of freedom at larger and larger length scales. The superscript $s\in\mathbb{N}$ labels the length scale of the entanglement renormalization layer. The full scale-invariant entanglement renormalization circuit can be succinctly written as 
\begin{equation} \label{eq:CpipProd}
\mathcal{C}_{p_{x}+ip_{y}}=\prod_{s\in\mathbb{N}}(\mathcal{C}_{p_{x}+ip_{y},\,y}^{s}\,\mathcal{C}_{p_{x}+ip_{y},\,x}^{s}),
\end{equation}
and the lattice $p_{x}+ip_{y}$ superconductor ground state is a fixed-point wavefunction under this circuit. Note that the product in Eq.~(\ref{eq:CpipProd}) is taken such that quasi-adiabatic circuits with greater $s$ appear on the left, i.e.\ act later. Therefore, we get an entanglement renormalization structure of Fig.~\ref{fig:layoutofMERAQLE} but without the auxiliary and discrete $Z_{2}^{f}$ lattice gauge theory circuit components. That is, we only have the quasi-local-evolution circuit components $\mathcal{C}_{{\rm ql},\,x}^{s}$ and $\mathcal{C}_{{\rm ql},\,y}^{s}$ in the subcircuits $\mathcal{C}_{x}^{s}$ and $\mathcal{C}_{y}^{s}$, respectively.

It is worth noting that even though the $p_{x}+ip_{y}$ topological superconductor model has a nonzero finite correlation length, it is still the fixed-point wavefunction of the entanglement renormalization circuits we constructed above. This is allowed because the quasi-adiabatic circuits are quasi-local (see Appendix~\ref{sec:appendix-quasiadiabatic}) and the formula $\ell'=\ell/b,\,b>1$ for conventional MERA circuits with strictly local gates, like the ones presented in Sec.~\ref{subsec:MERAforToricCode}, does not work here. Intuitively speaking, we can interpret the result as follows: even though the re-scaling procedure on the lattice shrinks the correlation length between sites by a factor $b$, the quasi-locality of the quasi-adiabatic circuit adds some correlation to the system to remedy that loss of correlation.

\section{Gauging Fermion Parity Symmetry and Bosonization}

In the previous section, we considered a simple non-interacting chiral topologically ordered model and its entanglement renormalization circuit. However, our goal in this paper is to construct circuits for interacting chiral topologically ordered models. In this section, we therefore introduce a formalism involving gauging the fermion parity symmetry to construct several exactly solvable interacting chiral topologically ordered models. The procedure can be conveniently simplified by a procedure called bosonization. In Sec.~\ref{subsec:Bosonization-Formalism}, we review the formalism of gauging the fermion parity symmetry and bosonization. Then, in Secs.~\ref{subsec:Gauged-trivial-fermionic}, \ref{subsec:Gauged-superconductor}, and \ref{subsec:Kitaev's-sixteen-foldway}, we use the formalism to construct interacting spin models, some of which have chiral topological order. The models will be presented in the order of increasing construction complexity. The exact solvability of these models will be used in Sec.~\ref{sec:MERAQLE} to analytically construct their entanglement renormalization circuits.

\label{sec:Gauging-Fermion-Parity}

\subsection{Formalism}

\label{subsec:Bosonization-Formalism}
In this subsection, we review how to obtain an interacting bosonic system from a 
two-dimensional fermionic lattice system by gauging the fermion parity symmetry. Even though a quadratic fermionic system with pairing does not conserve total fermion number ${N}_\mathrm{total}$,
it still conserves global $Z_2$ fermion parity $(-1)^{{N}_\mathrm{total}}$.  
We can gauge this $Z_2$ symmetry by coupling the fermionic system to a $Z_2$ gauge field (subject to a Gauss's law constraint), making the symmetry transformation local \cite{Kogut1975}.
Specifically, we will introduce new dynamical variables representing the gauge field and living on the edges connecting the original fermionic lattice sites such that the new system is invariant under local symmetry transformations. We will refer to these local symmetry transformations as local gauge transformations. In general,
gauging a symmetry of a gapped quantum system allows us to construct a new topological phase of matter \cite{Levin2012,Chen2016,Barkeshli2019}.
Using this approach, together with an additional ingredient of penalizing non-zero fluxes (to be discussed below), 
we will build in the following subsections a wide class of lattice models with nontrivial topological properties. In this subsection, we will also discuss how to reformulate 
the gauged theory in a purely bosonic language \cite{Chen2018a,Chen2019Bosonization3D,Chen2020} with spin-1/2 particles (or hard-core bosons). This shows how gauged fermionic theories naturally arise when studying quantum spin systems. 
If the original fermionic system is gapped, the resulting spin system is, in fact, a gapped quantum spin liquid  \cite{Kitaev2006,Savary2016}.

\begin{figure}[t]
\begin{centering}
\includegraphics[width=\columnwidth]{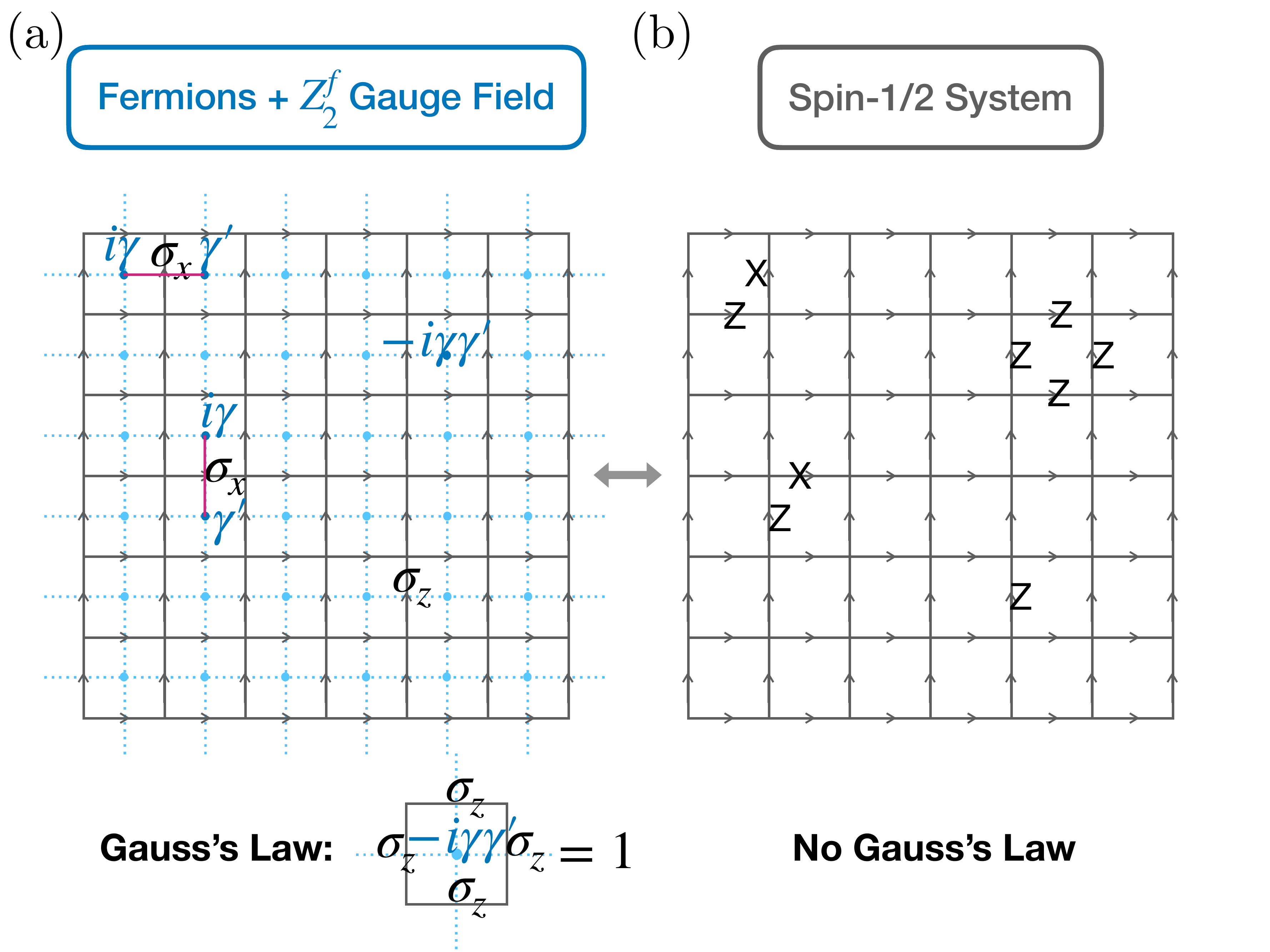}
\par\end{centering}
\centering{}\caption{
We associate an orientation with every edge of the square lattice. (a) shows fermions coupled to the $Z_{2}^{f}$ gauge field. The fermions live on the faces, while the $Z_{2}^{f}$ gauge field lives on the edges. The Gauss's law constraint is imposed on the Hilbert space. We can then rewrite the theory purely in terms of spins (qubits) living on the edges of the square lattice and remove the Gauss's law constraint, as shown in (b). The generators of the fermionic theory coupled to the $Z_{2}^{f}$ gauge field in (a) are mapped to the corresponding spin operators in (b). The ordering of the Majorana operators is defined in the text.
\label{fig:branchingsquare}}
\end{figure}

Before discussing the gauging of fermion parity, we first associate an orientation with every edge, as shown in Fig.~\ref{fig:branchingsquare}(a).
The fermions  live on the faces, i.e., on the sites of a dual lattice. Following the convention of Ref.~\cite{Chen2018a}, we decompose complex fermion operators into Majorana operators $c_{f}=(\gamma_{f}+i\gamma_{f}')/2$, $c_{f}^{\dagger}=(\gamma_{f}-i\gamma_{f}')/2$, where $f$ denotes faces. (Note that, throughout the manuscript, the symbol $f$ in a superscript denotes fermion-related objects. On the other hand, the symbol $f$ appearing as a subscript or in the normal line of type denotes a face on a square lattice.) The Majorana operators are Hermitian: $\gamma_{f}^{\dagger}=\gamma_{f}$ and $\gamma_{f}^{\prime\dagger}=\gamma_{f}^{\prime}$. Their anti-commutation relations are
\begin{align}
\{\gamma_{f},\gamma_{f'}\} & =2\delta_{f,f'},\nonumber \\
\{\gamma_{f},\gamma'_{f'}\} & =0.
\end{align}

Denoting by $N_{f}=c_{f}^{\dagger}c_{f}$ the fermion number operator on face $f$, the fermion parity operator on that face is $(-1)^{N_{f}}=-i\gamma_{f}\gamma_{f}^{\prime}$. We will refer to the operator $S_{e}=i\gamma_{L(e)}\gamma_{R(e)}^{\prime}$ as a Majorana hopping operator, where the edge $e$ is shared by the adjacent left face $L(e)$ and the adjacent right face $R(e)$ defined with respect to the edge orientation.
Note that the fermion parity operators on all faces and the Majorana hopping operators on all edges together generate the whole algebra that preserves the global fermion parity.

To gauge the $Z_2$ fermion parity symmetry of a fermionic system described by a Hamiltonian,  
we couple the fermions living on the faces to a $Z_{2}$ gauge field living on the edges. To emphasize that the $Z_{2}$ gauge field is related to fermion parity, we will refer to it as the $Z_{2}^{f}$ gauge field with an $f$ superscript. We will use $\sigma_{x,e}$, $\sigma_{y,e}$, and $\sigma_{z,e}$ to denote the Pauli matrices of the $Z_{2}^{f}$ gauge variables on edge $e$. By analogy with the ordinary $U(1)$ electromagnetism theory, we define the local gauge transformation operator acting on the fermion mode on face $f$ and the nearby $Z_{2}^{f}$ gauge variables as 
\[
G_{f}\equiv(-1)^{N_{f}}\prod_{e\in  f}\sigma_{z,e}.
\]
This operator flips the sign of the fermion mode $c_{f}$ and the surrounding gauge variables $\sigma_{x,e}$. Roughly speaking, $\sigma_{x,e}$ is the discrete (meaning a discrete gauge group) analog of $e^{iA(e)}$ in the ordinary $U(1)$ electromagnetism theory, and $\sigma_{z,e}$ the discrete analog of $e^{iE(e)}$ \cite{Kogut1975, Smit2002, Fradkin2013, Savary2016},
where $A(e)$ and $E(e)$ are, respectively, the vector potential and the electric field on edge $e$ with the lattice constant equal to one.
The local fermion parity operator $(-1)^{N_{f}}$ behaves as a local charge operator in this discrete theory. The presence of the operator $(-1)^{N_{f}}$ in $G_f$ is an indication that we are making the original global symmetry transformation $(-1)^{N_\mathrm{total}}$ local. Note that $G_{f}$ is both Hermitian and unitary. Now we demand that the physics should not change under gauge transformations, so all physical operators must commute with $G_{f}$ \cite{Kogut1975}. Therefore, to be invariant under all gauge transformations $G_f$, 
the original fermionic Hamiltonian must be modified by inserting gauge variables. Unless the fermion terms are on-site, for a generic term involving distant $n$-body ($n\in2\mathbb{N}$) fermionic interactions, we need to replace fermion operators with Wilson lines by inserting a string of gauge variables $\sigma_{x,e}$ along a path connecting the fermion operators. For example, a two-body operator $\gamma_{f_{1}}\gamma_{f_{2}}'$ with $f_{1}\neq f_{2}$ should be replaced with $\gamma_{f_{1}}\gamma_{f_{2}}'\left(\prod_{e\in\mathrm{path}(f_{1},f_{2})}\sigma_{x,e}\right)$. The notation $\mathrm{path}(f_{1},f_{2})$ denotes an unoriented path on the dual square lattice connecting faces $f_{1}$ and $f_{2}$. The notation $e\in\mathrm{path}(f_{1},f_{2})$ denotes edges on the square lattice that  $\mathrm{path}(f_{1},f_{2})$ crosses. Similarly, a four-body operator $\gamma_{f_{1}}\gamma_{f_{2}}'\gamma_{f_{3}}\gamma_{f_{4}}'$ with $f_{1}$, $f_{2}$, $f_{3}$, $f_{4}$ all being different (a sufficient but not necessary condition)
should be replaced with $\gamma_{f_{1}}\gamma_{f_{2}}'\gamma_{f_{3}}\gamma_{f_{4}}'\left(\prod_{e\in\mathrm{path}(f_{1},f_{2})}\sigma_{x,e}\right)\left(\prod_{e\in\mathrm{path}(f_{3},f_{4})}\sigma_{x,e}\right)$. We will use the notation 
$\mathcal{O}^{\mathrm{gauged}}$ or $\left\{ \mathcal{O}\right\}^{\mathrm{gauged}}$ to denote the gauge-invariant version (obtained using the above procedure) of a fermion-parity-conserving operator $\mathcal{O}$.

Let us now apply the above procedure to the generators of our fermion theory. Since the fermion parity operator $(-1)^{N_{f}}$ commutes with $G_f$, we can keep $(-1)^{N_{f}}$ unchanged. However, a Majorana hopping operator $S_{e}=i\gamma_{L(e)}\gamma_{R(e)}^{\prime}$ does not commute with $G_f$. So, we replace it with a gauge-invariant operator $S_{e}^{\mathrm{gauged}}=i\gamma_{L(e)}\sigma_{x,e}\gamma_{R(e)}^{\prime}$
instead, which is the shortest Wilson line. A generic Wilson line can be therefore decomposed into products of $(-1)^{N_{f}}$ and $S_{e}^{\mathrm{gauged}}$. In addition to $(-1)^{N_{f}}$ and $S_{e}^{\mathrm{gauged}}$, the operator $\sigma_{z,e}$ also commutes with the local gauge transformation operators $G_f$. In fact, all operators commuting with local gauge transformations are generated by $(-1)^{N_{f}}$, $S_{e}^{\mathrm{gauged}}$, and $\sigma_{z,e}$.

By analogy with the ordinary $U(1)$ electromagnetism theory, we now impose a discrete Gauss's law constraint on the system \cite{Kogut1975, Smit2002, Fradkin2013, Savary2016}:
\[
\prod_{e\in f}\sigma_{z,e}=(-1)^{N_{f}},\,\forall f.
\]
Note that both sides of this equation are gauge-invariant physical operators. This equation relates the parity of the local charge operator $N_f$ to the discrete electric field variables $\sigma_{z,e}$.
This is the exponentiation of the lattice discrete (discrete in the sense of the discrete gauge group)
analog of the familiar Gauss's law of the continuum $U(1)$ electromagnetism theory: $\nabla\cdot E=Q/\epsilon_{0}$. Equivalently, we can write the Gauss's law constraint as
\begin{equation}
(-1)^{N_{f}}\prod_{e\in f}\sigma_{z,e}\equiv G_{f}=1,\,\forall f.
\end{equation}

That is, the only allowed quantum states $\ket{\psi^{f, Z_2^f}}$
are those invariant under local gauge transformations: $G_f \ket{\psi^{f, Z_2^f}}=\ket{\psi^{f, Z_2^f}}$. (Note that here $f$ in the subscript of $G_f$ refers to face $f$, while the two instances of $f$ in the superscript of $\ket{\psi^{f, Z_2^f}}$ stand for fermions.)
Note that, due to this constraint, the generator $(-1)^{N_{f}}$ is no longer a fundamental generator of the operator algebra since it is equivalent to the composite operator $\prod_{e\in f}\sigma_{z,e}$, which is built from the four nearby $\sigma_{z,e}$ operators.

\begin{figure}[t]
\begin{centering}
\includegraphics[scale=0.4]{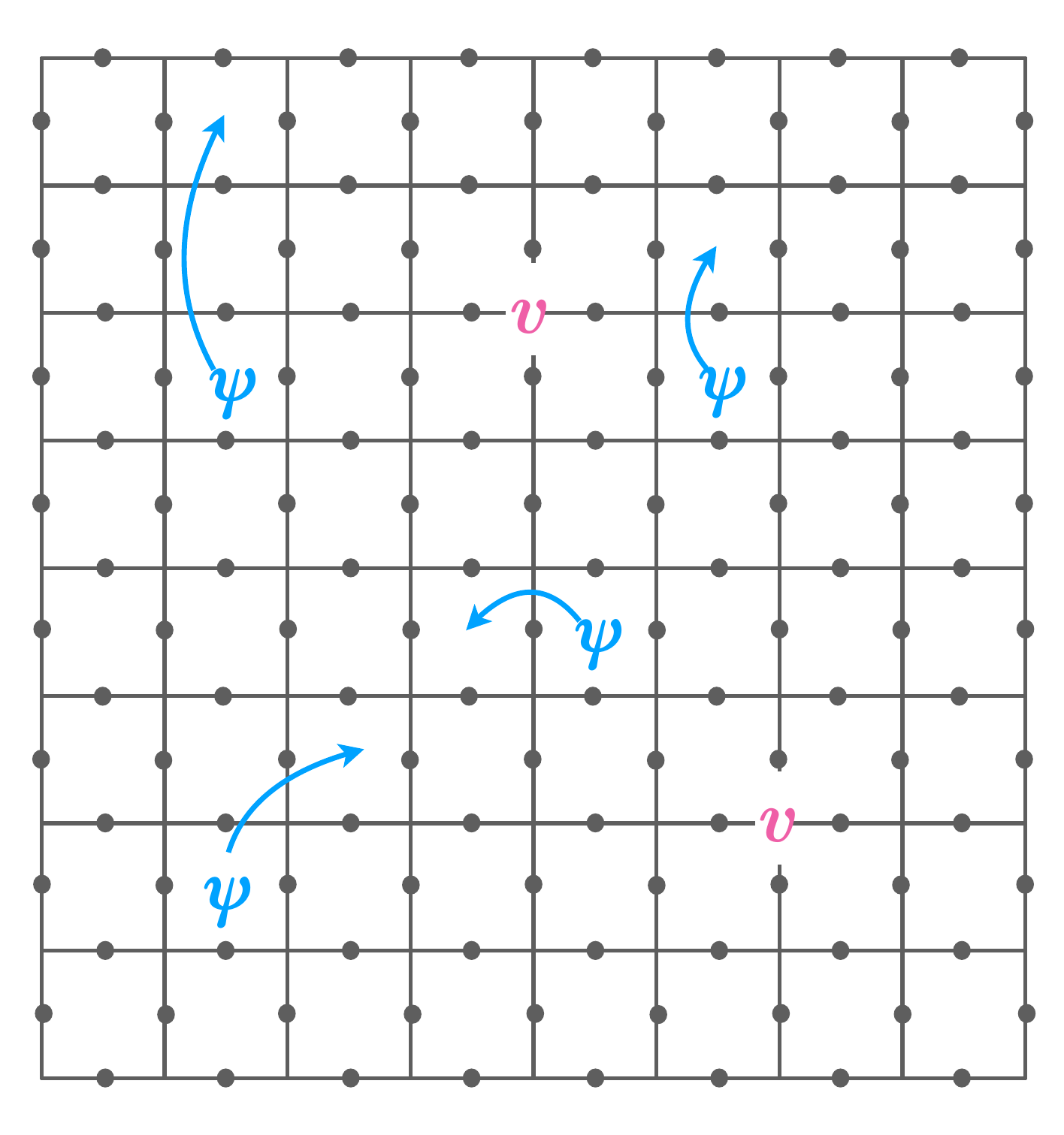}
\par\end{centering}
\centering{}\caption{We can have fermions $\psi$ on the faces and vortices $v$ on the vertices as emergent quasiparticles. Vortices come from the existence of 
$\pi$ fluxes of the $Z_{2}^{f}$ gauge field at the vertices. A vortex could be a composite object that is not strictly localized to a small region, so,  when we associate a vortex with a vertex, we only talk about its center-of-mass position or the position of the flux at the core.  A vortex and an anti-vortex might or might not be the same particle, depending on how the fermions are organized around a flux.
\label{fig:picturesforemergentparticles}}
\end{figure}

The $Z_{2}^{f}$ flux within the smallest loop encircling a vertex $v$ is measured by the gauge-invariant flux measuring operator
\begin{equation}
\Phi_{v}\equiv\prod_{e\in v}\sigma_{x,e}.\label{eq:fluxmeasure}
\end{equation}
It picks up a minus sign in the presence of a flux at the vertex $v$; otherwise, it gives $+1$. For historical reasons, we sometimes call a $Z_{2}^{f}$ flux a $\pi$ flux \footnote{This is because $\Phi_{v}\equiv\prod_{e\in v}\sigma_{x,e}$ can be roughly thought of as the analog of the quantity $\cos\left(\nabla\times A\right)=\cos\left(B\right)$ for $U(1)$ lattice gauge theory, where $A$ is the vector potential on links and $B$ is the magnetic flux through plaquettes. The quantity $\cos\left(\nabla\times A\right)$ becomes $-1$ when the magnetic flux $B$ is $\pi$.}. Note that flux measuring operators commute with each other and with Wilson lines. We now add a Hamiltonian term that energetically penalizes non-zero fluxes: 
\begin{equation}
H_{{\rm penalty}}^{\Phi}=-\Delta_{\Phi}\sum_{v}\Phi_{v}=-\Delta_{\Phi}\sum_{v}\prod_{e\in v}\sigma_{x,e},\label{eq:fluxpenalty}
\end{equation}
with $\Delta_{\Phi}>0$. If the flux energy parameter $\Delta_{\Phi}$ is large enough, the flux penalty Hamiltonian $H_{{\rm penalty}}^{\Phi}$ ensures that, in the low energy subspace, there is no flux anywhere: $\Phi_{v}=1,\,\forall v$. However, we can still consider
violations of this condition as vortex excitations of the theory. Note that a pair of fluxes can be created by applying a string of $\sigma_{z,e}$ operators, which is gauge-invariant and anti-commutes with the flux measuring operators $\Phi_{v}$ at the endpoints of the string. A vortex excitation is then typically found by solving for the ground state of fermions in the presence of a single $\pi$ flux with the other $\pi$ fluxes far away. Therefore, a vortex can be a composite object consisting of the flux and the response of the fermions to it. Hence, in this new theory, we have not only fermions living on faces but also vortex quasiparticles living on vertices, as shown in Fig.~\ref{fig:picturesforemergentparticles} \footnote{We follow the terminology of Ref.~\cite{Kitaev2006} to refer to the quasiparticles related to $\pi$ fluxes as vortices despite the fact that there might be no obvious definition in terms of a winding of a local order parameter.}.
In this paper, we will assume that the flux energy parameter $\Delta_{\Phi}$ is much greater than all the fermionic interactions, leading to a large energy gap for the vortices.

Here, we have to point out that, when we introduce gauge variables to gauge fermion operators, there can be many equivalent ways of writing down a gauge-invariant operator $\left\{ \mathcal{O}\right\} ^{\mathrm{gauged}}$ corresponding to the fermion operator $\mathcal{O}$ 
involving distant fermionic degrees of freedom. For example, the gauge-invariant operator $\left\{ \gamma_{f_{1}}\gamma_{f_{2}}'\right\} ^{\mathrm{gauged}}$ corresponding to the two-body operator $\gamma_{f_{1}}\gamma_{f_{2}}'$ with distant faces $f_{1}$ and $f_{2}$ can be written using any path $\mathrm{path}(f_{1},f_{2})$ on the dual lattice connecting the faces, even though in practice one might conveniently choose the shortest path. For a four-body operator, $\gamma_{f_{1}}\gamma_{f_{2}}'\gamma_{f_{3}}\gamma_{f_{4}}'$, there are even more equivalent choices for making the gauge-invariant operator. For example, instead of picking $\gamma_{f_{1}}\gamma_{f_{2}}'\gamma_{f_{3}}\gamma_{f_{4}}'\left(\prod_{e\in\mathrm{path}(f_{1},f_{2})}\sigma_{x,e}\right)\left(\prod_{e\in\mathrm{path}(f_{3},f_{4})}\sigma_{x,e}\right)$ as $\left\{ \gamma_{f_{1}}\gamma_{f_{2}}'\gamma_{f_{3}}\gamma_{f_{4}}'\right\} ^{\mathrm{gauged}}$, we can equivalently choose $\gamma_{f_{1}}\gamma_{f_{2}}'\gamma_{f_{3}}\gamma_{f_{4}}'\left(\prod_{e\in\mathrm{path}(f_{1},f_{3})}\sigma_{x,e}\right)\left(\prod_{e\in\mathrm{path}(f_{2},f_{4})}\sigma_{x,e}\right)$ or $\gamma_{f_{1}}\gamma_{f_{2}}'\gamma_{f_{3}}\gamma_{f_{4}}'\left(\prod_{e\in\mathrm{path}(f_{1},f_{4})}\sigma_{x,e}\right)\left(\prod_{e\in\mathrm{path}(f_{2},f_{3})}\sigma_{x,e}\right)$. A large flux energy parameter $\Delta_{\Phi}$ is important since it implies that different choices of $\mathcal{O}$ are equivalent at low energies due to the zero-flux condition $\Phi_{v}=1$.  
The equivalence between different choices simply comes from the lattice discrete 
analog of Stokes' theorem in electromagnetism.

\begin{figure}[t]
\begin{centering}
\includegraphics[width=\columnwidth]{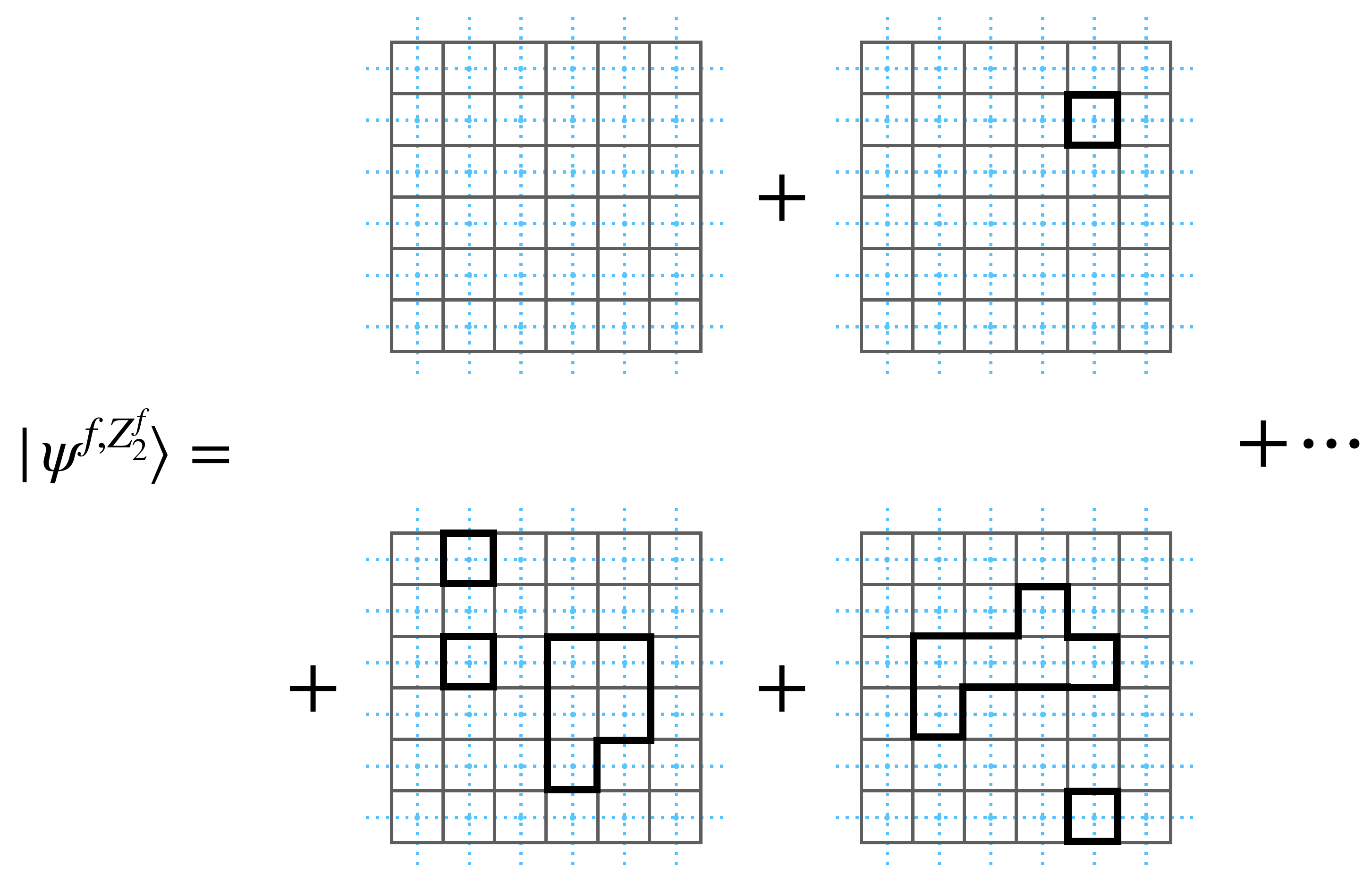}
\par\end{centering}
\centering{}\caption{
A schematic representation of the wavefunction $\ket{\psi^{f, Z_2^f}}$ in Eq.\ (\ref{eq:symmetrizedstate}). It is an equally weighted superposition of wavefunctions with different zero-flux gauge field configurations. We draw a solid black line along each edge with $\sigma_{x,e}=-1$. There is no solid black line along edges with $\sigma_{x,e}=1$. In this schematic picture, we omit the
modification of the fermionic wavefunction coming from the fermion parity operator in the local gauge transformation operator $G_f$. Therefore, the wavefunction $\ket{\psi^{f, Z_2^f}}$ can be thought of as a condensate of loops. 
We have dropped the overall normalization constant
in this plot.  \label{fig:loopcondensation}}
\end{figure}

A large flux energy parameter $\Delta_{\Phi}$ also allows us to solve the theory as if we only had fermions and no gauge field at low energies.
Under the zero-flux condition $\Phi_{v}=1$,
rather than directly dealing with the theory of fermions coupled to the gauge field,
we can observe the following. We first ignore the Gauss's law constraint and observe that any gauge field variable $\sigma_{x,e}$ commutes with all the flux measuring operators $\Phi_{v}$ and all the Wilson line operators derived from the original fermionic Hamiltonian \cite{Kitaev2006,Kitaev2009,Pachos2012}. Since $\sigma_{x,e}^{2}=1$, the Hamiltonian can be block-diagonalized into sectors labeled by the global gauge field configurations \{$\sigma_{x,e}\,|\, \sigma_{x,e}=\pm1$\}. We can satisfy the zero-flux condition by assigning proper gauge field configurations. The simplest gauge field configuration satisfying the zero-flux condition is $\sigma_{x,e}=1$ on every edges, in which case the Wilson lines now involve only the original fermionic operators. This means that, by this gauge fixing, we return the Hamiltonian of the gauged fermionic system back to the Hamiltonian of the original fermionic system up to a constant $-\Delta_{\Phi}\sum_{v}1$ coming from the flux penalty Hamiltonian. Therefore, we just need to solve the original fermionic theory. Denoting by $\ket{\psi^{f}}$ an eigenstate 
of the fermionic system, the corresponding eigenstate of the theory of fermions coupled to the $Z_2^f$ gauge field is simply $\ket{\psi^{f}}\bigotimes_{\forall e}\ket{\sigma_{x,e}=1}$, where we inserted gauge variables $\sigma_{x,e}=1$ of the edges into the state. However, due to the commutativity of local gauge transformations $G_{f}$ with other physical gauge-invariant operators, all states obtained by applying any product of local gauge transformations $\{G_{f}\}$ on $\ket{\psi^{f}}\bigotimes_{\forall e}\ket{\sigma_{x,e}=1}$ are also legitimate eigenstates. 
In each such legitimate eigenstate, the signs of gauge fields along some closed loops are flipped, and the fermionic state (written in terms of complex creation operators acting on the vacuum) has $c_f^\dagger \rightarrow -c_f^\dagger$ for faces $f$ inside the closed loops. 
All the above states corresponding to different choices of the product of $\{G_{f}\}$ are orthogonal to each other, so they span a large vector space. To remove the degeneracy, we now impose the Gauss's law constraint $G_{f}=1$. The only state satisfying this constraint within the large degenerate vector space defined above is an equally weighted superposition of all possible gauge-transformed states: 
\begin{align}
 \ket{\psi^{f, Z_2^f}}
=
\left(\prod_{f}\left(\frac{1+G_{f}}{\sqrt{2}}\right)\right)\ket{\psi^{f}}\bigotimes_{\forall e}\ket{\sigma_{x,e}=1}.
\label{eq:symmetrizedstate}
\end{align}
We present this state schematically in Fig.~\ref{fig:loopcondensation}.
To conclude, we can solve the gauged Hamiltonian at low energies with large enough $\Delta_{\Phi}$ by first solving the original fermionic system without the gauge field and then symmetrizing the wavefunction as in Eq.~(\ref{eq:symmetrizedstate}) by inserting gauge variables in the trivial states and summing over all states connected by local gauge transformations.

We will now show that the $Z_{2}^{f}$ gauge theory with fermions can be exactly rewritten 
purely in terms of spins (or hard-core bosons) on the edges of the square lattice with the same assignment of edge orientations. To do this, we demonstrate how the generators are mapped into the pure spin language. We map the shortest Wilson line involving nearest-neighbor Majorana hopping as follows:
\begin{equation}
S_{e}^{\mathrm{gauged}}=i\gamma_{L(e)}\sigma_{x,e}\gamma'_{R(e)}\rightarrow U_{e}:=X_{e}Z_{r(e)}.\label{eq:majoranahoppingmaptobosons}
\end{equation}
We have used the notation (shown in Fig.~\ref{fig:branchingsquare})
that, if $e$ is oriented east, $r(e)$ is the north-oriented edge  whose arrowhead is at the tail of the $e$ arrow.
If $e$ is oriented north, $r(e)$ is the east-oriented edge whose arrowhead is at the tail of the $e$ arrow. 
We have chosen a different notation ($X_{e}$ and $Z_{e}$) for the operators in the pure spin systems to distinguish them from the operators ($\sigma_{x,e}$ and $\sigma_{z,e}$) of the $Z_{2}^{f}$ gauge field, even though they are related. Note that the commutation relations between the operators $U_{e}$ on different edges are the same as those of $S_{e}^{\mathrm{gauged}}$. For a face $f$, the fermion parity operator is mapped as follows: 
\begin{equation}
(-1)^{N_{f}}=-i\gamma_{f}\gamma_{f}'\rightarrow W_{f}=\prod_{e\in f}Z_{e}.\label{eq:fermionparitymaptobosons}
\end{equation}
We will call $W_f$ an emergent fermion parity operator and a bosonized fermion parity operator interchangeably. The word ``bosonized" will be explained later. 
The flux creation operator remains the same:
\begin{equation}
\sigma_{z,e}\rightarrow Z_{e}.\label{eq:vortexcreationmap}
\end{equation}
The mapping is summarized in Fig.~\ref{fig:branchingsquare}. One can verify that the mapping above gives an algebra isomorphism 
by checking that it induces an algebra homomorphism (preserves the algebraic structure of the physical operators) and is injective (by construction) and surjective (all the operators on the pure spin side can be generated by $U_{e}$, $W_{f}$, and $Z_{e}$). Note that the Gauss's law constraint for the operators in the gauged fermion theory is trivially satisfied in the pure spin language.

Note that any long Wilson line can be decomposed into $S_{e}^{\mathrm{gauged}}$ and $(-1)^{N_{f}}$, so it can always be mapped to the pure spin language. As a special case, the original flux measuring operator in Eq.~(\ref{eq:fluxmeasure}) can be decomposed into a product of Wilson line segments and fermion parity operators. Using Eq.~(\ref{eq:majoranahoppingmaptobosons}) and Eq.~(\ref{eq:fermionparitymaptobosons}), we can see that the flux measuring operator is mapped as follows: 
\begin{equation}
\Phi_{v}\equiv\prod_{e\in v}\sigma_{x,e}\rightarrow F_{v}\equiv W_{NE(v)}\prod_{e\in v}X_{e},\label{eq:penalizationmaptobosons}
\end{equation}
where $NE(v)$ indicates the face directly to the northeast of vertex $v$, and the $W$ operator is defined in Eq.~(\ref{eq:fermionparitymaptobosons}).
The flux penalty Hamiltonian thus becomes 
\[
H_{{\rm penalty}}^{F}=-\Delta_{\Phi}\sum_{v}W_{NE(v)}\prod_{e\in v}X_{e}.
\label{eq:Ffluxpenalty}
\]
The zero-flux condition $\Phi_{v}=1,\,\forall v$ at low energies is translated into $F_{v}=1,\,\forall v$.

To make sure there is a duality
between the two theories, we also have to know the mapping of the Hilbert spaces. This is done by requiring that the quantum state that is an eigenvalue-one eigenstate of all $(-1)^{N_{f}}$, $\Phi_{v}$, and $G_{f}$ operators (i.e.~the fermion vacuum state with zero flux and Gauss's law constraints) is mapped to the eigenvalue-one eigenstate of all the $W_{f}$, $F_{v}$ 
\footnote{To be precise, assuming an infinite 2D lattice, we also require that the quantum state that is being mapped here is an eigenvalue-one eigenstate of any infinitely long Wilson line consisting of a string of $\sigma_{x,e}$. We also require that the quantum state that it is mapped to by Eq.~(\ref{eq:majoranahoppingmaptobosons}) and Eq.~(\ref{eq:fermionparitymaptobosons})  is an eigenvalue-one eigenstate of
the operators corresponding to these infinitely long Wilson lines. These requirements are imposed since we want to rule out the possibility of a pair (or pairs) of fluxes at two points at infinity.}. The former state is nothing but a loop condensate of the gauge field shown in Fig.~\ref{fig:loopcondensation} with fermions in the vacuum state, while the latter has a similar loop-condensate picture in the basis of $X_{e}$ but without fermions. Knowing the mapping of a single state, we can derive the mapping of the other states by applying the generators to the states on both sides. One can verify that the number of degrees of freedom on both sides matches. For the gauged fermion theory, we have $1\,(\mathrm{fermion})+2\,(\mathrm{gauge\,field})-1\,(\mathrm{Gauss's\,law})=2$ spins  for each vertex, which matches what we have in the pure spin system.

Let us now summarize what we have done. We have successfully mapped a fermionic theory coupled to a $Z_{2}^{f}$ gauge field on a lattice (subject to the Gauss's law constraint)
to a theory of spins \footnote{In fact, conversely, any theory of $S=1/2$ spins on a square lattice can be dual to a theory of fermions coupled to a $Z_{2}^{f}$ gauge field (subject to Gauss's law constraint). 
However, this duality might not be especially useful if the gauge flux for each vertex is not $0$ or $\pi$ at low energies, in which case we have to deal with a difficult problem of fermions coupled to a fluctuating gauge field with the Gauss's law constraint.}. Notice the similarity of $\sigma_{e,x}$ and $\sigma_{e,z}$ to $X_{e}$ and $Z_{e}$. We can interpret the mapping as ``integrating out'' the fermions, where we remove the Gauss's law constraint and rewrite the whole theory, including the fermions and the gauge field, in terms of the gauge field with purely bosonic $Z_{2}$ variables. This kind of statistical transmutation between fermions and bosons by coupling fermions to a gauge field 
can be traced back at least to the composite fermion story in fractional quantum Hall systems \cite{Jain2007,Jain2020}. What is interesting here is that the statistical transmutation is done exactly on a lattice instead of working at the continuum field-theory level. In hindsight, it is also physically clear why a toric-code-like spin theory should be able to represent fermionic degrees of freedom \footnote{We observe that a bosonized model is toric-code-like since it involves a large number of mutually commuting projectors $F_{\nu}$ serving as good quantum numbers of the Hamiltonian, similar to the plaquette operators or vertex operators in the toric code model. In fact, each $F_{v}$ is composed of a plaquette operator and a neighboring vertex operator. The reader will also see the connection between bosonization and the toric code in the following subsections.}. Recall (see Sec.~\ref{sec:Toric-Code}) that, in the toric code model, even though the theory is made of spins (hard-core bosons), we have fermions as stationary emergent quasiparticles, each a combination of an $e$ particle and an $m$ particle. By deforming the toric code model to introduce interactions between the emergent fermions, one should be able to rewrite a fermionic theory fully in terms of spins  (hard-core bosons). Any unpaired $e$ or $m$ particle can be interpreted as a $Z_{2}^{f}$ flux.

In the following sections, we will consider physical systems fundamentally defined in terms of spins and work out quantum circuits that coarse-grain the spin systems and leave behind some disentangled spins, even though the spin theories will be dual to 
theories with fermions coupled to a $Z_{2}^{f}$ gauge field. We will treat the dual picture of fermions with the $Z_2^f$ gauge field as a helpful interpretation of the theory and use it to inspire certain circuit operations on the spin systems. In later sections, we will sometimes use the word ``emergent fermions'' for the dual fermions because they are anyons  
emergent in the spin models. They have to be created in pairs and not one at a time.

In particular, in this paper, we will take advantage of gauging the fermion parity of some well-understood free fermionic theories to generate new topological theories, which will then be mapped to the spin language, where they will become exactly solvable interacting chiral topologically ordered theories. To be specific, we will be interested in gauging non-interacting fermionic models made of layers of lattice $p_{x}+ip_{y}$ topological superconductors to construct sixteen inequivalent chiral bosonic topologically ordered theories classified by Kitaev \cite{Kitaev2006}. 
These bosonic models are exactly solvable precisely because they are dual to free fermionic theories under the zero-flux condition. 
We will use this idea to write down lattice spin models with progressively increasing complexity.

As we mentioned above, for excitation energies much lower than $\Delta_{\Phi}$, we are effectively in the zero-flux sector. Furthermore, in the remainder of the paper, we will only be interested in working on entanglement renormalization circuits that only operate on ground states, which contain zero flux.  For our purposes, the Hamiltonians constructed using the above approach 
simply illustrate that the corresponding ground states can have parent Hamiltonians with anyonic excitations and thus have topological order. Therefore, for the purposes of studying the ground states, instead of adding the flux penalty Hamiltonian, for the rest of the paper, we can conveniently consider another class of Hamiltonians that have the zero-flux condition as a hard constraint directly on the Hilbert space \cite{Chen2018a,Chen2019Bosonization3D,Chen2020}. In other words, we will not include $H^{\Phi}_{\mathrm{penalty}}$ described by Eq.~(\ref{eq:fluxpenalty}) into the gauged fermion Hamiltonians, and we will not include $H^{F}_{\mathrm{penalty}}$ described by Eq.~(\ref{eq:Ffluxpenalty}) into the corresponding spin Hamiltonians. We simply follow the procedure of replacing operators in the fermion Hamiltonian with Wilson lines and doing the algebra isomorphism and require that $\Phi_v=1$ on the gauged-fermion side, and that $F_v=1$ on the spin side.

\begin{figure*}[t]
\begin{centering}
\includegraphics[scale=0.33]{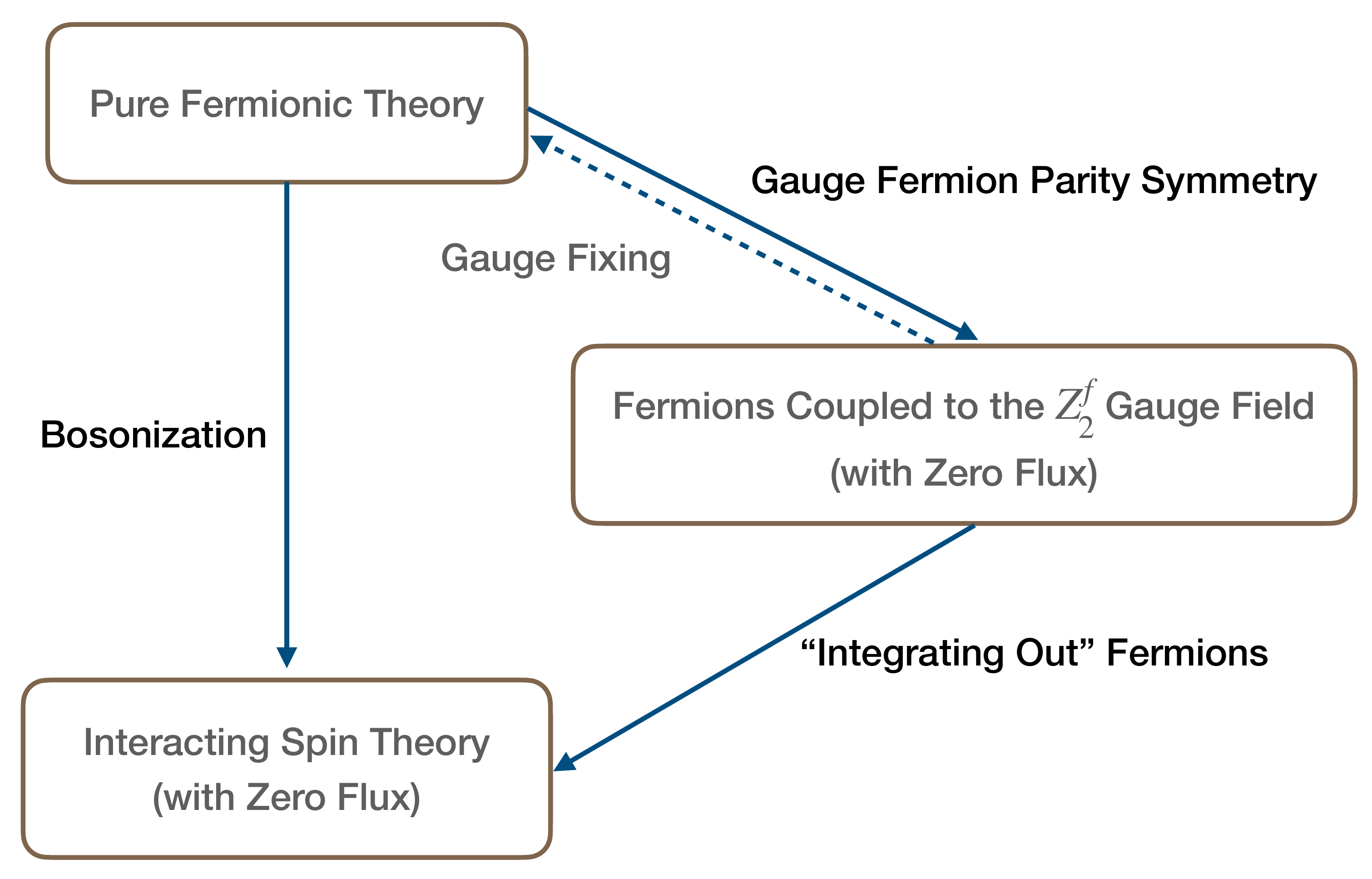}
\par\end{centering}
\centering{}
\caption{Bosonization is composed of two steps. The first step is gauging the fermion parity of the original fermionic theory under the zero-flux condition as a constraint, $\Phi_v=1$. Due to the zero-flux constraint, different choices of Wilson lines for a fermion operator become equivalent. Under the constraint and with the gauge choice $\sigma_{x,e}=1$ (gauge fixing), we can recover the original fermionic theory from the gauged theory. The second step of bosonization is integrating out the fermions by the algebra isomorphism depicted in Fig.~\ref{fig:branchingsquare}. The non-uniqueness of Wilson lines in the first step results in the non-uniqueness of bosonization (one-to-many mapping). However, different choices of the bosonized operator for a given fermion operator are equivalent under the zero-flux constraint on the spin theory, $F_v=1$.
\label{fig:bosonizationdiagram}}
\end{figure*}

\begin{figure}[t]
\begin{centering}
\includegraphics[width=\columnwidth]{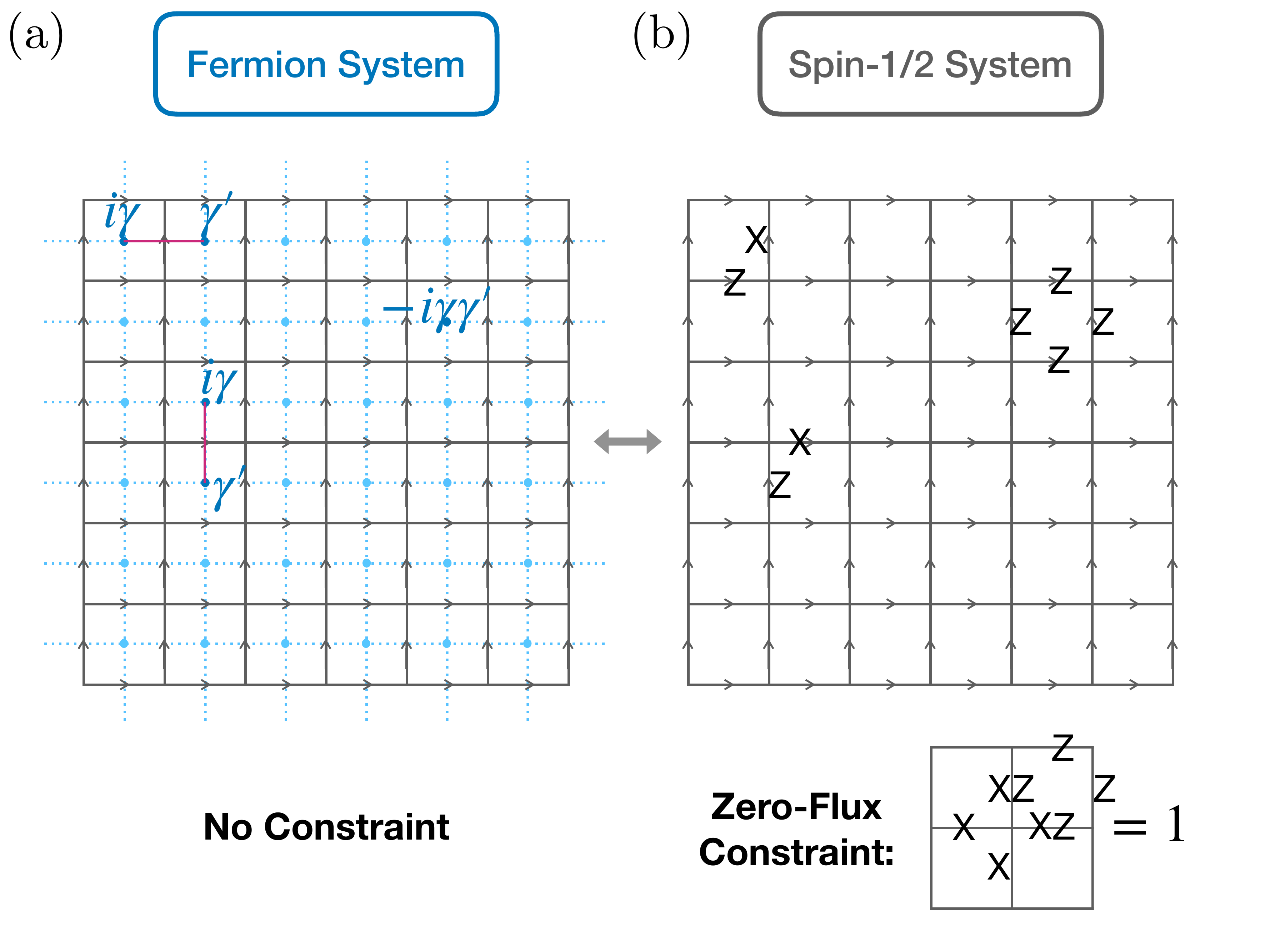}
\par\end{centering}
\centering{}\caption{Here we have a bosonization duality between (a) the pure fermionic theory and (b) the pure spin theory by imposing the zero-flux condition $F_v=1$ as a constraint on the spin side. The fermions live on the faces of the square lattice, while the spins live on the edges of the square lattice. The three fermionic operators on the left are mapped to the three spin operators on the right. The ordering of the Majorana operators is defined in
the text.
\label{fig:fermionstospins}}
\end{figure}

Under the zero-flux constraint, we refer to the successive procedures of gauging a fermionic Hamiltonian with a $Z_{2}^{f}$ gauge field and then integrating out the fermions as bosonization \cite{Chen2018a,Chen2019Bosonization3D,Chen2020} since one can also view the spin-$1/2$ degrees of freedom in the final spin theory as hard-core bosons, where infinitely strong on-site repulsion between bosons renders each site either unoccupied or occupied by a single boson. The whole bosonization procedure turns a purely fermionic Hamiltonian into a purely (hard-core) bosonic Hamiltonian with a constraint (the zero-flux constraint). The flow chart of bosonization to arrive at a bosonic Hamiltonian from a purely fermionic Hamiltonian is shown in Fig.~\ref{fig:bosonizationdiagram}.
In addition, we can also define bosonization for any fermionic operator, not just for a fermionic Hamiltonian. This is done similarly by inserting gauge variables into fermion operators to obtain Wilson lines and integrating the fermions out by the algebra isomorphism under the zero-flux constraint. 
We will use the notation $\left\{ \mathcal{O}^{f}\right\} ^{\mathrm{bosonized}}$ to denote the bosonized operator of the fermionic operator $\mathcal{O}^{f}$ that conserves fermion parity. The reader should distinguish the bosonization in two spatial dimensions presented here from the traditional bosonization for the Luttinger liquid in one dimension \cite{Fradkin2013, Shankar2017}. 
Once again, the bosonization procedure of fermionic operators $\left\{ \cdot\right\} ^{\mathrm{bosonized}}$ is not unique due to different choices of Wilson lines in the gauging procedure $\left\{ \cdot\right\} ^{\mathrm{gauged}}$; however, when we use bosonization to construct bosonic physical systems, under the zero-flux constraint they will be equivalent. Note that bosonization gives rise to an exact duality directly between spins with a constraint and the original fermions. The bosonization duality isomorphism is determined by the following mapping of the generators:
\begin{align}
    S_{e}&\leftrightarrow U_{e},\\
    (-1)^{N_f}&\leftrightarrow W_f.
\end{align}
The mapping of the operators is depicted in Fig.~\ref{fig:fermionstospins}.
Also, the fermion vacuum state is mapped to the simultaneous $+1$ eigenstate of all the $W_f$ operators on the spin side with the zero-flux constraint $F_v=1$ being satisfied. Not surprisingly, the duality resembles the duality between fermions coupled to the $Z_2^f$ gauge field and the spin theory without a constraint. The slight difference is that, compared to the duality in Fig.~\ref{fig:branchingsquare}, Figure~\ref{fig:fermionstospins} does not have the $Z_2^f$ gauge field on the fermion side, and thus we forbid the $Z_e$ generators that create fluxes on the spin side and that do not commute with the zero-flux constraint. Notice that the exact solvability of the spin models with the flux penalty Hamiltonian at low energies previously mentioned now turns into the exact duality relating the spin theory with a constraint to the original non-interacting fermion theory.

Let us emphasize once again that this bosonization perspective with the zero-flux constraint on the Hilbert space merely provides a convenient way to describe the ground states by constructing the corresponding parent Hamiltonians in the presence of the zero-flux constraint. This simplification will help us elucidate the construction of the entanglement renormalization circuits in the remainder of the paper. However, in order to realize these systems experimentally, we do not have to impose the hard constraint on the total spin Hilbert space or the ground states. Instead, if we want, we are free to replace the zero-flux constraint back with the flux penalty Hamiltonian. Therefore, when we discuss  entanglement of the spin system, the total Hilbert space is still assumed to have the structure of a tensor product of local spin Hilbert spaces.

\subsection{Gauging trivial insulator $\Longrightarrow$ pure $Z_{2}^{f}$ lattice gauge theory}

\label{subsec:Gauged-trivial-fermionic}

Starting from this subsection, we will construct several spin models based on non-interacting fermionic models using the bosonization technique introduced in the previous subsection. We will construct models with increasing complexity so the readers can become gradually familiar with the formalism of gauging the fermion parity symmetry and the formalism of bosonization.

The first non-interacting fermionic model we consider is the following fermionic Hamiltonian with a trivial insulating ground state:
\begin{equation}
H_{{\rm trivial\,insulator}}=-\sum_{\mathbf{r}}(1-2c_{\mathbf{r}}^{\dagger}c_{\mathbf{r}})\label{eq:trivialfermionicHamiltonian}.
\end{equation}
We can easily see that each term measures the local fermion parity $(-1)^{N_{\mathbf{r}}}$. Clearly, the ground state is the vacuum, and the chiral central charge is zero. 
It is straightforward to gauge the fermion parity symmetry of the theory since the Hamiltonian $H_{{\rm trivial\,insulator}}$ does not involve interactions among different sites. We simply introduce the Hilbert space of $Z_2^f$ gauge variables and do not have to replace the fermionic interacting terms in the Hamiltonian with Wilson lines. In addition, we can derive the bosonized Hamiltonian with spins living on the edges of a square lattice using bosonization rules shown in Fig.~\ref{fig:fermionstospins}:
\begin{align}
H_{Z_{2}^{f}} & =\left\{ H_{{\rm trivial\,insulator}}\right\} ^{\mathrm{bosonized}}=-\sum_{f}\prod_{e\in f}Z_{e},\label{eq:pure-Z2f-lattice-gauge-Hamiltonian}
\end{align}
with the zero-flux condition $F_v=1$ as a constraint.
The original lattice sites for the  trivial fermionic insulator labeled by $\mathbf{r}$ are sitting at the centers of the faces $f$ of the new square lattice when we perform the bosonization.
The Hamiltonian simply comes from the emergent fermion parity operators $W_{f}$, the bosonization of Eq.~(\ref{eq:trivialfermionicHamiltonian}) by using Eq.~(\ref{eq:fermionparitymaptobosons}).
As mentioned previously in Sec.~\ref{subsec:Bosonization-Formalism}, we can alternatively include the flux penalty Hamiltonian $H^{F}_{\mathrm{penalty}}$ to penalize softly the sectors with nonzero fluxes rather than imposing the zero-flux condition as a constraint. If we do that, one can see that this model is a commuting-projector model and behaves almost like the toric code model with a slight modification of the Hamiltonian. 

\begin{figure}[t]
\begin{centering}
\includegraphics[scale=0.35]{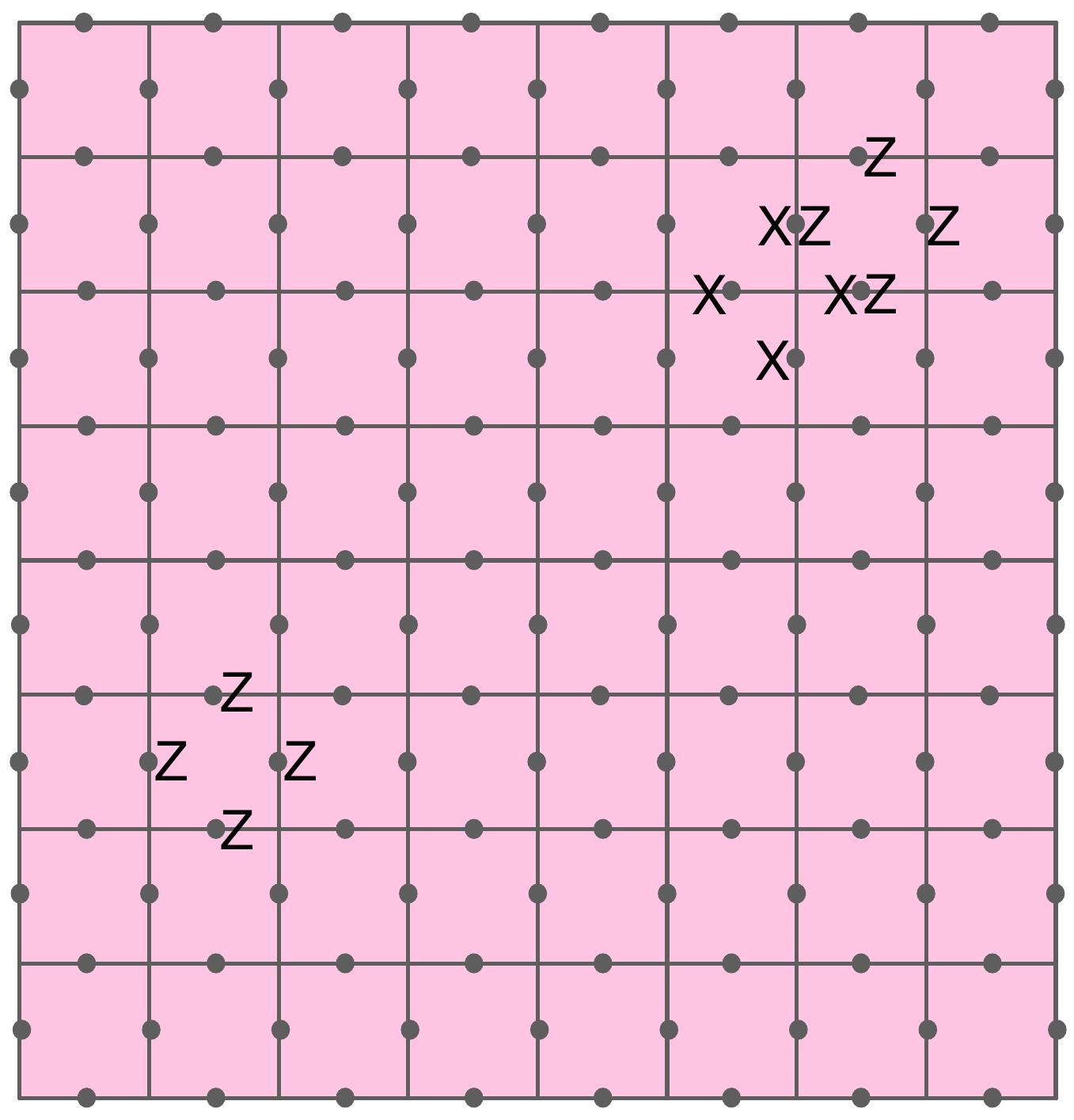}
\par\end{centering}
\centering{}\caption{The stabilizer generators of the pure $Z_{2}^{f}$ lattice gauge theory: the emergent fermion parity operator $W_f$ (the 4-qubit operator on the left) and the flux measuring operator $F_v$ (the 6-qubit operator on the right). The $X$ operators are written slightly to the left of the qubits, while the $Z$ operators are written slightly to the right of the qubits. When an $X$ operator and a $Z$ operator are both associated with the same qubit, the $Z$ operator acts first. \label{fig:bosonizedtrivialstate}}
\end{figure}

First, we observe that the emergent fermion parity operators $W_f$ in the Hamiltonian $H_{Z_2^f}$ defined in Eq.~(\ref{eq:pure-Z2f-lattice-gauge-Hamiltonian}) together with the flux measuring operators $F_v$ in the flux penalty Hamiltonian $H^{F}_{\mathrm{penalty}}$ define a stabilizer group that stabilizes the ground state, and the stabilizer group is the same as that of the toric code in Sec.~\ref{subsec:toriccodemodel}, even though we have a different choice of the stabilizer generators: $\left\{ W_f=\prod_{e\in f}Z_{e},\,\forall f\right\} \cup\left\{ F_{v}=\Bigl(\prod_{e'\in NE(v)}Z_{e'}\Bigr) \Bigl(\prod_{e\in v}X_{e} \Bigr),\,\forall v \right\}$. The stabilizer generators are shown in  Fig.~\ref{fig:bosonizedtrivialstate}.
Therefore, the ground state stabilized by the stabilizer group is the same as that of the toric code.
Therefore, the ground state is non-chiral and has a zero correlation length. 

Second, we can see that the the topological data is the same as that of the toric code. We can either check this from the dual picture of fermions coupled to the $Z_2^f$ gauge field or by working directly with spins. Notice that a single violation from the first set of stabilizer generators $\left\{ W_f,\,\forall f\right\}$ with no nearby violation of the second set of generators $\left\{ F_v,\,\forall v\right\}$ implies the existence of a fermion $\psi$ living on the face whose stabilizer is violated,
whereas a single violation of the second set of stabilizer generators $\left\{ F_{v},\,\forall v\right\}$ without violation of the first set of stabilizer generators $\left\{ W_f,\,\forall f\right\}$ nearby implies the existence of a vortex boson $m$. The vortex here is simply a flux since, in the dual picture of fermions coupled to the $Z_2^f$ gauge field, there are no interactions between fluxes and fermions. The fusion of a fermion with a nearby vortex particle gives rise to a boson, which we call an $e$ particle. The vacuum $1$ together with the $e$ $m$ $\psi$ particles 
constitute 
the same topological data as in the toric code.

By going from Eq.~(\ref{eq:trivialfermionicHamiltonian}) to Eq.~(\ref{eq:pure-Z2f-lattice-gauge-Hamiltonian}), we turned an non-interacting fermionic model into an interacting bosonic spin model. We call this model the pure $Z_2^f$ lattice gauge theory. This is because, in the dual fermionic picture, if we included the flux penalty Hamiltonian $H^{\Phi}_{\mathrm{penalty}}$ to penalize fluxes softly, and if we made the coefficient in front of $(-1)^{N_\mathbf{r}}$ much greater than $\Delta_\Phi$, the model at low energies would have no matter field excitations, and it would be described by the pure $Z_2^f$ lattice gauge field. This is also why we add the subscript $Z_2^f$ to the Hamiltonian in Eq.~(\ref{eq:pure-Z2f-lattice-gauge-Hamiltonian}).

\subsection{Gauging lattice $p_{x}+ip_{y}$ topological superconductor $\Longrightarrow$ lattice Ising TQFT}

\label{subsec:Gauged-superconductor}

In this subsection, we consider the bosonization of a less trivial non-interacting fermionic model, arriving again at an exactly solvable interacting chiral spin model. The fermionic model we want to bosonize is the lattice $p_{x}+ip_{y}$ topological superconductor model $H_{p_{x}+ip_{y}}$ in Eq.~(\ref{eq:TopologicalSuperconductor-Hamiltonian}) in Sec.~\ref{subsec:modelfortopological superconductor}. Even though a $p_{x}+ip_{y}$ topological superconductor does not have intrinsic topological order, it is well-known that, if we gauge the fermion parity symmetry of a $p_{x}+ip_{y}$ topological superconductor, we will have intrinsic chiral topological order with the Ising topological quantum field theory (Ising TQFT) description at low energies \cite{Kitaev2006,Bernevig2013,Nayak2008}.

For the sake of gauging the Hamiltonian $H_{p_{x}+ip_{y}}$, we first put the fermionic degrees of freedom of the Hamiltonian $H_{p_{x}+ip_{y}}$ onto the faces of a square lattice and rewrite the Hamiltonian in terms of the edge orientation assignments in Fig.~\ref{fig:branchingsquare}:
\begin{widetext}
\begin{align}
  H_{p_{x}+ip_{y}}= & \sum_{e_{y}}\left(-t\,c_{R(e_{y})}^{\dagger}c_{L(e_{y})}-t\,c_{L(e_{y})}^{\dagger}c_{R(e_{y})}+\Delta\,c_{R(e_{y})}^{\dagger}c_{L(e_{y})}^{\dagger}+\Delta\,c_{L(e_{y})}c_{R(e_{y})}\right)-\mu\sum_{f}c_{f}^{\dagger}c_{f}\nonumber\\
 & +\sum_{e_{x}}\left(-t\,c_{R(e_{x})}^{\dagger}c_{L(e_{x})}-t\,c_{L(e_{x})}^{\dagger}c_{R(e_{x})}-i\Delta c_{R(e_{x})}c_{L(e_{x})}+i\Delta c_{L(e_{x})}^{\dagger}c_{R(e_{x})}^{\dagger}\right), 
 \label{eq:superconductorbeforegauging}
\end{align}
where $e_{y}$ labels the vertical edges, $e_{x}$ labels the horizontal edges, and $f$ labels the faces. 
We can then rewrite the theory in the language of Majorana operators, gauge the theory by using the shortest Wilson lines, and decompose the Wilson lines into the generators $S_{e}^{\mathrm{gauged}}=i\gamma_{L(e)}\sigma_{x,e}\gamma'_{R(e)}$ and $(-1)^{N_{f}}=-i\gamma_{f}\gamma_{f}'$. The result is the gauged Hamiltonian
\begin{align}
& \left\{ H_{p_{x}+ip_{y}}\right\} ^{\mathrm{gauged}} = \nonumber \\
=&  \sum_{e_{y}}\Bigg[-\left(\frac{t+\Delta}{2}\right)\,(-i\gamma_{L(e_{y})}\gamma_{L(e_{y})}^{\prime})(i\gamma_{L(e_{y})}\sigma_{x,e_{y}}\gamma_{R(e_{y})}^{\prime})(-i\gamma_{R(e_{y})}\gamma_{R(e_{y})}^{\prime})-\left(\frac{t-\Delta}{2}\right)\,(i\gamma_{L(e_{y})}\sigma_{x,e_{y}}\gamma_{R(e_{y})}^{\prime})\Bigg]\nonumber \\
 & +\sum_{e_{x}}\Bigg[-\frac{t}{2}\,(-i\gamma_{L(e_{x})}\gamma_{L(e_{x})}^{\prime})(i\gamma_{L(e_{x})}\sigma_{x,e_{x}}\gamma_{R(e_{x})}^{\prime})(-i\gamma_{R(e_{x})}\gamma_{R(e_{x})}^{\prime})-\frac{t}{2}(\,i\gamma_{L(e_{x})}\sigma_{x,e_{x}}\gamma_{R(e_{x})}^{\prime})\nonumber \\
 & +i\frac{\Delta}{2}(-i\gamma_{L(e_{x})}\gamma_{L(e_{x})}^{\prime})(i\gamma_{L(e_{x})}\sigma_{x,e_{x}}\gamma_{R(e_{x})}^{\prime})-\frac{i\Delta}{2}(i\gamma_{L(e_{x})}\sigma_{x,e_{x}}\gamma_{R(e_{x})}^{\prime})(-i\gamma_{R(e_{x})}\gamma_{R(e_{x})}^{\prime})\Bigg]-\mu\sum_{f}\left(1+i\gamma_{f}\gamma_{f}^{\prime}\right),
 \label{eq:gaugedsuperconductor}
\end{align}
where Gauss's law is imposed onto the Hilbert space.
In order to obtain the dual spin model of the gauged fermionic theory under the zero flux constraint, we can either ``integrate out'' the fermions in Eq.~(\ref{eq:gaugedsuperconductor}) to get rid of the Gauss's law constraint using Eqs.~(\ref{eq:majoranahoppingmaptobosons},\ref{eq:fermionparitymaptobosons}) or directly apply the bosonization mapping in Fig.~\ref{fig:fermionstospins} to the fundamental generators, $S_e$ and $(-1)^{N_f}$, that generate Eq.~(\ref{eq:superconductorbeforegauging}) with the shortest paths. Either way, the resulting spin Hamiltonian is given by
\begin{align}
H_{\mathrm{Ising\,TQFT}}= & \left\{ H_{p_{x}+ip_{y}}\right\} ^{\mathrm{bosonized}} \nonumber \\
= & \sum_{e_{y}}\Bigg[-\left(\frac{t+\Delta}{2}\right)\,(W_{L(e_{y})})(X_{e_{y}}Z_{r(e_{y})})(W_{R(e_{y})})-\left(\frac{t-\Delta}{2}\right)\,X_{e_{y}}Z_{r(e_{y})}\Bigg]\nonumber\\
 & +\sum_{e_{x}}\Bigg[-\frac{t}{2}\,(W_{L(e_{x})})(X_{e_{x}}Z_{r(e_{x})})(W_{R(e_{x})})-\frac{t}{2}\,(X_{e_{x}}Z_{r(e_{x})})\nonumber \\
 & +i\frac{\Delta}{2}(W_{L(e_{x})})(X_{e_{x}}Z_{r(e_{x})})-\frac{i\Delta}{2}(X_{e_{x}}Z_{r(e_{x})})(W_{R(e_{x})})\Bigg]-\mu\sum_{f}\left(1-W_{f}\right).\label{eq:IsingTQFTHamiltonian}
\end{align}
\end{widetext}
As before, we have imposed the zero-flux condition $F_v=1$ as a constraint. We have turned a non-interacting fermionic model $H_{p_{x}+ip_{y}}$ into an interacting spin model  $H_{\mathrm{Ising\,TQFT}}$. The parameters $(t,\,\mu,\,\Delta)$ here are chosen to be again in the regime described in Sec.~\ref{subsec:modelfortopological superconductor}. Since the Hamiltonian in Eq.~(\ref{eq:IsingTQFTHamiltonian}) with the zero-flux constraint is dual to the non-interacting lattice $p_{x}+ip_{y}$ topological superconductor, we can understand the properties of its ground state very well. In particular the ground state is chiral with $c=1/2$ and has a nonzero finite correlation length. Therefore, we have obtained an exactly solvable chiral spin liquid Hamiltonian. 

The presence of the zero-flux condition $F_v=1$ as a constraint on the spin system allowed us to quickly obtain the parent Hamiltonian that describes the ground state of the interacting spin system by using the bosonization mapping in Fig.~\ref{fig:fermionstospins}. This simplification is enabled by the fact that the ground state happens to be in the sector $F_v=1$. However, the zero-flux constraint should not be viewed as something intrinsic to the actual Hilbert space of the spin (qubit) system. Therefore, instead of imposing the zero-flux condition as a hard constraint, we can alternatively include the flux penalty Hamiltonian $H_{{\rm penalty}}^{F}$ with a large flux energy parameter $\Delta_{\Phi}$ as a soft constraint to penalize fluxes. This will allow the constraint to be violated if we add energy to the system. A pair of $\pi$ fluxes will create a pair of vortex quasiparticles, and each will bind a Majorana zero mode. The existence of a Majorana zero mode around each $\pi$ flux is typically shown in the continuum limit \cite{Alicea2012}; however, it still exists when we introduce a lattice structure \cite{Kitaev2006}. With the bound Majorana zero modes, the vortices are non-Abelian Ising anyons \cite{Fradkin2013}. Therefore, we expect that the effective description of the low-energy behavior near the ground state is the Ising TQFT. Hence, we introduce a subscript ``Ising TQFT" for the Hamiltonian in Eq.~(\ref{eq:IsingTQFTHamiltonian}). We may also think of this model as a lattice regularization of the continuum Ising TQFT, so, in the following, we will sometimes call it the lattice Ising TQFT model. As the model consists of spins and is gapped and topologically nontrivial, we can regard this model as a chiral spin liquid, whether we impose the zero-flux condition as a hard constraint or as a flux penalty Hamiltonian.

\subsection{Gauging layers of $p_x+ip_y$ superconductors $\Longrightarrow$ Kitaev's sixteenfold way chiral spin liquids}

\label{subsec:Kitaev's-sixteen-foldway}
After introducing the lattice Ising TQFT model as an example of a chiral spin liquid in the previous subsection, in this subsection, we are going to introduce more exactly solvable chiral spin liquids by using bosonization introduced in Sec.~\ref{subsec:Bosonization-Formalism}.

In Ref. \cite{Kitaev2006}, Kitaev proved that any spin theory that is dual to non-interacting fermions with a spectral Chern number $\nu$ coupled to a $Z_{2}^{f}$ gauge field should fall into a sixteenfold way classification under certain assumptions  
\cite{Bernevig2015}. From the bulk perspective, we should obtain 16 different kinds of topological order determined by $\nu\,(\mathrm{mod }\, 16)$. The periodicity in $\nu$ means that a spin system corresponding to the spectral Chern number $\nu$ should be topologically indistinguishable from a spin system with the spectral Chern number $\nu +16$. Note, however, that, from the boundary perspective, the chiral central charge $c$ should be determined by $\nu$ via the formula $c=\nu/2$ without periodicity. Some topological data of the sixteenfold way classification in the bulk is provided in Table \ref{tab:sixteenfoldwayfusions}. A review of the sixteenfold way classification is provided in Ref.\ \cite{Bernevig2015}. 

\begin{table*}
\begin{centering}
\begin{tabular}{|c|c|c|}
\hline 
$\nu$ (number of $p_{x}+ip_{y}$ superconductors) & anyons & fusion rules\tabularnewline
\hline 
0 (mod 4) & $\left\{ 1,e,m,\psi\right\} $ & $e\times e=1$, $m\times m=1$, $\psi\times\psi=1$, $\psi\times e=m$, $\psi\times m=e$, $e\times m=\psi$\tabularnewline
\hline 
1 (mod 4) or 3 (mod 4) & $\left\{ 1,\sigma,\psi\right\} $ & $\psi\times\psi=1$, $\psi\times\sigma=\sigma$, $\sigma\times\sigma=1+\psi$\tabularnewline
\hline 
2 (mod 4) & $\left\{ 1,a,\bar{a},\psi\right\} $ & $a\times a=\psi$, $\bar{a}\times\bar{a}=\psi$, $\psi\times\psi=1$, $a\times\bar{a}=1$, $a\times\psi=\bar{a}$, $\bar{a}\times\psi=a$\tabularnewline
\hline 
\end{tabular}
\par\end{centering}
\caption{Part of the topological data of the sixteenfold way classification \cite{Kitaev2006}. The symbol $1$ stands for the vacuum, and $\psi$ stands for an emergent fermion. The symbols $e$, $m$, $\sigma$, $a$, and $\bar{a}$ represent vortices induced by the $Z_2^f$ fluxes. The vortices can be either Abelian or non-Abelian. For $\nu = 0 \, (\mathrm{mod}\,4)$ and $\nu=2 \,(\mathrm{mod} \,4)$, the vortices are Abelian anyons.  For $\nu = 1\, (\mathrm{mod} \,4)$ and  $\nu = 3 \,(\mathrm{mod} \,4)$,  vortex $\sigma$ is a non-Abelian anyon due to an unpaired Majorana zero mode at its core. The topological spins of all the vortices are determined by the formula $\theta=\exp(i \pi \nu/8)$ that reflects the $\nu\, (\mathrm{mod} 16)$ periodicity. More data is included in the $S$ and $T$ matrices and other quantities, which we do not list here. \label{tab:sixteenfoldwayfusions}}
\end{table*}

In Ref. \cite{Kitaev2006}, Kitaev introduced a spin model on a honeycomb lattice ($B$ phase in a magnetic field) whose dual is a $p_x+ip_y$ topological superconductor ($\nu=1$) coupled to a $Z_{2}^{f}$ gauge field. Here we construct spin models corresponding to other values of $\nu$. Instead of working with spins on a honeycomb lattice, we will work with the formalism on the square lattice discussed in Sec.~\ref{subsec:Bosonization-Formalism} since the operator duality between the spin theory and the fermionic theory with the $Z_2^f$ gauge field is more obvious to us on the square lattice.

A simple way to construct a fermionic system with a higher spectral Chern number is to consider a stack of $p_x+ip_y$ topological superconductors. Hence, for a fermionic system with spectral Chern number $\nu$, we can simply consider $\nu$ layers of the lattice $p_x+ip_y$ topological superconductors in Sec.~\ref{sec:-Topological-Superconductor}, each of which contributes a chiral central charge $c=1/2$. Note that, for each value of $\nu$, we can freely add an arbitrary number of the trivial insulators $H_\mathrm{trivial\,insulator}$ in Sec.~\ref{subsec:Gauged-trivial-fermionic} since they carry zero chiral central charge.

\begin{figure*}[t]
\centering{}\includegraphics[scale=0.40]{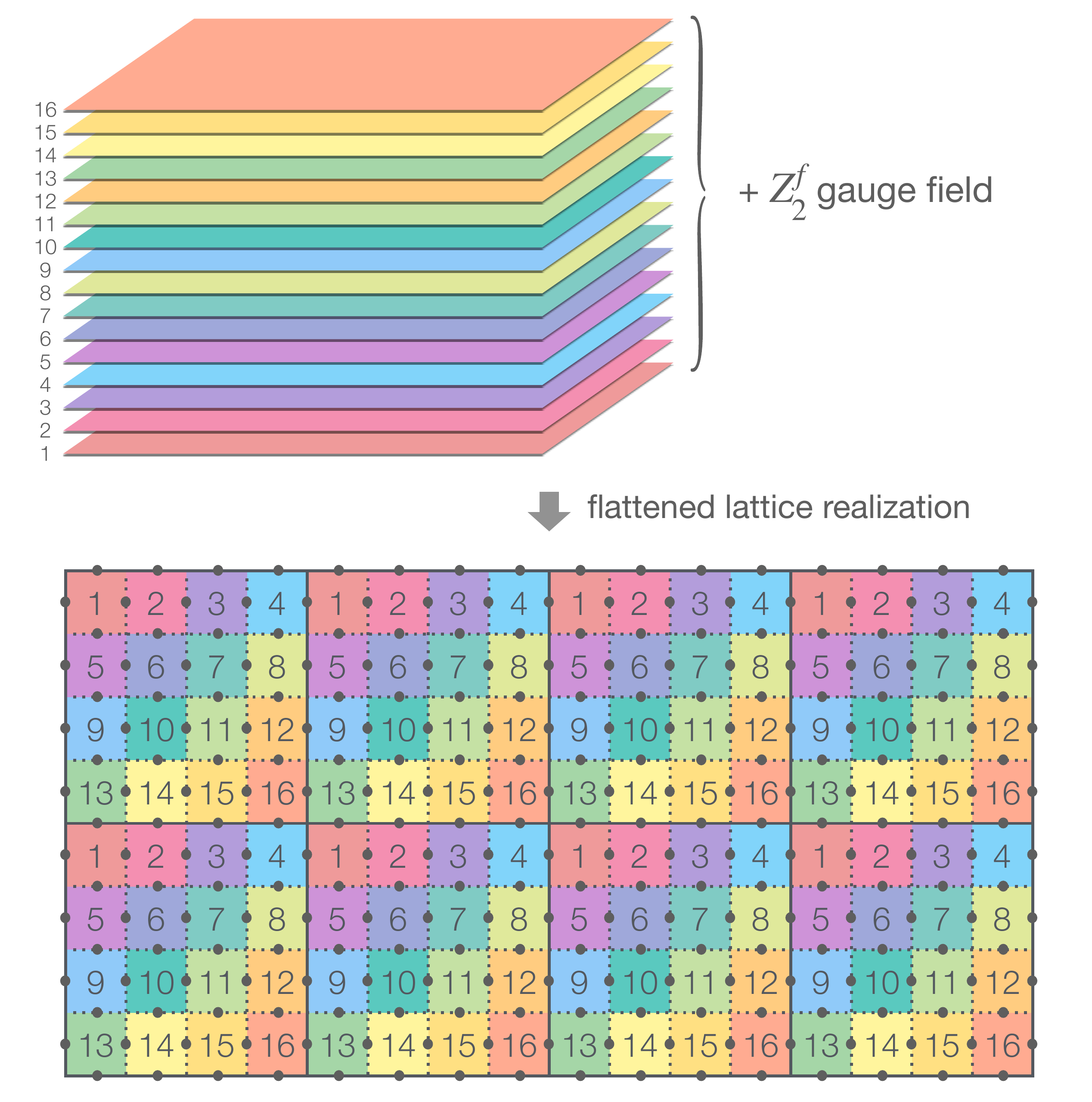}\caption{A way to realize the sixteenfold way states on a square lattice. We flatten 16 fermionic layers onto a superlattice defined on a single square lattice
and subsequently apply the bosonization procedure.
Fermions live at the centers of the faces. A single unit cell for the 
fermions consists of 16 faces. A sublattice of
fermions consists of faces with the same color (and hence same layer number label). Each sublattice (i.e.\ layer/color) describes either a layer of the lattice $p_{x}+ip_{y}$ topological superconductor or a layer of the trivial insulator. Therefore, we introduce hoppings and pairings of
fermions only among sites with the same color (and hence same layer number label).
The bosonic degrees of freedom are spin-$1/2$'s living on the edges surrounding the faces. 
In order to obtain the interacting spin model, we bosonize all the terms of the fermionic Hamiltonian, as described earlier in Sec.~\ref{subsec:Bosonization-Formalism}. Here, we only draw a small portion of the superlattice, and the reader should extend the periodic structure horizontally and vertically. \label{fig:put16foldway}}
\end{figure*}

We will now systematically construct all the states in Kitaev's sixteenfold way classification using the technique (introduced in Sec.~\ref{subsec:Bosonization-Formalism}) of gauging fermion parity and integrating out fermions on the square lattice. First, we introduce a superlattice structure with lattice periodicity determined by large $4\times4$ unit cells as shown in Fig.~\ref{fig:put16foldway}
to have multiple layers of the lattice $p_{x}+ip_{y}$ topological superconductors and trivial insulators flattened to a square lattice. Here, we assign a fermionic degree of freedom to the center of each face. We associate different faces within a unit with different colors corresponding to different layer numbers $1\leq i \leq 16$ (which we also write on the faces) while periodically extending the pattern to the other unit cells. (Notice that we use the phrase ``layer number'' to refer to two different things: one is the layer number $s$ labeling the scale of the entanglement renormalization operation; the other is the layer number $i$ labeling the layers of the fermionic degrees of freedom.) 
We then use faces with the same color (and hence same layer number) to represent the degrees of freedom of either a single layer of the lattice $p_{x}+ip_{y}$ topological superconductor or a single layer of the topologically trivial insulator. For example, if we want the blue faces (layer number $i = 4$) to describe a single layer of the lattice $p_{x}+ip_{y}$ topological superconductor, we will add chemical potential terms to all blue ($i=4$) faces and introduce horizontal and vertical hopping and pairing terms that directly (and hence remotely) couple blue ($i=4$) faces in adjacent unit cells.
As another example, if we want the green faces (layer number $i = 7$) to  describe a single layer of the trivial insulator, we will only add chemical potential terms as in Eq.~(\ref{eq:trivialfermionicHamiltonian}) to those faces. Therefore, in order to study a system with spectral Chern number $\nu$, we will use the first $\nu$ sublattices (i.e.\ layers/colors) numbered $i=1,\cdots,\nu$ to describe $p_{x}+ip_{y}$ superconductors and let the remaining $16-\nu$ ($i=\nu+1,\cdots,16$) sublattices describe trivial insulators. We can write this as the following fermionic Hamiltonian:
\begin{align}
 & H_{\nu^{f}} = \nonumber \\
= & \sum_{i=1}^{\nu}H_{i,\,p_{x}+ip_{y}}+\sum_{i=\nu+1}^{16}H_{i,\,\mathrm{trivial\:insulator}}\nonumber \\
= & \sum_{i=1}^{\nu}\Bigg[-t\sum_{\mathbf{r}}\left(c_{i,\,\mathbf{r}+\hat{\mathbf{x}}}^{\dagger}c_{i,\,\mathbf{r}}+c_{i,\,\mathbf{r}+\hat{\mathbf{y}}}^{\dagger}c_{i,\,\mathbf{r}}\right)-\mu\sum_{\mathbf{r}}c_{i,\,\mathbf{r}}^{\dagger}c_{i,\,\mathbf{r}}\nonumber \\
 & +\sum_{\mathbf{r}}\left(\Delta\,c_{i,\,\mathbf{r}+\hat{\mathbf{x}}}^{\dagger}c_{i,\,\mathbf{r}}^{\dagger}+i\Delta\,c_{i,\,\mathbf{r}+\hat{\mathbf{y}}}^{\dagger}c_{i,\,\mathbf{r}}^{\dagger}\right)+{\rm h.c.}\Bigg]\nonumber \\
 & +\sum_{i=\nu+1}^{16}\left[-\sum_{\mathbf{r}}(1-2c_{i,\,\mathbf{r}}^{\dagger}c_{i,\,\mathbf{r}})\right].
\end{align}
The label $i$ running from 1 to 16 indicates the layer (i.e.\ color/sublattice) number. We used the subscript $\nu^{f}$ for the Hamiltonian to denote that we have $\nu$ layers of fermions in topological superconducting states. Once again, the parameters $(t,\,\mu,\,\Delta)$ here are chosen to be in the regime described in Sec.~\ref{subsec:modelfortopological superconductor}.

We arrive at the dual interacting spin models, with spins on the edges of the square lattice shown at the bottom of Fig.\ \ref{fig:put16foldway}, by coupling the fermions to the $Z_2^f$ gauge field on the edges of this lattice in the presence of a flux penalty term,
and subsequently integrating the fermions out.
As a result, we obtain all the sixteenfold way states $0\leq\nu\leq16$ as the ground states of the 
interacting spin models. Since the ground states of these 
interacting spin models are in the sector $F_v=1$, instead of introducing a flux penalty term, we can alternatively impose the zero-flux condition $F_v=1$ as a hard constraint and derive another class of dual parent spin Hamiltonians $H_{\nu}=\left\{H_{\nu^{f}}\right\}^\mathrm{bosonized}$ that describe the sixteenfold way states using the bosonization technique in Fig.~\ref{fig:fermionstospins}. We will not repeat the bosonization here as the principle
is the same as in Secs.~\ref{subsec:Gauged-trivial-fermionic} and \ref{subsec:Gauged-superconductor}. The only difference is that the Hamiltonian $H_{\nu^{f}}$  now involves long-range hoppings and pairings. Since the
fermions we started with were non-interacting, the spin models constructed here are exactly solvable at all energies with the hard zero-flux constraint and at low energies with a soft flux penalty.
The chiral central charge of the spin models will be $c=\nu/2$, coming from the $p_{x}+ip_{y}$ topological superconductors we started with.

We now discuss properties of the sixteenfold way spin states obtained using this construction. When $\nu=0$, from the zero chiral central charge, we learn that the theory is non-chiral. It is nothing but the pure $Z_2^f$ lattice gauge theory discussed in Sec.~\ref{subsec:Gauged-trivial-fermionic}. For $\nu \geq 1$, we obtain several exactly solvable chiral spin liquids with nonzero finite correlation lengths. When $\nu=1$, we get an Ising TQFT Hamiltonian similar  to Eq.~(\ref{eq:IsingTQFTHamiltonian}) in Sec.~\ref{subsec:Gauged-superconductor}. Their topological properties are the same even though they are different lattice realizations of the Ising TQFT. When $\nu=2$, we obtain a system with topological order equivalent to Laughlin's fractional quantum Hall state at filling fraction $1/4$ \cite{Khan2017}.
When $\nu=3$, the ground state of our construction belongs to the universality class of the bosonic Moore-Read fractional quantum Hall state at filling fraction one 
\cite{Yao2011,Nakai2012,Ma2019}. An intuitive way to understand this is to see that the Moore-Read state has an edge mode composed of a chiral Dirac fermion and a chiral Majorana fermion. Since a chiral Dirac fermion can be decomposed into two chiral Majorana fermions, the edge mode effectively has three chiral Majorana fermions and carries a chiral central charge $c=3/2$, which is reminiscent of the chiral Majorana fermions from the three layers of $p_{x}+ip_{y}$ topological superconductors from our construction. Since the boundary conformal field theory of the bosonic Moore-Read state and that of the $\nu=3$ sixteenfold way state match, we expect the two bulk theories to be in the same universality class via the bulk-boundary correspondence. However, our construction loses the $U(1)$ charge conservation symmetry appearing in the Moore-Read state. It will be interesting to see whether one can connect the two wavefunctions directly by a constant-depth quantum circuit respecting some locality constraints. When $\nu=16$, the bulk topological properties should be the same as when $\nu=0$, i.e., the pure $Z_{2}^{f}$ lattice gauge theory, even though the boundary chiral central charge is nonzero.

From the topological data listed in Table \ref{tab:sixteenfoldwayfusions}, we can learn that, when $\nu$ is even, the topological order is Abelian. When $\nu$ is odd, the topological order becomes non-Abelian since the fusion of two vortices outputs two possibilities. The topological properties of the vortices are similar to the topological properties of the vortices of the Ising TQFT at $\nu=1$ \cite{Kitaev2006, Bernevig2015}. For later convenience, we refer to all spin-model ground states derived using this construction as Kitaev's sixteenfold way chiral spin liquids, while keeping in mind that the non-chiral case $\nu=0$ is a trivial special case.

\section{Entanglement Renormalization Circuits for Kitaev's Sixteenfold Way Chiral Spin Liquids}

\label{sec:MERAQLE}

In this section, we will present the scale-invariant entanglement renormalization circuits for a class of spin states constructed in the previous section using the technique of gauging fermion parity or bosonization. Many of these states are chiral. In Sec.~\ref{subsec:MERAQLE-for-pure}, we will present the circuit for the ground state of the non-chiral pure $Z_2^f$ lattice gauge theory model.  In Sec.~\ref{subsec:MERAQLE-for-Ising}, we will present the circuit for the ground state of the chiral lattice Ising TQFT model. In Sec.~\ref{subsec:MERAQLE-for-16foldway}, our discussion will culminate with a presentation of the scale-invariant entanglement renormalization circuits for the ground states of all Kitaev's sixteenfold way chiral spin liquids. In all cases, each layer of the entanglement renormalization circuit will contain two subcircuits: (1)~a single step of horizontal entanglement renormalization in the $x$ direction and (2)~a single step of vertical entanglement renormalization in the $y$ direction. We will build the entanglement renormalization circuits by combining the conventional MERA circuit reviewed in Sec.~\ref{sec:Toric-Code} and  the quasi-local evolution discussed in Sec.~\ref{sec:-Topological-Superconductor}. We refer to these types of circuits as MERA with quasi-local evolution (MERAQLE).

\subsection{MERAQLE for the pure $Z_{2}^{f}$ lattice gauge theory}

\label{subsec:MERAQLE-for-pure}
Before showing the scale-invariant MERAQLE circuits for
all Kitaev's sixteenfold way chiral spin liquids, we start with the simplest state among them to get a glimpse of the entanglement structure of the models constructed from the bosonization technique. Specifically, we discuss the scale-invariant MERAQLE circuit for the ground state of the pure $Z_{2}^{f}$ lattice gauge theory Hamiltonian in Eq.~(\ref{eq:pure-Z2f-lattice-gauge-Hamiltonian}), which is obtained by gauging the fermion parity symmetry of a trivial insulator. Since the ground state of the pure $Z_{2}^{f}$ lattice gauge theory is the same as that of the toric code, the MERAQLE circuit should be the same as the conventional strictly-local MERA circuit shown in Sec.~\ref{sec:Toric-Code}. Even though the theory is non-chiral, and the MERAQLE circuit here has no quasi-local components like the MERAQLE circuits presented in the following subsections, it is still illuminating to see the action of the circuit on the stabilizer generators. The insight gained from the calculations done on the stabilizer generators will be useful for the circuit constructions for other models that don't have the simple toric code interpretation and that will be discussed in later subsections.

\begin{figure}[t]
\begin{centering}
\includegraphics[width=\columnwidth]{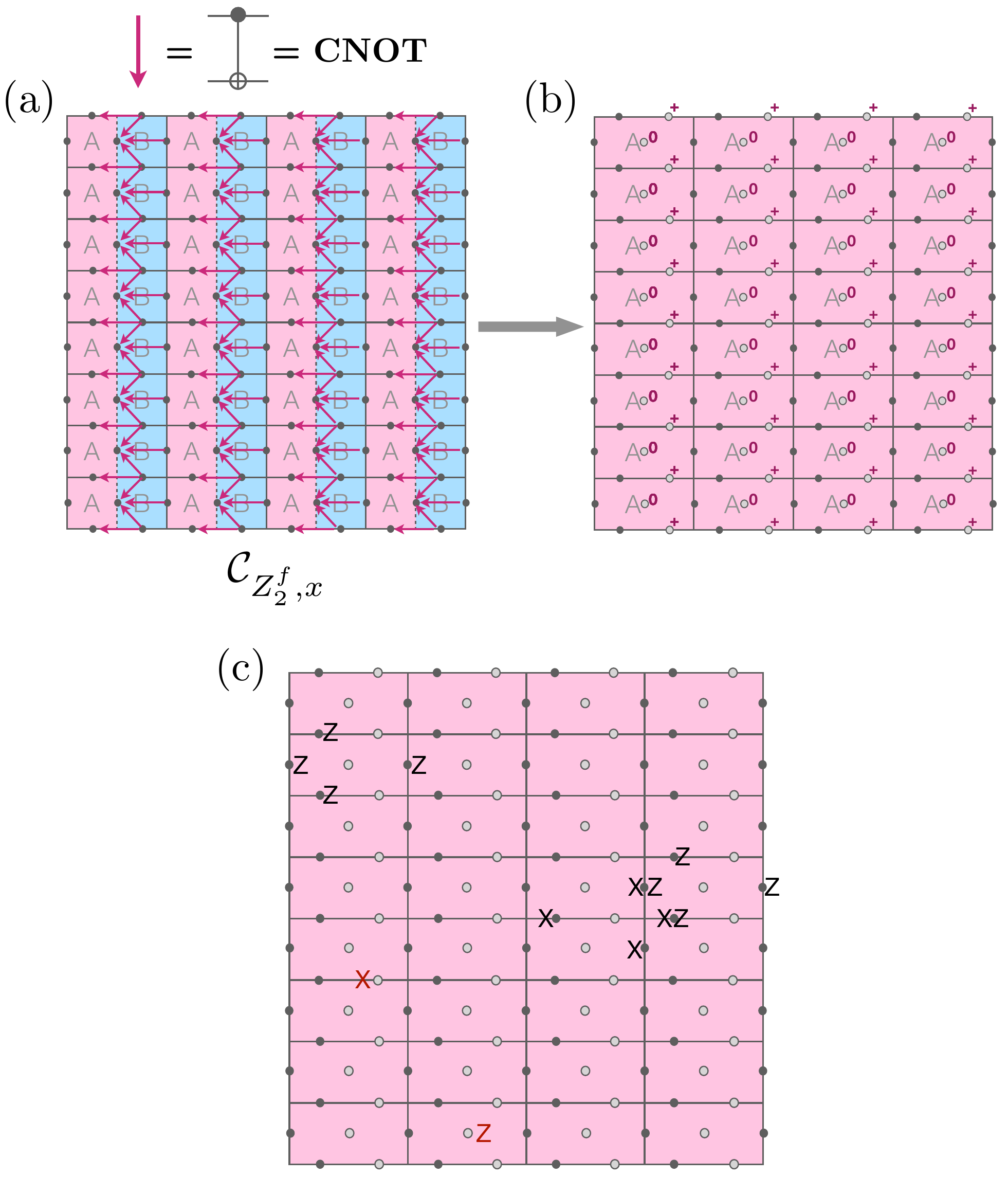}
\par\end{centering}
\centering{}\caption{(a) The circuit $\mathcal{C}_{Z_{2}^{f},x}$ for a single step of horizontal entanglement renormalization, which happens to be the same as $\mathcal{C}_{Z_{2},x}$ in Fig.~\ref{fig:MERA-ToricCode-hori}(a). The filled circles represent qubits (spins) constituting the toric code model. (b,c) The state of the system after the circuit has been applied. Unfilled circles are the qubits (spins) that have been disentangled by the circuit into $\ket{0}$ and  $\ket{+}$ states, as indicated by the labels in (b). (c) shows the new stabilizer generators.  The red single-site $Z$ and $X$ generators stabilize the disentangled qubits, while the black generators stabilize the pure $Z_2^f$ lattice gauge theory defined on the new horizontally elongated square lattice.  The derivation of the new stabilizer generators is presented in Fig.~\ref{fig:MERA-Z2f-hori-calulation}. \label{fig:MERA-Z2f-hori}}
\end{figure}

\begin{figure}[t]
\begin{centering}
\includegraphics[width=\columnwidth]{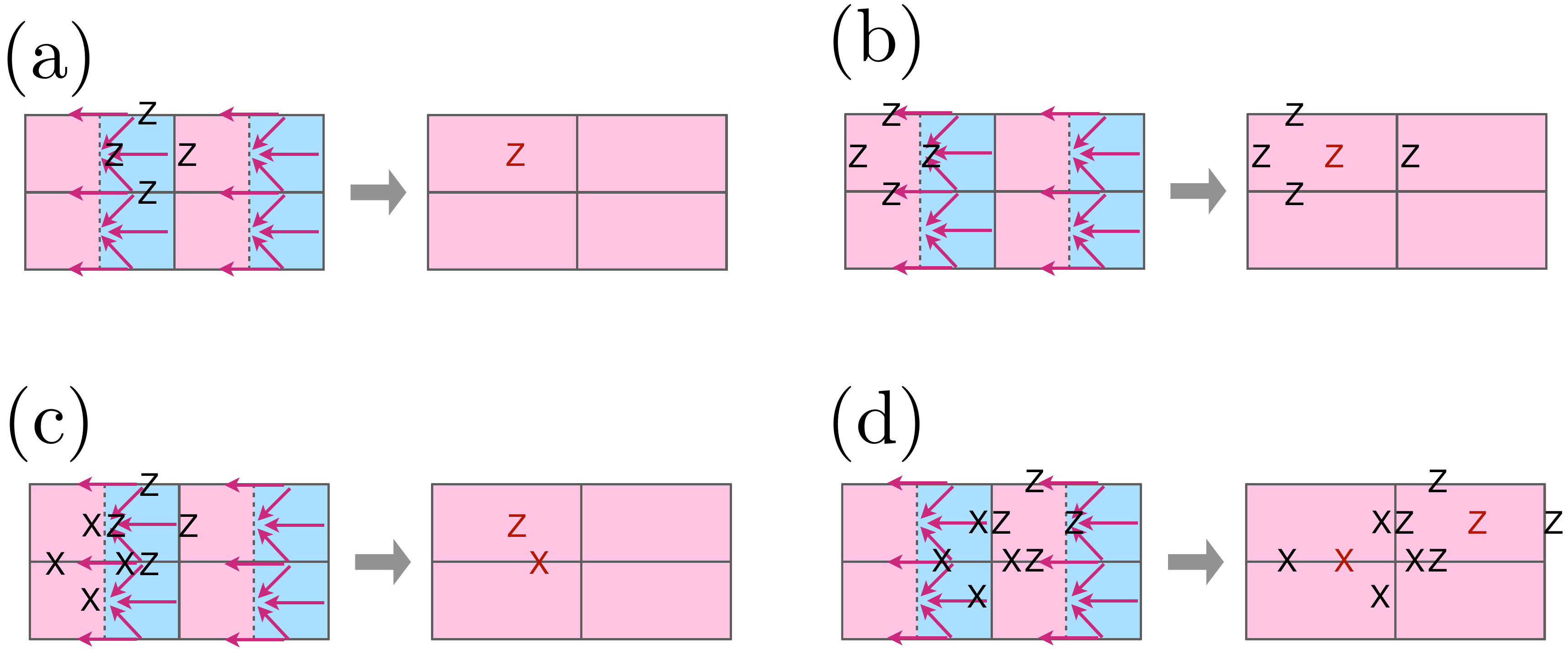}
\par\end{centering}
\centering{}\caption{
Transformation of the stabilizer generators of the pure $Z_{2}^{f}$ lattice gauge theory under conjugation by the horizontal entanglement renormalization subcircuit $\mathcal{C}_{Z_{2}^{f},x}$ in Fig.~\ref{fig:MERA-Z2f-hori}(a). (a)(b) Transformation of the emergent fermion parity operators $W_{f}$. (c)(d) Transformation of the flux measuring operators $F_{v}$. The new stabilizer group generated by the operators on the right-hand sides of the subfigures is the same as the stabilizer group generated by the operators in Fig.~\ref{fig:MERA-Z2f-hori}(c). The red Pauli operators in the subfigures are the red single-qubit stabilizer generators in Fig.~\ref{fig:MERA-Z2f-hori}(c) acting on the disentangled qubits. \label{fig:MERA-Z2f-hori-calulation}}
\end{figure}

A single step of horizontal entanglement renormalization in this case is implemented by the subcircuit  $\mathcal{C}_{Z_{2}^{f},x}$ shown in Fig.~\ref{fig:MERA-Z2f-hori}. Following Sec.~\ref{subsec:MERAforToricCode}, we again use the stabilizer formalism to analyze the transformation of the ground state under $\mathcal{C}_{Z_{2}^{f},x}$. The generators of the original stabilizer group of the ground state are shown in Fig.~\ref{fig:bosonizedtrivialstate}(a). The generators of the transformed stabilizer group
are shown in Fig.~\ref{fig:MERA-Z2f-hori}(c). Even though the subcircuit $\mathcal{C}_{Z_{2}^{f},x}$ is an exact copy of the subcircuit $\mathcal{C}_{Z_{2},x}$ in Fig.~\ref{fig:MERA-ToricCode-hori}, the interpretation is different. Here, on each face, there is an emergent fermion mode (one can think of the fermions as emergent if one takes the spin model as the original model of interest). It is convenient to introduce an $AB$ sublattice structure for the emergent fermions modes. The structure of the sublattices is shown in Fig.~\ref{fig:MERA-Z2f-hori}(a), where the $A$ sublattice is associated with the pink faces, and the $B$ sublattice is associated with the blue faces. Recall that a product of four $Z$ operators around a face measures the emergent fermion parity of the face. From the stabilizer computation in Fig.~\ref{fig:MERA-Z2f-hori-calulation}(b), we learn that the emergent fermion parity operator $W_f$ of a blue $B$ face is transformed into a single $Z$ operator under conjugation by the subcircuit $\mathcal{C}_{Z_{2}^{f},x}$. We can intuitively say that, after the renormalization procedure, the emergent fermion degrees of freedom on the blue $B$ faces are effectively shifted to the ancillary qubits, and the corresponding fermion parities become single-qubit Pauli-$Z$ operators. Since the ground state originally has all the emergent fermions frozen in the vacuum state, the qubits associated with the red single-qubit Pauli-$Z$ operators will be transformed under $\mathcal{C}_{Z_{2}^{f},x}$ to state $\ket{0}$, decoupled from the rest of the system. The change of the stabilizer generator corresponding to an emergent fermion parity operator $W_f$ on a pink $A$ face is computed in Fig.~\ref{fig:MERA-Z2f-hori-calulation}(a). One can also compute the change of the flux measuring operators $F_v$ in Fig.~\ref{fig:MERA-Z2f-hori-calulation}(c,d). One can recombine the results of the conjugation of the stabilizer generators under the subcircuit $\mathcal{C}_{Z_{2}^{f},x}$ in Fig.~\ref{fig:MERA-Z2f-hori-calulation} to get a new set of stabilizer generators shown in Fig.~\ref{fig:MERA-Z2f-hori}(c). For the transformed state stabilized by the new set of stabilizer generators shown in Fig.~\ref{fig:MERA-Z2f-hori}(c), we can see that, in addition to the $\ket{0}$ states to the left of the blue $B$ faces corresponding to the stabilizer generators equal to the red single-qubit Pauli-$Z$ operators, we also get disentangled qubits in the $\ket{+}$ states at the bottom of the blue $B$ faces corresponding to the stabilizer generators equal to the red single-qubit Pauli-$X$ operators. The red single-qubit Pauli-$X$ operators are a result of the original emergent fermion parity operators $W_f$ acting on the blue $B$ faces and the flux measuring operators $F_v$ with $NE(v)$ being blue $B$ faces. The disentangled qubits are represented by the unfilled circles in Fig.~\ref{fig:MERA-Z2f-hori}(b).  For the remaining entangled qubits, the stabilizer generators are colored in black in Fig.~\ref{fig:MERA-Z2f-hori}(c). These are nothing but the emergent fermion parity operator $W_{f}$ and flux measuring operator $F_{v}$ defined on the new horizontally elongated square lattice.
If one follows the algebra transformation carefully, one can find that the new emergent fermion parity operator on the new lattice composed of the remaining entangled qubits originally comes from the emergent fermion parity operator of a pink $A$ face together with the emergent fermion parity operator of a blue $B$ face next to it. In this case, since all emergent fermionic modes of the ground state on the original lattice are empty, the emergent fermion parity operator on the new lattice takes eigenvalue one for the transformed ground state.


\begin{figure}[t]
\begin{centering}
\includegraphics[width=\columnwidth]{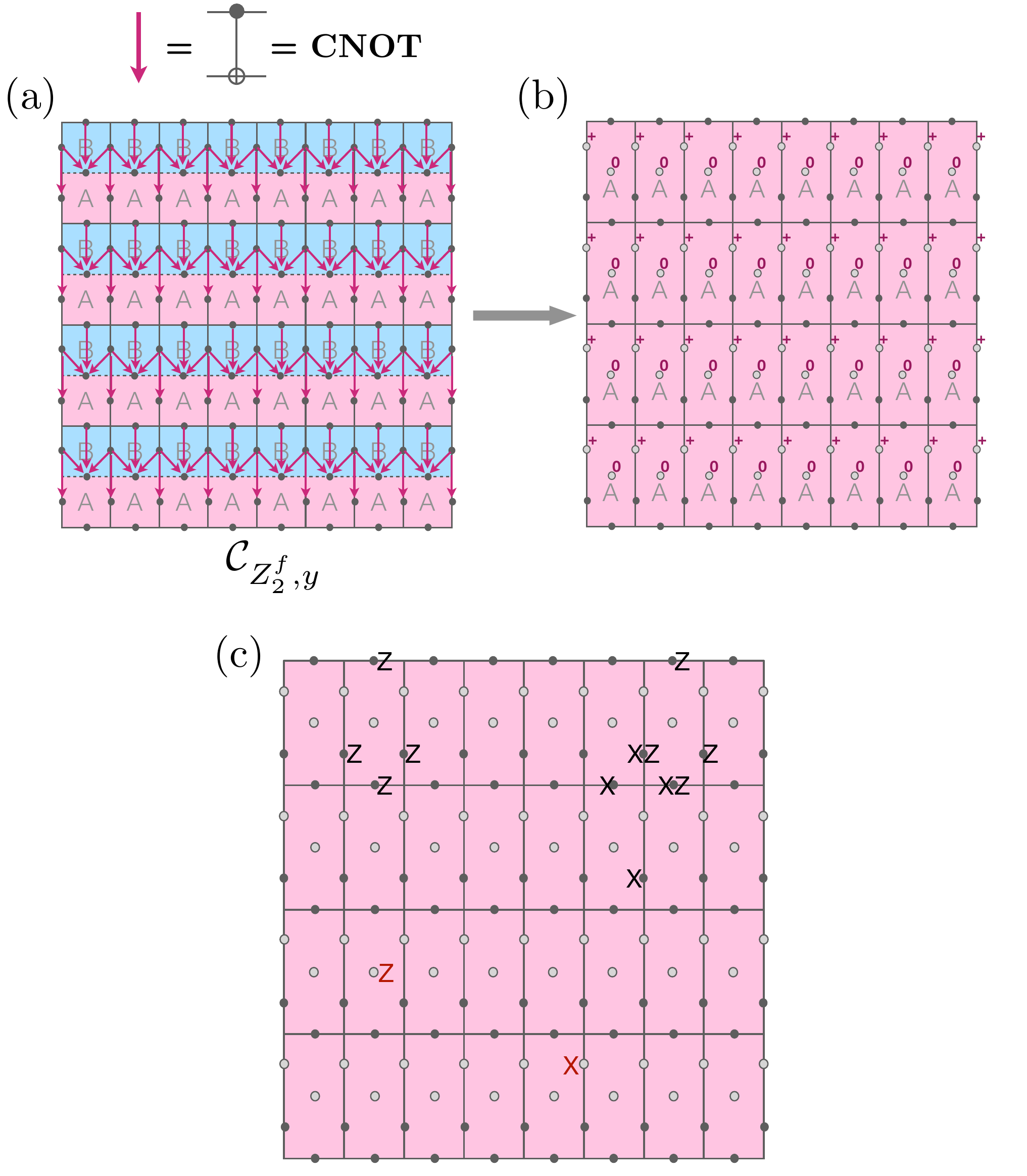}
\par\end{centering}
\centering{}\caption{
(a) The circuit $\mathcal{C}_{Z_{2}^{f},y}$ for a single step of vertical entanglement renormalization, which happens to be the same as $\mathcal{C}_{Z_{2},y}$ in Fig.~\ref{fig:MERA-ToricCode-verti}(a). The filled circles represent qubits (spins) constituting the toric code model. (b,c) The state of the system after the circuit has been applied. Unfilled circles are the qubits (spins) that have been disentangled by the circuit into $\ket{0}$ and  $\ket{+}$ states, as indicated by the labels in (b). (c) shows the new stabilizer generators.  The red single-site $Z$ and $X$ generators stabilize the disentangled qubits, while the black generators stabilize the pure $Z_2^f$ lattice gauge theory defined on the new vertically elongated square lattice.  The derivation of the new stabilizer generators is presented in Fig.~\ref{fig:MERA-Z2f-vert-calculation}.\label{fig:MERA-Z2f-vert}}
\end{figure}

\begin{figure}[t]
\begin{centering}
\includegraphics[width=\columnwidth]{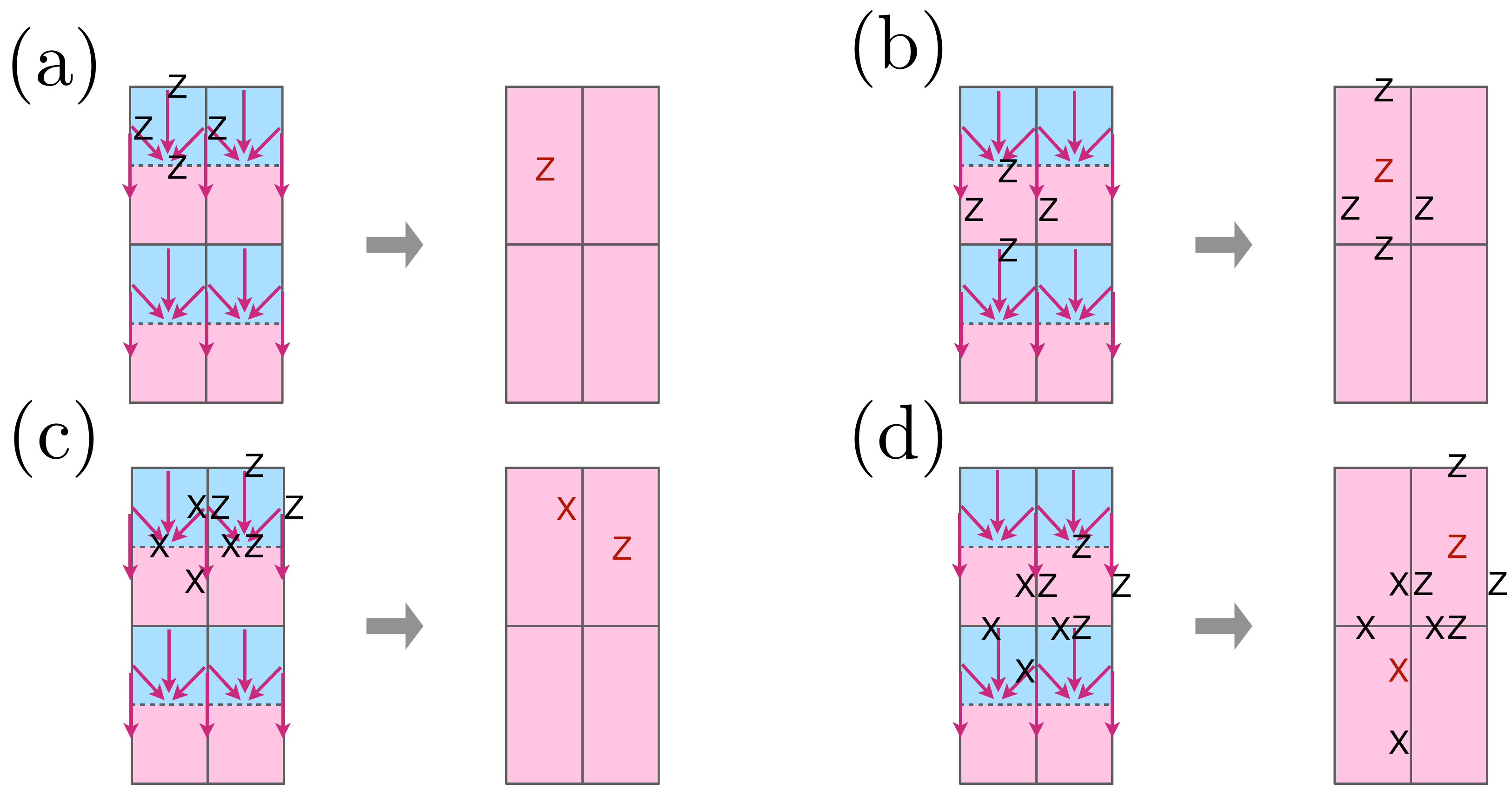}
\par\end{centering}
\centering{}\caption{
Transformation of the stabilizer generators of the pure $Z_{2}^{f}$ lattice gauge theory under conjugation by the vertical entanglement renormalization subcircuit $\mathcal{C}_{Z_{2}^{f},y}$ in Fig.~\ref{fig:MERA-Z2f-vert}(a). (a)(b) Transformation of the emergent fermion parity operators $W_{f}$. (c)(d) Transformation of the flux measuring operators $F_{v}$. The new stabilizer group generated by the operators on the right-hand sides of the subfigures is the same as the stabilizer group generated by the operators in Fig.~\ref{fig:MERA-Z2f-vert}(c). The red Pauli operators in the subfigures are the red single-qubit stabilizer generators in Fig.~\ref{fig:MERA-Z2f-vert}(c) acting on the disentangled qubits. \label{fig:MERA-Z2f-vert-calculation}}
\end{figure}

A single step of vertical entanglement renormalization is implemented by the subcircuit $\mathcal{C}_{Z_{2}^{f},y}$ shown in Fig.~\ref{fig:MERA-Z2f-vert}(a), which is the same as the subcircuit in Fig.~\ref{fig:MERA-ToricCode-verti}. We can perform analysis similar to horizontal entanglement renormalization. After vertical entanglement renormalization, we obtain a new set of stabilizer generators shown in Fig.~\ref{fig:MERA-Z2f-vert}(c) and computed by recombining the results of the transformation of the old stabilizer generators in Fig.~\ref{fig:MERA-Z2f-vert-calculation}. The new generators in Fig.~\ref{fig:MERA-Z2f-vert}(c) are red single-qubit Pauli operators stabilizing the states $\ket{0}$ and $\ket{+}$ of disentangled qubits between and below blue $B$ faces and emergent fermion parity operators and flux measuring operators (colored in black) with respect to the new vertically elongated lattice defined by the remaining entangled qubits. The disentangled qubits are represented by the unfilled circles in Fig.~\ref{fig:MERA-Z2f-vert}(b). 


If we alternately apply the subcircuits $\mathcal{C}_{Z_{2}^{f},x}$ and $\mathcal{C}_{Z_{2}^{f},y}$ at different length scales labeled by $s$, we arrive at the scale-invariant entanglement renormalization circuit tower shown in Fig.~\ref{fig:layoutofMERAQLE}. However, it contains only the strictly-local circuit components $\mathcal{C}_{Z_{2}^{f},x}^s$ and $\mathcal{C}_{Z_{2}^{f},y}^s$ and still does not have the quasi-local evolution and the auxiliary circuit components. The ground state of the pure $Z_{2}^{f}$ lattice gauge theory is a fixed-point wavefunction throughout the whole scale-invariant entanglement renormalization procedure.

\subsection{MERAQLE for the lattice Ising TQFT}

\label{subsec:MERAQLE-for-Ising}
So far, we have not harnessed the full power of the MERAQLE circuit as promised in Sec.~\ref{sec:Introduction}. We have only used either the conventional MERA layers involving only strictly local gates illustrated as $\mathcal{C}_{Z_{2}^{f},x}^{s}$ ($\mathcal{C}_{Z_{2}^{f},y}^{s}$) for the toric code in Sec.~\ref{subsec:MERAforToricCode} \footnote{Recall that $\mathcal{C}_{Z_{2},x}^{s}=\mathcal{C}_{Z_{2}^{f},x}^{s}$.} and the pure $Z_{2}^{f}$ lattice gauge theory in Sec.~\ref{subsec:MERAQLE-for-pure} or the quasi-local evolution circuit components illustrated as $\mathcal{C}_{\mathrm{ql},\,x}^{s}$ ($\mathcal{C}_{\mathrm{ql},\,y}^{s}$) in Fig.~\ref{fig:layoutofMERAQLE} for the entanglement renormalization of the lattice $p_{x}+ip_{y}$ topological superconductor in Sec.~\ref{subsec:QLEfortopological superconductor}. We have not combined the concept of MERA and the concept of quasi-local evolution yet. In addition, the models we have renormalized so far were either chiral but non-interacting or interacting but non-chiral. Starting from this subsection, we are going to see the power of MERAQLE circuits to entanglement-renormalize interacting chiral models.

\begin{figure*}[t]
\begin{centering}
\includegraphics[scale=0.44]{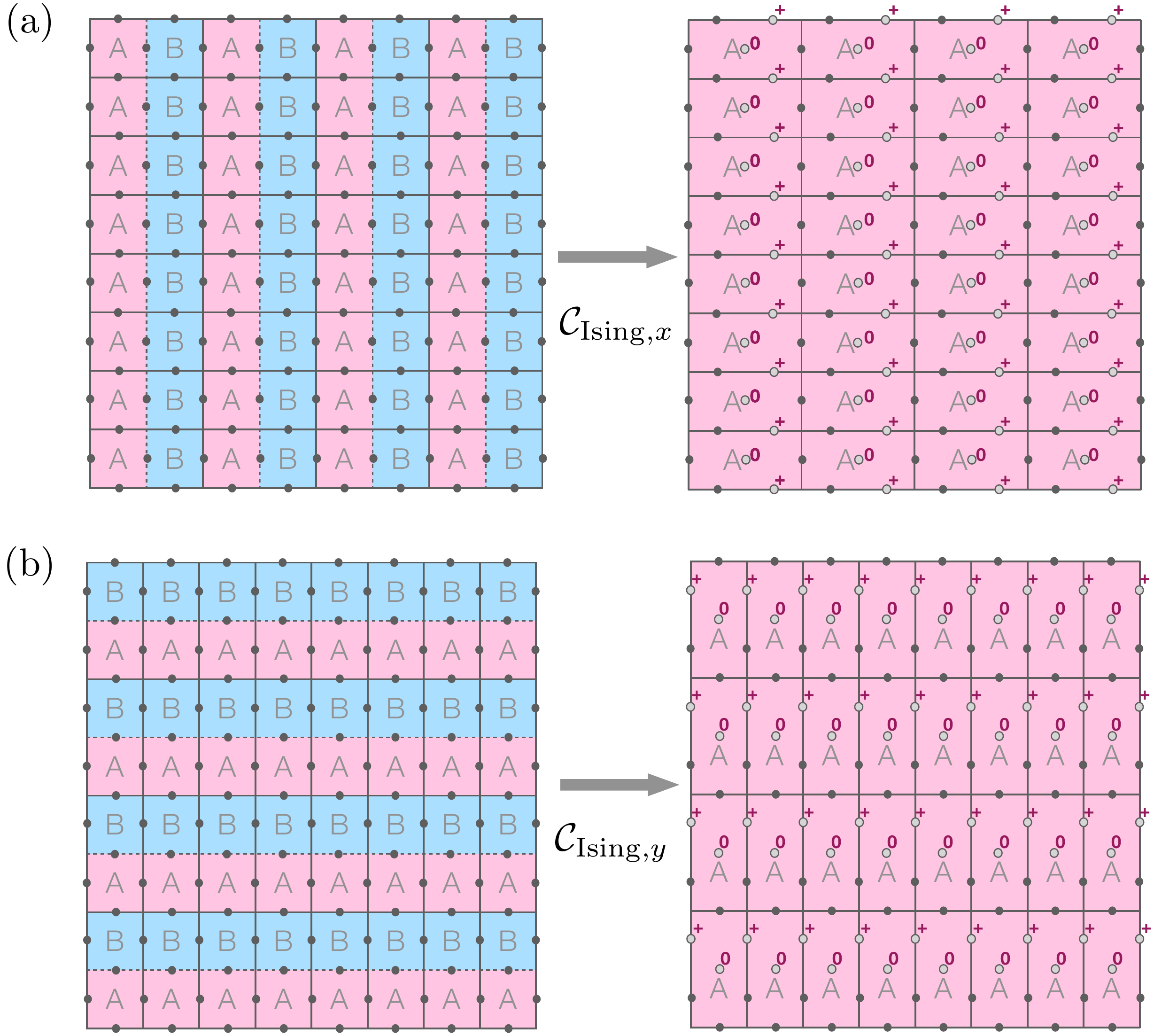}
\par\end{centering}
\centering{}\caption{We aim to (a) horizontally and (b) vertically entanglement renormalize the Ising TQFT model. We want to design subcircuits  ${\cal C}_{{\rm Ising},\,x}$ and  ${\cal C}_{{\rm Ising},\,y}$ such that we effectively have a new (elongated) square lattice with unit cells doubled in size in the horizontal direction and the vertical direction, respectively. In the final state, the spins represented by unfilled circles are disentangled spins in the $\ket{0}$ or $\ket{+}$ states. The remaining entangled spins of the final state represented by filled circles form the edges of the new coarse-grained square lattice. In addition, we want to have the initial and the final spin state having the interpretation of emergent fermions in the lattice $p_{x}+ip_{y}$ topological superconductor state. The details of the subcircuit ${\cal C}_{{\rm Ising},\,x}$ are provided in Fig.~\ref{fig:Isinghorizontalrenormalizaitondetail}. The details of the subcircuit ${\cal C}_{{\rm Ising},\,y}$ are provided in Fig.~\ref{fig:Isingverticalrenormalizaitondetail}. \label{fig:IsingTQFTrenormalizationgoal}}
\end{figure*}

In this subsection, we describe the MERAQLE circuit for the lattice Ising TQFT model constructed in Sec.~\ref{subsec:Gauged-superconductor}. We will formulate the subcircuit for a single step of horizontal entanglement renormalization and the subcircuit for a single step of vertical entanglement renormalization separately. Our goal for a single step of horizontal entanglement renormalization of the lattice Ising TQFT is shown in Fig.~\ref{fig:IsingTQFTrenormalizationgoal}(a).\begin{figure*}[t]
\begin{centering}
\includegraphics[scale=0.50]{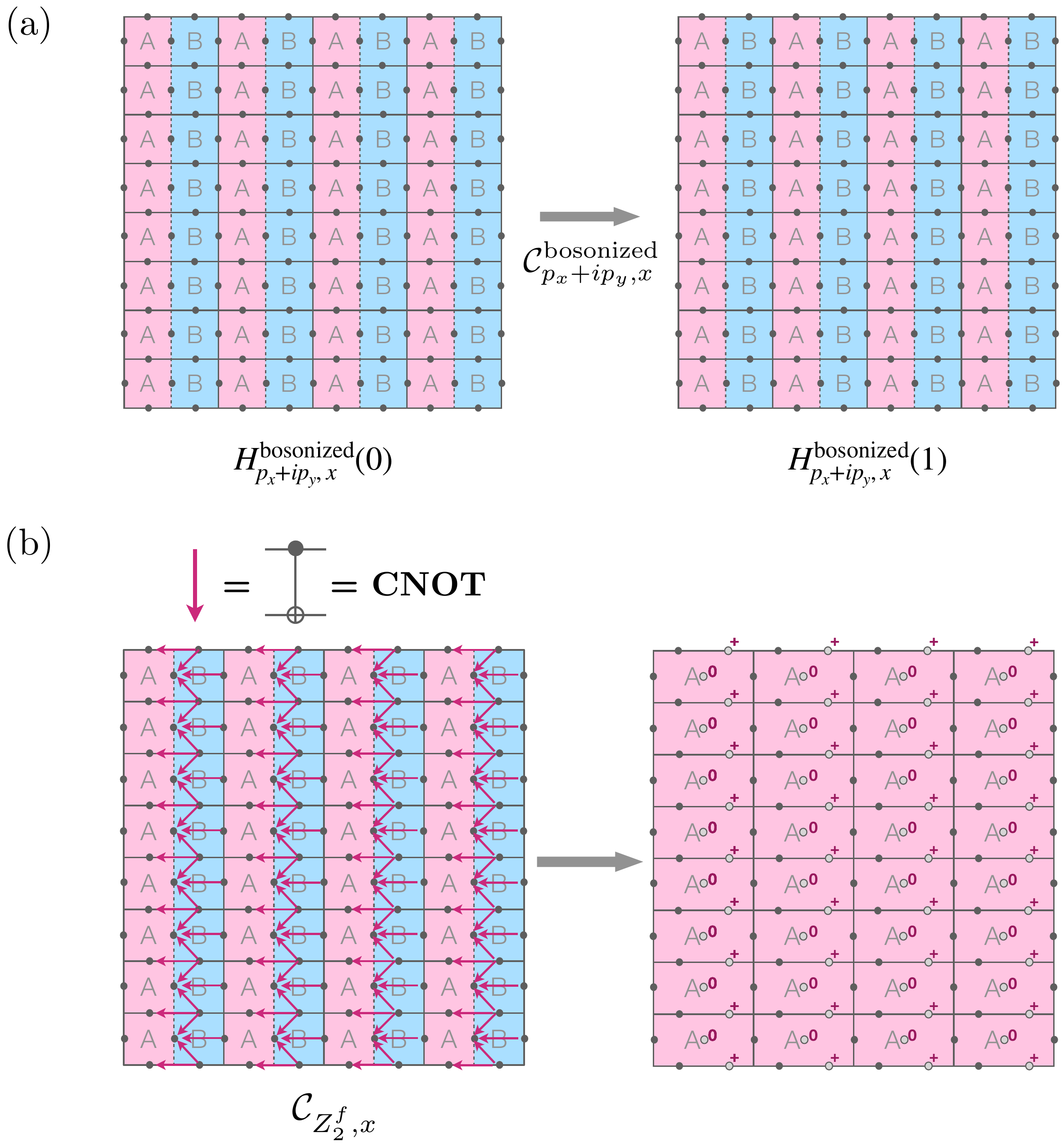}
\par\end{centering}
\centering{}\caption{The horizontal entanglement renormalization subcircuit $\mathcal{C}_{\mathrm{Ising},\,x}$ is separated into two circuit components $\mathcal{C}_{\mathrm{Ising},\,x}=\mathcal{C}_{Z_{2}^{f},x}\mathcal{C}_{p_{x}+ip_{y},x}^{\mathrm{bosonized}}$. (a) The circuit component $\mathcal{C}_{p_{x}+ip_{y},x}^{\mathrm{bosonized}}$ based on quasi-local evolution illustrated on the right of Fig.~\ref{fig:strictlocalityandquasilocality}(a). It can be written as the bosonization of the quantum circuit ${\cal C}_{p_{x}+ip_{y},\,x}$ constructed in Sec.~\ref{subsec:QLEfortopological superconductor}. The circuit component $\mathcal{C}_{p_{x}+ip_{y},x}^{\mathrm{bosonized}}$ takes the emergent fermions from a $p_{x}+ip_{y}$ topological superconductor state on both the pink $A$ faces and the blue $B$ faces to the $p_{x}+ip_{y}$ topological superconductor state on the pink $A$ faces only, leaving the blue $B$ faces with empty emergent fermionic modes.  (b) The circuit component $\mathcal{C}_{Z_{2}^{f},x}$, i.e.~the horizontal entanglement renormalization subcircuit $\mathcal{C}_{Z_{2}^{f},x}$ described in Fig.~\ref{fig:MERA-Z2f-hori}.  Initially, the emergent fermionic modes on the blue $B$ faces are empty and have fermion parity $+1$. After we apply the circuit component $\mathcal{C}_{Z_{2}^{f},x}$, the bottom and the left spins of the $B$ faces become disentangled, as in Fig.~\ref{fig:MERA-Z2f-hori}(b). The disentangled spins are shown as unfilled circles. The new lattice is defined by the remaining entangled qubits, represented by filled circles. Effectively, we have larger faces horizontally. The state of the new emergent fermions on the new lattice will be the ground state of the $p_{x}+ip_{y}$ topological superconductor. \label{fig:Isinghorizontalrenormalizaitondetail}}
\end{figure*}\begin{figure*}[t]
\begin{centering}
\includegraphics[scale=0.50]{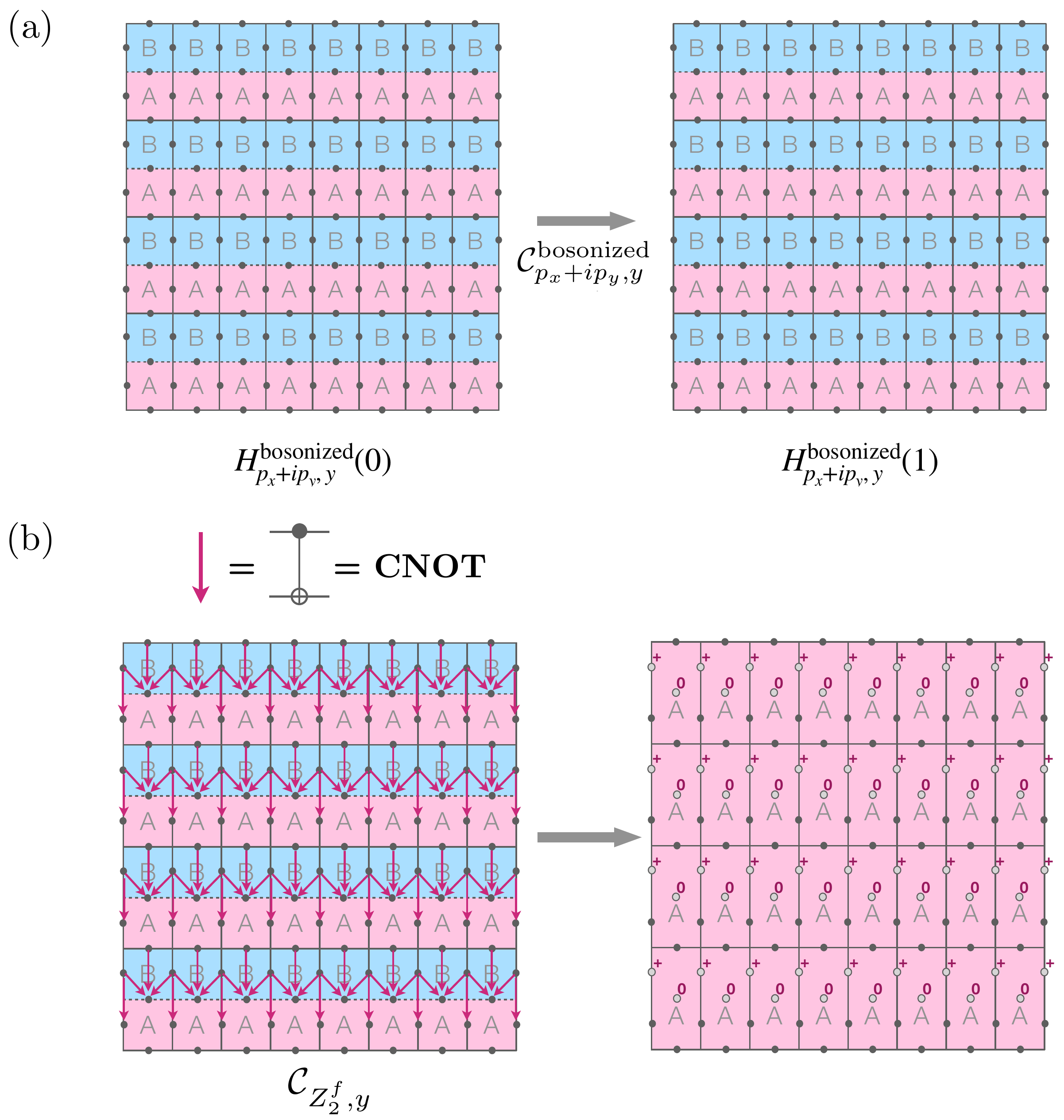}
\par\end{centering}
\centering{}\caption{The vertical entanglement renormalization subcircuit $\mathcal{C}_{\mathrm{Ising},\,y}$ is separated into two circuit components $\mathcal{C}_{\mathrm{Ising},\,y}=\mathcal{C}_{Z_{2}^{f},y}\mathcal{C}_{p_{x}+ip_{y},y}^{\mathrm{bosonized}}$. (a) The circuit component $\mathcal{C}_{p_{x}+ip_{y},y}^{\mathrm{bosonized}}$ based on quasi-local evolution illustrated on the right of Fig.~\ref{fig:strictlocalityandquasilocality}(a).  It can be written as the bosonization of the quantum circuit ${\cal C}_{p_{x}+ip_{y},\,y}$ constructed in Sec.~\ref{subsec:QLEfortopological superconductor}. The circuit component $\mathcal{C}_{p_{x}+ip_{y},\,y}^{\mathrm{bosonized}}$ takes the emergent fermions from a $p_{x}+ip_{y}$ topological superconductor state on both the pink $A$ faces and the blue $B$ faces to the $p_{x}+ip_{y}$ topological superconductor state on the pink $A$ faces only, leaving blue $B$ faces with empty emergent fermionic modes. (b) The circuit component $\mathcal{C}_{Z_{2}^{f},y}$, i.e.~the vertical entanglement renormalization subcircuit $\mathcal{C}_{Z_{2}^{f},y}$ described in Fig.~\ref{fig:MERA-Z2f-vert}. Initially, the emergent fermionic modes on the blue $B$ faces are empty and have fermion parity $+1$. After we apply the circuit $\mathcal{C}_{Z_{2}^{f},y}$, the bottom and the left spins of the blue $B$ faces become disentangled, as in Fig.~\ref{fig:MERA-Z2f-vert}(b). The disentangled spins are shown as unfilled circles. The new lattice is defined by the remaining entangled qubits, represented by filled circles. Effectively, we have larger faces vertically. The state of the new emergent fermions on the new lattice will be the ground state of the $p_{x}+ip_{y}$ topological superconductor. \label{fig:Isingverticalrenormalizaitondetail}}
\end{figure*}
We aim to construct a quantum circuit ${\cal {\cal C}}_{{\rm Ising},\,x}$ such that, when we apply it to the ground state of the lattice Ising TQFT Hamiltonian $H_{{\rm Ising\,TQFT}}$ in Eq.~(\ref{eq:IsingTQFTHamiltonian}) under the zero-flux condition $F_v=1$, the subcircuit disentangles half of the spins, effectively generating a horizontally coarse-grained lattice with the size of the unit cells doubled horizontally. Furthermore, the state becomes the ground state of $H_{{\rm Ising\,TQFT}}$ defined on the new (elongated) square lattice under a new zero-flux condition. It means that we again have an interpretation of emergent fermions defined on the faces of the new lattice in the ground state of the lattice $p_{x}+ip_{y}$ topological superconductor. Our goal of a single step of vertical entanglement renormalization is similarly shown in Fig.~\ref{fig:IsingTQFTrenormalizationgoal}(b).

The central idea that we will use to realize this goal is that, for our lattice construction of the Ising TQFT, we have the dual interpretation in terms of emergent fermions living on the faces and prepared in the lattice $p_{x}+ip_{y}$ topological superconducting state with the quantum $Z_{2}^{f}$ gauge field background containing no flux. We will use as our inspiration the lessons learned in Sec.~\ref{subsec:MERAQLE-for-pure} for the pure $Z_{2}^{f}$ lattice gauge theory, which is the bosonization of a trivial insulator. Specifically, if we can empty half of the emergent fermion modes, then we can disentangle half of the constituent spins corresponding to those empty modes using the subcircuits $\mathcal{C}_{Z_{2}^{f},x}$ and $\mathcal{C}_{Z_{2}^{f},y}$ for the pure $Z_{2}^{f}$ lattice gauge theory. This is expected because, intuitively, the area around the emptied fermionic modes locally behaves like the ground state of the pure $Z_{2}^{f}$ lattice gauge theory. To be mathematically more precise, recall that, from the calculations in Fig.~\ref{fig:MERA-Z2f-hori-calulation}(a) and Fig.~\ref{fig:MERA-Z2f-vert-calculation}(a), we know that, intuitively, the emergent fermionic degrees of freedom in the empty modes of the blue $B$ faces can be shifted to the disentangled qubits in state $\ket{0}$ on the left of the blue $B$ faces under $\mathcal{C}_{Z_{2}^{f},x}$ or at the bottom of the blue $B$ faces under $\mathcal{C}_{Z_{2}^{f},y}$. Furthermore, from the zero-flux condition $F_{v}=1$ with $NE(v)$ being blue $B$ faces and from the emptiness of the emergent fermionic modes on the blue $B$ faces, through the calculations in Fig.~\ref{fig:MERA-Z2f-hori-calulation}(a,c) (Fig.~\ref{fig:MERA-Z2f-vert-calculation}(a,c)), we deduced that the qubits sitting at the bottom (on the left) of the blue $B$ faces should be in the $\ket{+}$ state after the circuit $\mathcal{C}_{Z_{2}^{f},x}$ ($\mathcal{C}_{Z_{2}^{f},y}$). To achieve a similar goal for the lattice Ising TQFT model, we here again introduce an $AB$ sublattice structure to the emergent fermions on the faces of the lattice Ising TQFT model. The sublattice structures for horizontal renormalization and vertical renormalization are both shown on the left-hand sides of the subfigures in Fig.~\ref{fig:IsingTQFTrenormalizationgoal}. We require that the blue $B$ faces here play the same role as the blue $B$ faces in the pure $Z_{2}^{f}$ lattice gauge theory and want to disentangle the spins to the left and at the bottom of the blue $B$ faces, as claimed in Fig. \ref{fig:IsingTQFTrenormalizationgoal}. Therefore, we need to find a way to empty the emergent fermionic modes on the blue $B$ faces. However, the way to empty half of the fermionic modes from a lattice $p_{x}+ip_{y}$ topological superconductor is nothing but the quasi-adiabatic evolution we introduced in Sec.~\ref{subsec:QLEfortopological superconductor}! Recall that the quasi-adiabatic evolution empties the blue $B$ fermionic modes while keeping the remaining modes in the lattice $p_{x}+ip_{y}$ topological superconducting state. The only difference between Sec.~\ref{subsec:QLEfortopological superconductor} and the present section is that, in the former, the fermions are the fundamental constituents, whereas in the latter they serve as the emergent degrees of freedom on the faces. Thus, we need to adapt the quasi-adiabatic evolution from Sec.~\ref{subsec:QLEfortopological superconductor} to the spin system studied here. To construct the corresponding spin circuit, we will use the bosonization duality, with the zero-flux condition $F_v=1$ as a hard constraint on the spin side.

We first study the process that will horizontally entanglement-renormalize the emergent $p_{x}+ip_{y}$ superconducting fermions.  Here, we only want to work in the zero-flux sector $F_v=1$ and change only the emergent fermion configuration, so, for our convenience, we temporarily impose a hard zero-flux constraint on the Hilbert space. Since there exists a gapped adiabatic process between the $p_{x}+ip_{y}$ superconductor Hamiltonians before and after the coarse-graining process in Sec.~\ref{subsec:QLEfortopological superconductor}, there should also be a corresponding gapped adiabatic process for spins under the zero-flux constraint by choosing the interpolating spin Hamiltonian as the bosonization of the interpolating fermionic Hamiltonian $H_{p_{x}+ip_{y},\,x}(\lambda)$ in Eq.~(\ref{eq:p+ipadiabaticHamiltonian}). The bosonization duality ensures that the energy spectra on both sides are the same. With this gapped adiabatic path, we are able to construct the quasi-adiabatic quantum circuit that takes the ground state of the bosonized $p_{x}+ip_{y}$ topological superconductor Hamiltonian, i.e., $H_{\mathrm{Ising\,TQFT}}$ in Eq.~(\ref{eq:IsingTQFTHamiltonian}), to the ground state of the bosonized Hamiltonian of the $p_{x}+ip_{y}$ superconductor with active fermionic hoppings and pairings on every other site horizontally. The resulting state is nothing but the desired state with $F_v=1$ and an alternating pattern of pink $A$ sites being in the $p_{x}+ip_{y}$ topological superconducting state and blue $B$ sites representing empty emergent fermionic modes on the faces. (Note that here we only entanglement-renormalize the emergent fermionic modes. The underlying spins are still entangled.)

To be mathematically more precise, the adiabatic gapped path for spins with $\lambda\in[0,1]$ is described by the interpolating spin Hamiltonian
\begin{align}
H_{p_{x}+ip_{y},\,x}^{\mathrm{bosonized}}(\lambda) & \equiv \left\{ H_{p_{x}+ip_{y},\,x}(\lambda)\right\} ^{\mathrm{bosonized}}.\label{eq:bosonizedp+ipadiabaticHamiltonian}
\end{align}
We define the quasi-adiabatic circuit corresponding to this gapped path of spins as ${\cal C}_{p_{x}+ip_{y},\,x}^{{\rm bosonized}}$ with the quasi-adiabatic continuation operator given by
\begin{align}
&\mathcal{D}_{p_{x}+ip_{y},\,x}^{\mathrm{bosonized}}(\lambda)  \equiv-i\int_{-\infty}^{\infty}dt\,F(E_\mathrm{gap}t)\,\times\nonumber \\ 
& e^{iH_{p_{x}+ip_{y},\,x}^{\mathrm{bosonized}}(\lambda)t}\,\partial_{\lambda}H_{p_{x}+ip_{y},\,x}^{\mathrm{bosonized}}(\lambda)\,e^{-iH_{p_{x}+ip_{y},\,x}^{\mathrm{bosonized}}(\lambda)t}.
\label{eq:quasi-adiabatic-continuation-operator-for-bosonizedp+ipadiabaticHamiltonian}
\end{align}
The quasi-adiabatic continuation operator is quasi-local in the sense of Fig.~\ref{fig:strictlocalityandquasilocality}. From the bosonization duality, the quasi-adiabatic circuit is just the  bosonization of the unitary operator ${\cal C}_{p_{x}+ip_{y},\,x}^{{\rm }}$ discussed in Sec.~\ref{subsec:QLEfortopological superconductor} (with the layer number  $s$  superscript dropped):
\begin{align}
{\cal C}_{p_{x}+ip_{y},\,x}^{{\rm bosonized}}=\left\{ {\cal C}_{p_{x}+ip_{y},\,x}^{{\rm }}\right\} ^{\mathrm{bosonized}}.\label{eq:bosoniziedp+ipadiabaticcircuit}
\end{align}

We constructed the circuit ${\cal C}_{p_{x}+ip_{y},\,x}^{{\rm bosonized}}$ under the zero-flux constraint. However, as we mentioned before, the actual spin Hilbert space, composed of a tensor product of spins, does not inherently have the zero-flux constraint \footnote{In fact, the MERA subcircuit $C_{Z_2^f,\,x}$, which will be discussed in the following paragraphs, does not commute with the zero-flux constraint. In fact, $C_{Z_2^f,\,x}$ creates a new zero-flux constraint on the new coarse-grained square lattice.}. The constraint is a convenient tool for deriving the desired circuit using the bosonization duality.
One might ask whether the circuit constructed here is not a unitary operator if the zero-flux constraint is lifted. In general, an operator being unitary under a constraint does not necessarily imply that it remains unitary when the constraint is lifted. However, in our case, this is still true. This is because the operator generating the circuit described in Eq.~(\ref{eq:quasi-adiabatic-continuation-operator-for-bosonizedp+ipadiabaticHamiltonian}) remains Hermitian even in the absence of the zero-flux constraint since the bosonization mapping described in Fig.~\ref{fig:fermionstospins} preserves Hermiticity of operators.

Using a similar construction, we can obtain a quasi-adiabatic circuit ${\cal C}_{p_{x}+ip_{y},\,y}^{{\rm bosonized}}$ for vertical entanglement renormalization of the emergent $p_{x}+ip_{y}$ topological superconducting fermions by directly bosonizing the fermionic quasi-adiabatic unitary operator ${\cal C}_{p_{x}+ip_{y},\,y}$ for entanglement renormalization in the $y$ direction in Sec.~\ref{subsec:QLEfortopological superconductor}.

Experimentally, instead of using the quasi-adiabatic circuit, we can  implement the adiabatic evolution itself over a long but finite time period and incur an error. In that case, instead of having the zero-flux constraint, we can add back the flux penalty Hamiltonian $H^F_{\mathrm{penalty}}$ with a strong enough flux energy parameter $\Delta_{\Phi}$ such that the vortices are gapped along the entire path of interpolating Hamiltonians and $F_{v}=1$ is satisfied for their ground states. At the same time, we would like to keep $\Delta_{\Phi}$ small enough to be experimentally realizable 
\footnote{In fact, even though we say that we need the flux penalty Hamiltonian $H^F_{\mathrm{penalty}}$ and a vortex gap from a large flux energy parameter $\Delta_{\Phi}$ for adiabatic evolution, they are not strictly necessary. Indeed, the interpolating Hamiltonian commutes with the flux measuring operator $F_v$ and will therefore keep the quantum state in the zero flux sector throughout the evolution, independently of whether we have a flux penalty Hamiltonian or not.
However, for experimental realizations, it might be hard to perfectly engineer a Hamiltonian commuting with the flux measuring operators $F_v$. In such cases, the addition of the flux penalty Hamiltonian with a large flux energy parameter might be helpful, as the vortex gap would ensure the stability of the adiabatic evolution to experimental imperfections.}.

As we claimed above, the situation of the unoccupancy of the blue $B$ faces subsequent to the application of ${\cal C}_{p_{x}+ip_{y},\,x}^{{\rm bosonized}}$ (${\cal C}_{p_{x}+ip_{y},\,y}^{{\rm bosonized}}$) is similar to blue $B$ faces of the ground state of the trivial fermionic insulator $H_{{\rm trivial\,insulator}}$, which in turn is the state of the emergent fermions in the pure $Z_{2}^{f}$ lattice gauge theory described by $H_{Z_{2}^{f}}$. It is therefore tempting to apply the MERA subcircuits $\mathcal{C}_{Z_{2}^{f},x}$ and $\mathcal{C}_{Z_{2}^{f},y}$ used for the pure $Z_{2}^{f}$ lattice gauge theory in Sec.~\ref{subsec:MERAQLE-for-pure} as our next step of the entanglement renormalization procedure to disentangle the qubits on the left and at the bottom of the blue $B$ faces. We now confirm that $\mathcal{C}_{Z_{2}^{f},x}$ and $\mathcal{C}_{Z_{2}^{f},y}$ are indeed the right circuits. Notice that the emptiness of the blue $B$ faces guarantees that the ground state is stabilized by the emergent fermion parity operator $W_{f}$ on the blue $B$ faces. In addition, the ground state is stabilized by all the flux measuring operators $F_{v}$: $F_v=1$. Hence, we can use the transformation of the stabilizer generators shown in Fig.~\ref{fig:MERA-Z2f-hori-calulation}(a,c,d) and Fig.~\ref{fig:MERA-Z2f-vert-calculation}(a,c,d)for the case of the pure $Z_{2}^{f}$ lattice gauge theory to deduce how the ground state transforms in the present case. 
Putting together the results of the transformations in Fig.~\ref{fig:MERA-Z2f-hori-calulation}(a,c) and Fig.~\ref{fig:MERA-Z2f-vert-calculation}(a,c), we find that the red single-qubit Pauli operators in Fig.~\ref{fig:MERA-Z2f-hori}(c) and Fig.~\ref{fig:MERA-Z2f-vert}(c) are again stabilizer generators in the present case. These single-qubit Pauli operators require again that the qubits on the left and at the bottom of the $B$ faces are disentangled into the $\ket{0}$ and $\ket{+}$ states, as claimed in Fig.~\ref{fig:IsingTQFTrenormalizationgoal}. Here, the remaining entangled qubits form a new (elongated) square lattice with qubits defined on the edges. By following the transformation of the flux measuring operators in Fig.~\ref{fig:MERA-Z2f-hori-calulation}(d) and Fig.~\ref{fig:MERA-Z2f-vert-calculation}(d), we obtain new flux measuring operators defined on the new elongated square lattice up to single-qubit Pauli operators colored in red. Since those single-qubit Pauli operators are already stabilizer generators, we can safely remove them and conclude that the new flux measuring operators are indeed stabilizer generators.

\begin{figure}[t]
\begin{centering}
\includegraphics[width=\columnwidth]{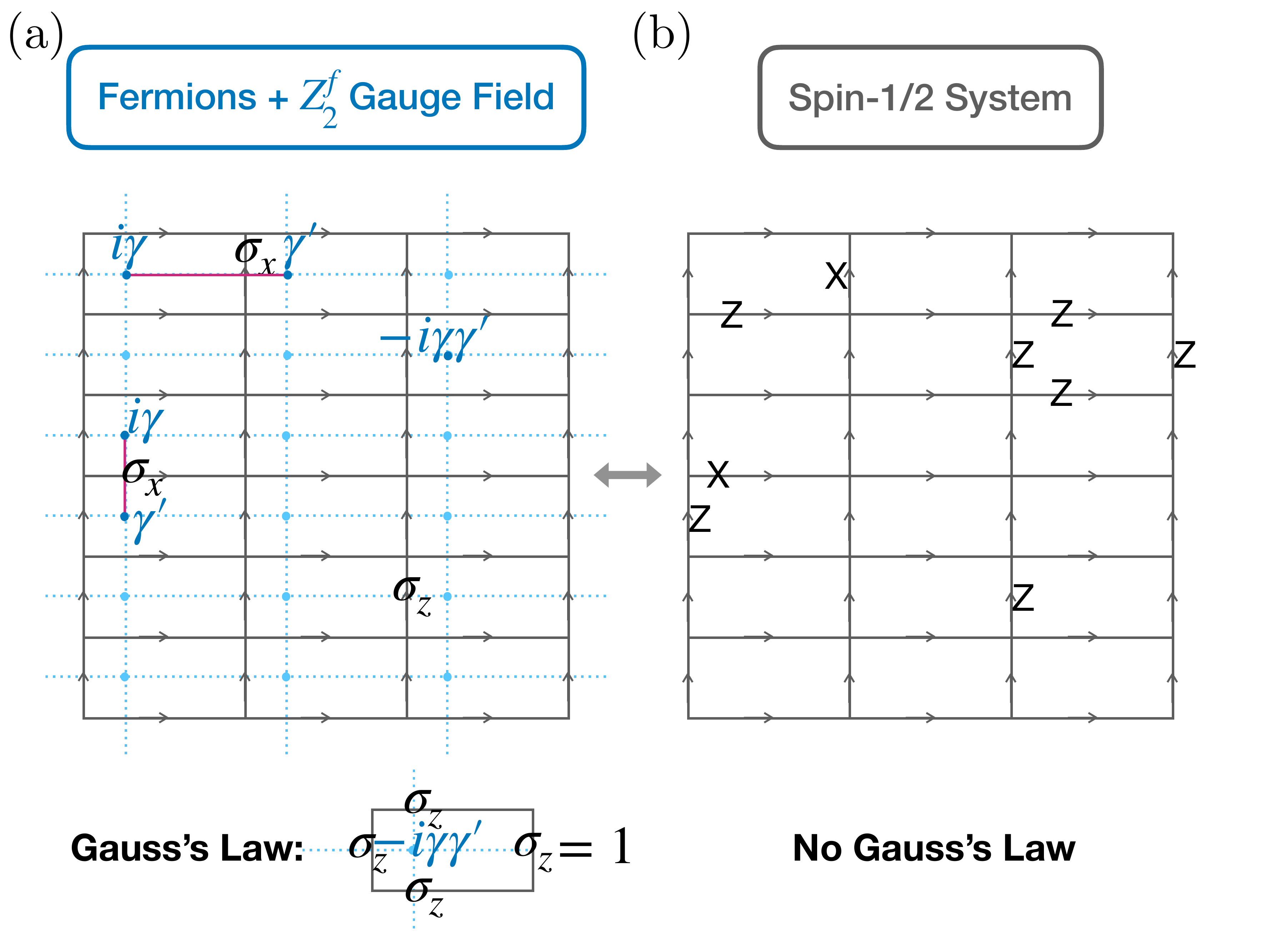}
\par\end{centering}
\centering{}\caption{The edge orientation assignments and the corresponding algebra isomorphism between (a) the theory of fermions coupled to a $Z_{2}^{f}$ gauge field and (b) the theory of spins for the new horizontally coarse-grained lattice. After the horizontal entanglement renormalization circuit component $\mathcal{C}_{Z_{2}^{f},x}$, some spins are disentangled, as shown in Fig.~\ref{fig:IsingTQFTrenormalizationgoal}(a). We associate edge orientation assignments with the new lattice edges formed by the remaining spins. Compared to Fig.~\ref{fig:branchingsquare}, the lattice is elongated horizontally. We intentionally put the operators on the horizontal edges to the left of the midpoint to remember the original positions of the corresponding spins on the old lattice before coarse-graining. 
\label{fig:newbranchingsquare}}
\end{figure}

\begin{figure}[t]
\begin{centering}
\includegraphics[width=\columnwidth]{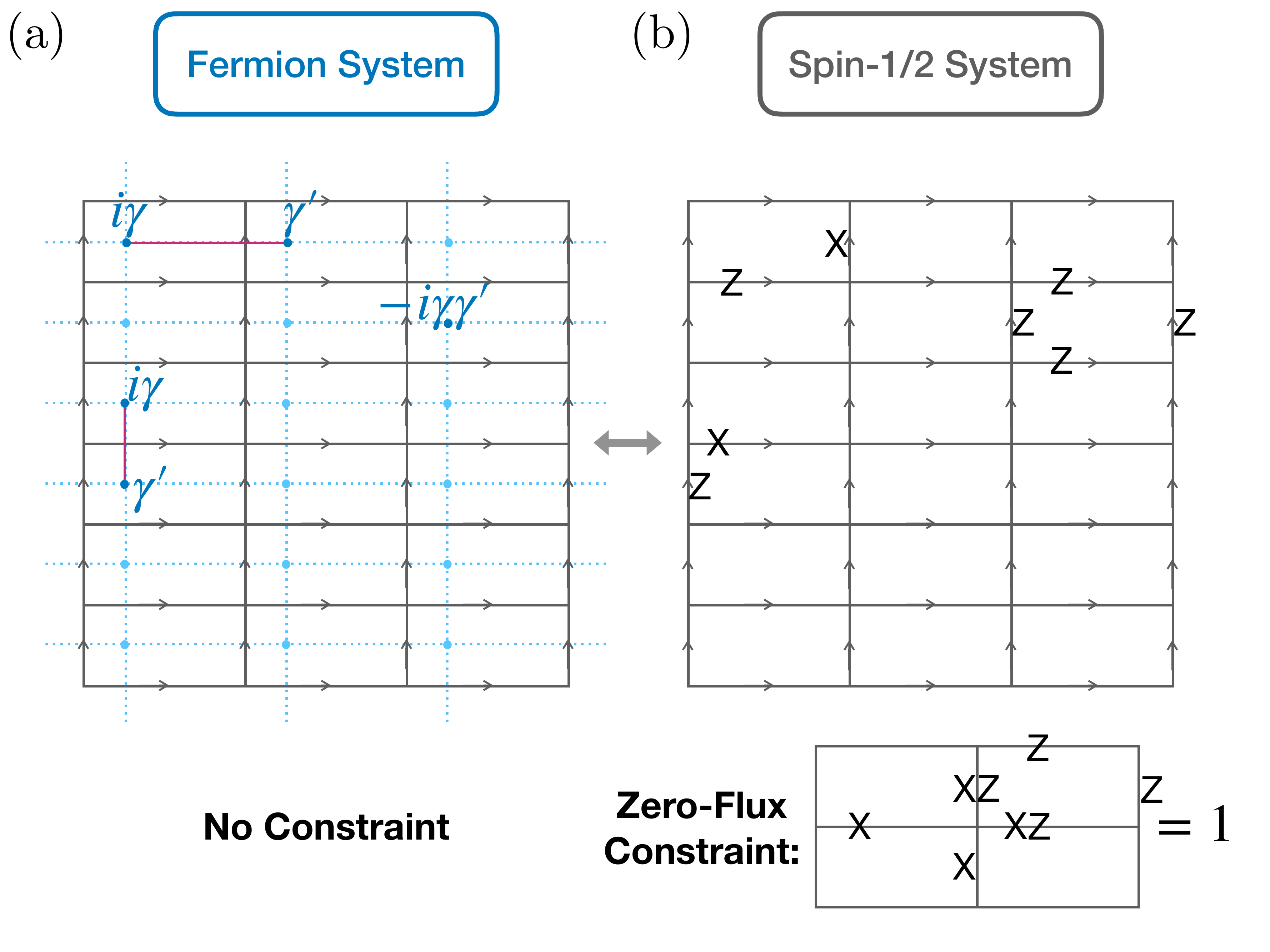}
\par\end{centering}
\centering{}\caption{Here we have a new bosonization duality between (a) the pure fermionic theory and (b) the pure spin theory on the new horizontally coarse-grained (elongated) square lattice by imposing a new zero-flux constraint $F_v=1$ on the spin side. The fermions live on the horizontally elongated faces of the coarse-grained square lattice, while the spins live on its edges. \label{fig:newfermionstospins}}
\end{figure}

Having obtained, as a result of horizontal entanglement renormalization, a spin model for the remaining spins defined on a new coarse-grained (elongated) square lattice, let us reinterpret this model in the dual picture consisting of fermions coupled to a $Z_{2}^{f}$ gauge field. We again assign edge orientations to the new coarse-grained lattice system, where all horizontal edges are oriented east, and all vertical edges are all oriented north. These edge orientation assignments are shown in Fig.~\ref{fig:newbranchingsquare},
where the emergent fermionic modes live on the new horizontally elongated faces and the $Z_2^f$ gauge variables live on the edges of the new horizontally elongated lattice. The existence of the new flux-measuring operators as stabilizer generators indicates that the new state satisfies the new zero-flux condition on the coarse-grained lattice, which allows us to define an associated new bosonization duality between pure fermions and pure spins for the coarse-grained lattice as shown in Fig.~\ref{fig:newfermionstospins},
provided that we impose the new zero-flux constraint $F_v=1$ on the spin side.  The edge orientation assignments and the bosonization duality under the new zero-flux constraint of the vertically coarse-grained square lattice can be drawn similarly. So, what is the behavior of the fermions of the horizontally coarse-grained lattice in Fig.~\ref{fig:newbranchingsquare} or Fig.~\ref{fig:newfermionstospins} and the behavior of the fermions of the vertically coarse-grained lattice? Since our goal is to have horizontal and vertical entanglement renormalization subcircuits with the lattice Ising TQFT ground state as a fixed-point wavefunction, we hope that the emergent fermions defined on the new faces of the horizontally or vertically coarse-grained lattice are still in the lattice $p_{x}+ip_{y}$ topological superconducting state. In Appendix~\ref{sec:appendix-compatibility}, we show that this is indeed the case. In fact, Appendix~\ref{sec:appendix-compatibility} shows a stronger statement. There, we show that, with the $AB$ sublattice structure on the faces of the original square lattice, whether horizontal or vertical, if the emergent fermionic modes on the blue $B$ faces are empty, and if the zero-flux condition $F_{v}=1$ is satisfied, the emergent fermion on a given elongated face of the new lattice will be just the original emergent fermion on the pink $A$ face of the old square lattice enclosed by that elongated face. Effectively, we can say that the coarse-graining operation $\mathcal{C}_{Z_{2}^{f},x}(\mathcal{C}_{Z_{2}^{f},y})$ makes the blue $B$ faces disappear and elongates the pink $A$ faces. Figuratively speaking, the pink $A$ faces ``consume" the blue $B$ faces \footnote{We can also see this phenomenon via the following argument.
Fig.~\ref{fig:MERA-Z2f-hori-calulation}(b) and Fig.~\ref{fig:MERA-Z2f-vert-calculation}(b) show how the emergent fermion parity operators on the pink $A$ faces transform under coarse-graining, while Fig.~\ref{fig:MERA-Z2f-hori-calulation}(a) and Fig.~\ref{fig:MERA-Z2f-vert-calculation}(a) show the corresponding transformation for the blue $B$ faces. We see that the emergent fermion parity operator of a new elongated face $f$ is equal to the product of transformed parity operators (i.e.\ the total parity operator) on the old pink $A$ face and the old blue $B$ face within face $f$. Since the old emergent fermion parity of the blue $B$ face is even for the old ground state, the emergent fermion parity operator of the new face $f$ corresponds to the emergent fermion parity on the old pink $A$ face.}. Since, before the application of $\mathcal{C}_{Z_{2}^{f},x}(\mathcal{C}_{Z_{2}^{f},y})$, the emergent fermions on the pink $A$ faces are in the lattice $p_{x}+ip_{y}$ topological superconducting state, the final emergent fermionic modes on the faces of the new elongated lattice are also in the lattice $p_{x}+ip_{y}$ topological superconducting state.

To summarize, the subcircuit $\mathcal{C}_{\mathrm{Ising},\,x}$ for a single step of horizontal entanglement renormalization of the lattice Ising TQFT model  is composed of two circuit components, $\mathcal{C}_{\mathrm{Ising},\,x}=\mathcal{C}_{Z_{2}^{f},x}\mathcal{C}_{p_{x}+ip_{y},x}^{\mathrm{bosonized}}$, whereas the vertical counterpart $\mathcal{C}_{\mathrm{Ising},\,y}$ is composed of $\mathcal{C}_{\mathrm{Ising},\,y}=\mathcal{C}_{Z_{2}^{f},y}\mathcal{C}_{p_{x}+ip_{y},y}^{\mathrm{bosonized}}$. The entanglement renormalization subcircuits are schematically shown in Fig.~\ref{fig:Isinghorizontalrenormalizaitondetail} and Fig.~\ref{fig:Isingverticalrenormalizaitondetail}. Roughly speaking, the reason why the subcircuits have such decompositions is that the bosonic spin model $H_{\mathrm{Ising\,TQFT}}$ is dual to superconducting fermions in the $Z_{2}^{f}$ gauge field with no flux. Thus, in hindsight, it is natural to have separate circuit components related to the fermions and to the $Z_{2}^{f}$ gauge field.

The whole MERAQLE circuit will be a repeated application of the circuits ${\cal C}_{{\rm Ising},\,x}$ and ${\cal C}_{{\rm Ising},\,y}$. We disentangle half of the spins each time when we apply either ${\cal C}_{{\rm Ising},\,x}$ or ${\cal C}_{{\rm Ising},\,y}$. To formalize this construction, we use $s\in\mathbb{N}$ as a label for the scale (more precisely, the logarithm of the length scale) at which the system resides. Then the whole circuit can be written as 
\[
{\cal C}_{{\rm Ising}}=\prod_{s\in\mathbb{N}}\left({\cal C}_{{\rm Ising},\,y}^{s}\,{\cal C}_{{\rm Ising},\,x}^{s}\right).
\]
Therefore, we obtain a scale-invariant entanglement renormalization circuit for the lattice Ising TQFT system, which has intrinsic chiral topological order. The whole entanglement renormalization circuit ${\cal C}_{{\rm Ising}}$ follows the structure shown in Fig.~\ref{fig:layoutofMERAQLE}. The quasi-local evolution circuit components $\mathcal{C}_{\mathrm{ql},\,x}^{s}$ and $\mathcal{C}_{\mathrm{ql},\,y}^{s}$ are $\mathcal{C}_{p_{x}+ip_{y},\,x}^{\mathrm{bosonized}}$ and $\mathcal{C}_{p_{x}+ip_{y},\,y}^{\mathrm{bosonized}}$, respectively; the auxiliary circuit components are trivial here, $\mathcal{C}_{\mathrm{aux},\,x}^{s}=\mathcal{C}_{\mathrm{aux},\,y}^{s}=I$.

It is worth noting that, much like its dual $p_{x}+ip_{y}$ topological superconductor in Sec.~\ref{sec:-Topological-Superconductor}, the ground state of the lattice Ising TQFT Hamiltonian does not have a zero correlation length. However, this ground state nevertheless serves as the fixed-point wavefunction of the MERAQLE circuit we constructed. This is enabled by the quasi-locality of the MERAQLE circuit which increases the range of correlations reduced by the lattice coarse-graining procedure.

\subsection{MERAQLE for all the sixteenfold way chiral spin liquids}

\label{subsec:MERAQLE-for-16foldway}

Having discussed entanglement renormalization of the lattice Ising TQFT as the simplest demonstration of the power of MERA with quasi-local evolution to renormalize interacting chiral topologically ordered states, we are now ready to present the MERAQLE circuits for all Kitaev's sixteenfold way chiral spin liquids introduced in Sec.~\ref{subsec:Kitaev's-sixteen-foldway}.

\begin{figure*}[t]
\begin{centering}
\includegraphics[scale=0.46]{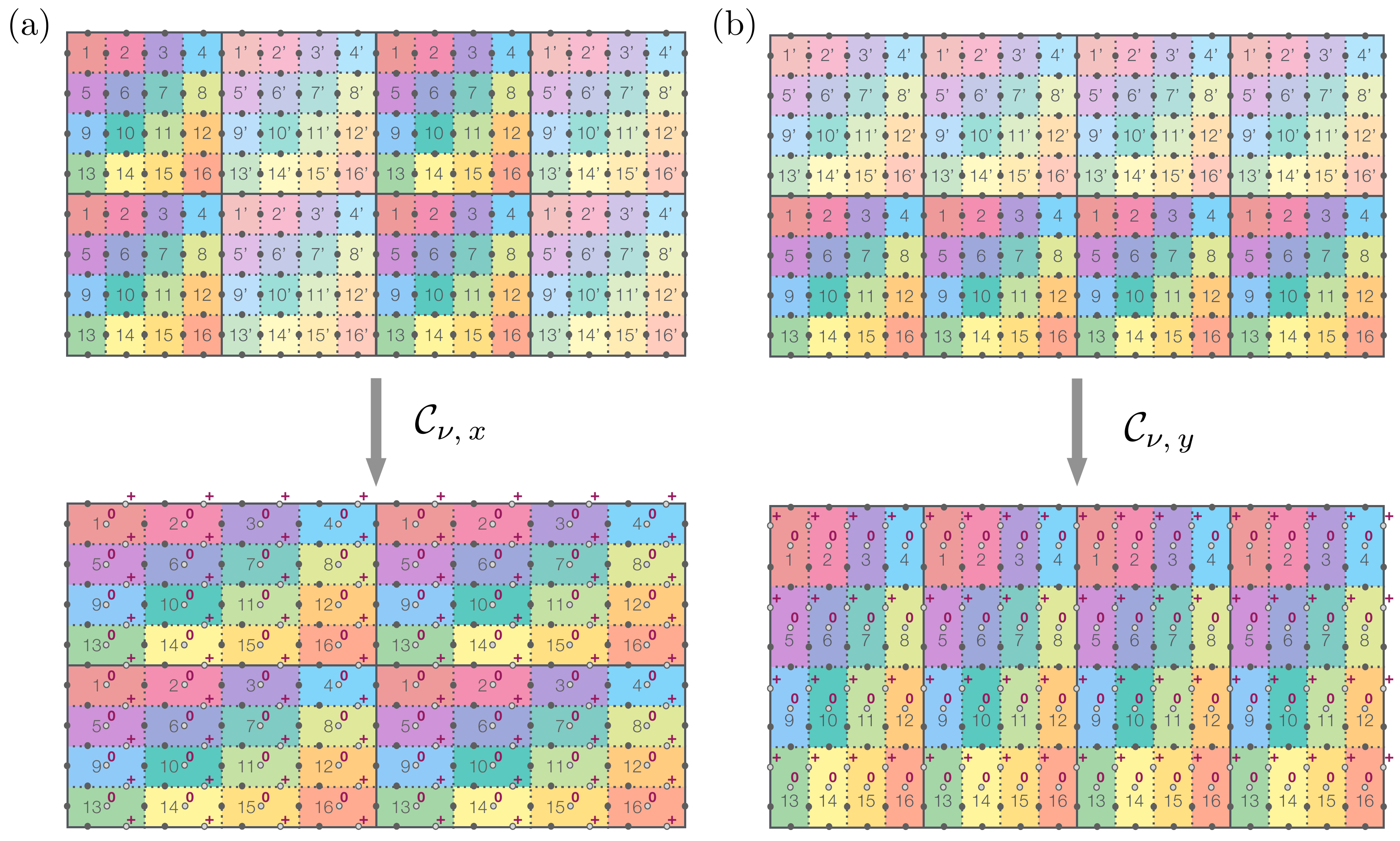}
\par\end{centering}
\caption{Subfigure\,(a) shows a single step, $\mathcal{C}_{\nu,\,x}$, of horizontal entanglement renormalization of the $\nu$-th Kitaev's sixteenfold way chiral spin liquid. Subfigure\,(b) shows a single step, $\mathcal{C}_{\nu,\,y}$, of vertical entanglement renormalization of the $\nu$-th Kitaev's sixteenfold way chiral spin liquid. After a single step of horizontal or vertical renormalization, half of the spins are disentangled into states $\ket{0}$ or $\ket{+}$. The disentangled spins are represented by unfilled circles. After renormalization, the faces of the square lattice are elongated horizontally or vertically and are defined by the remaining entangled spins represented by the filled circles. The remaining spins forming the new elongated square lattice are again in the same Kitaev's sixteenfold way chiral spin liquid state. The primed emergent fermion layer numbers mark emergent fermionic modes to be disentangled and removed by the renormalization. We also use dimmer colors on primed faces to further distinguish them from other (unprimed) faces within the same emergent fermion layer. The subcircuit ${\cal C}_{\nu,\,x}$ for horizontal entanglement renormalization is described in detail in Fig.~\ref{fig:16foldwayhorizontalrenormalizationdetail}, while the subcircuit ${\cal C}_{\nu,\,y}$ for vertical entanglement renormalization is described in detail in Fig.~\ref{fig:16foldwayverticalrenormalizationdetail}. \label{fig:16foldwayrenormalizationgoal}}
\end{figure*}

Recall that Kitaev's sixteenfold way chiral spin liquids can be constructed by bosonizing a stack of trivial fermionic insulators and lattice $p_{x}+ip_{y}$ topological superconductors under the zero-flux condition. Hence, the MERAQLE circuits for these chiral spin liquids will be generalizations of the MERAQLE circuit [Sec.~\ref{subsec:MERAQLE-for-pure}]  for the pure $Z_{2}^{f}$ lattice gauge theory as the bosonized trivial fermionic insulator and of the MERAQLE circuit [Sec.~\ref{subsec:MERAQLE-for-Ising}] for the lattice Ising TQFT as the bosonized lattice $p_{x}+ip_{y}$ topological superconductor. Our goal, shown in Fig.~\ref{fig:16foldwayrenormalizationgoal}, is to realize single steps of entanglement renormalization of the $\nu$-th chiral spin liquid horizontally and vertically.
Similar to Sec.~\ref{subsec:MERAQLE-for-Ising}, in order to ensure that the chiral spin liquid is a fixed-point wavefunction, we want to find coarse-graining operations that disentangle half of the spins and leave the remaining spins in the same chiral spin liquid state defined on a new elongated square lattice. For the new elongated square lattice, we have a redefinition of how the spins are associated with the edges such that the face lengths are doubled horizontally or vertically. For the horizontally coarse-grained lattice, the dual picture of fermions coupled to a $Z_{2}^{f}$ gauge field is defined by the new duality mapping through the new edge orientation assignments depicted in Fig.~\ref{fig:newbranchingsquare}. The dual picture for vertically coarse-grained lattice is defined similarly. After horizontal or vertical renormalization, the emergent fermions live on the newly defined elongated faces.

Let us now describe this procedure in more detail. As in Sec.~\ref{subsec:MERAQLE-for-Ising}, we start with the observation that, if some of the emergent fermionic modes on the faces were empty, then some of the spins on the edges adjacent to these fermionic modes would behave like the ground state of the pure $Z_{2}^{f}$ lattice gauge theory. This means that we can apply the quantum circuit $\mathcal{C}_{Z_{2}^{f},x}$ or $\mathcal{C}_{Z_{2}^{f},y}$, initially designed for the pure $Z_{2}^{f}$ lattice gauge theory, to disentangle these spins. Hence, we can directly apply circuits $\mathcal{C}_{Z_{2}^{f},x}$ or $\mathcal{C}_{Z_{2}^{f},y}$ to disentangle some of the spins bordering half of the faces associated with trivial fermionic insulator layers.  On the other hand, for fermionic modes associated with each $p_{x}+ip_{y}$ topological superconductor layer, as in Sec.~\ref{subsec:MERAQLE-for-Ising}, we make use of the quasi-adiabatic evolution circuit developed in Sec.~\ref{subsec:QLEfortopological superconductor} to get a renormalized state with half of the emergent fermionic modes empty, at which point we can again disentangle half of the adjacent spins. 
Therefore, as shown at the top of Fig.~\ref{fig:16foldwayrenormalizationgoal}(a,b), we introduce a sublattice structure for each layer of emergent fermions, including the layers with emergent trivial insulators. 
To avoid clutter associated with $A$ and $B$ symbols, we instead use the prime symbol and dimmer colors (instead of label $B$) to label the fermionic modes to be emptied and removed.
We do this for all sixteen fermionic layers, independently of whether a given layer is associated with a superconductor or a trivial insulator.

We now work out more precisely the interpolating spin Hamiltonian for emptying half of the fermionic modes associated with superconductor layers. We first consider the (quasi-)adiabatic evolution for horizontal entanglement renormalization. As before, since we want to work in the zero-flux sector, we impose the zero-flux condition $F_v=1$ as a hard constraint on the spin system. Adapting the notation in Sec.~\ref{subsec:Kitaev's-sixteen-foldway}, we start by writing down an interpolating fermionic Hamiltonian for 16 layers of fermions:
\begin{align}
&H_{\nu^{f},\,x}(\lambda) = \nonumber\\
=\sum_{i=1}^{\nu}&H_{i,\,p_{x}+ip_{y},\,x}(\lambda)+\sum_{i=\nu+1}^{16}H_{i,\,\mathrm{trivial\:insulator}}
\end{align}
with $\lambda\in[0,1]$. For each fermion layer $i=1,\cdots,\nu$, $H_{i,\,p_{x}+ip_{y},\,x}(\lambda)$ is nothing but the original $H_{p_{x}+ip_{y},\,x}(\lambda)$ in Eq.~(\ref{eq:p+ipadiabaticHamiltonian}), where the $A$ and $B$ labels are now replaced by the absence and  presence of the prime symbol, respectively. For the layers $i=\nu+1,\cdots,16$ with trivial insulator Hamiltonians, we also impose horizontally alternating patterns of fermionic modes labeled by the absence and presence of the prime symbol. Therefore, we can express the trivial insulator Hamiltonians as 
\begin{align}
&H_{i,\,\mathrm{trivial\:insulator}} = \nonumber\\
=-&\sum_{\mathbf{r}}\left[(1-2c_{i,\,\mathbf{r}}^{\dagger}c_{i,\,\mathbf{r}})+(1-2c_{i,\,\mathbf{r}}^{\prime\dagger}c_{i,\,\mathbf{r}}^{\prime})\right].
\end{align}

The Hamiltonian $H_{\nu^{f},\,x}(\lambda)$ is periodic in $\mathbf{r}$. For each $\mathbf{r}$, we have a unit cell of $16\,(\mathrm{number\,of\,layers})\times2\,(\mathrm{with\,and\,without\,prime\,symbols})=32$ fermionic sites. Notice that the trivial insulator part of the Hamiltonian does not depend on the parameter $\lambda$. The gap of $H_{\nu,\,x}^{f}(\lambda)$ is guaranteed by the gap of each individual term. For the initial state of $\nu$ layers of lattice $p_x+ip_y$ topological superconducting states and $16-\nu$ layers of trivial insulating states, after the adiabatic evolution under $H_{\nu^{f},\,x}(\lambda)$, the primed fermionic modes of the superconducting layers are emptied. Note that the primed fermionic modes of the trivial insulating layers are empty throughout the whole adiabatic evolution.

\begin{figure*}[t]
\begin{centering}
\includegraphics[scale=0.70]{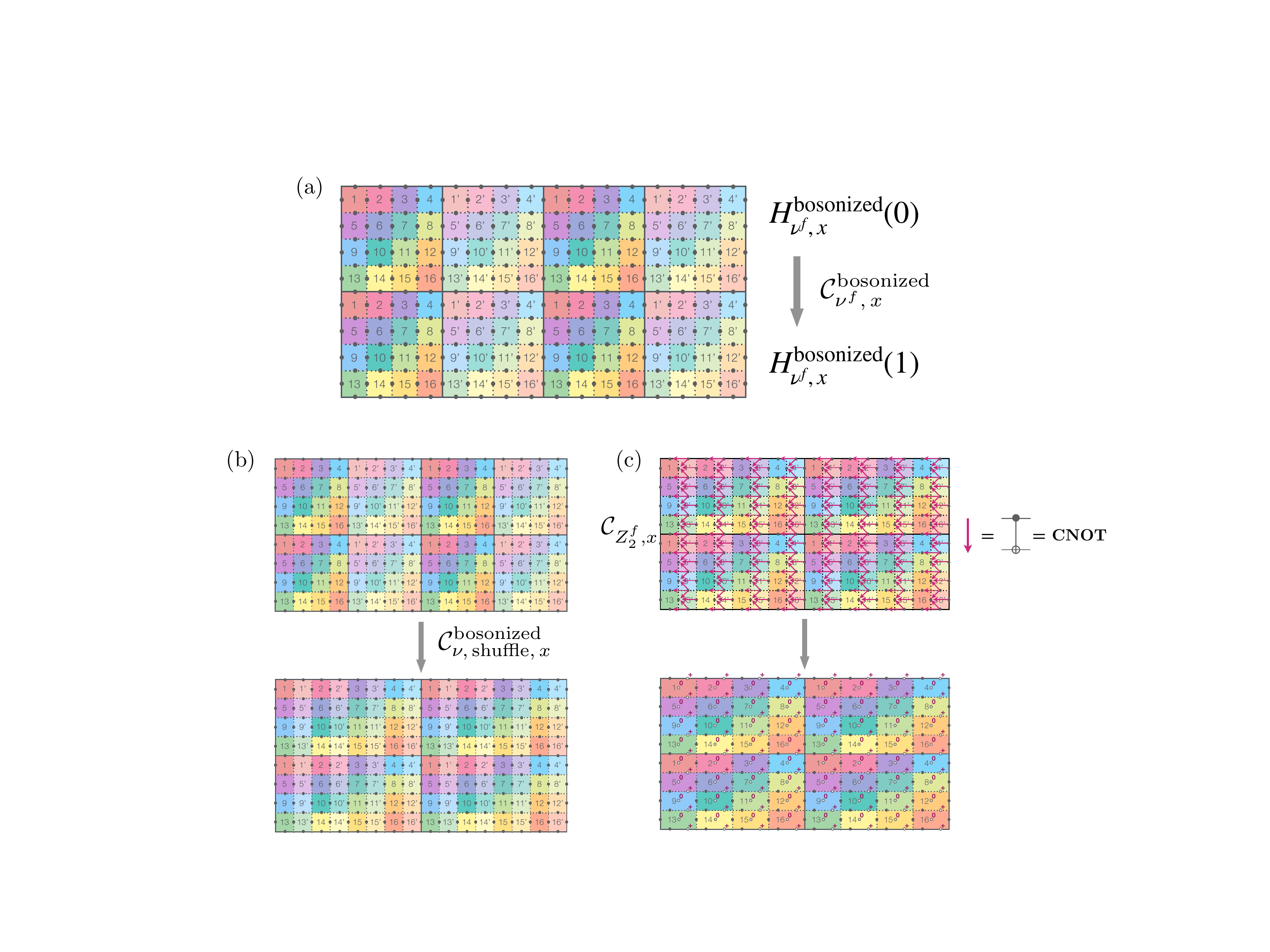}
\par\end{centering}
\centering{}\caption{The horizontal entanglement renormalization subcircuit ${\cal C}_{\nu,\,x}$ for the $\nu$-th Kitaev's sixteenfold way chiral spin liquid can be decomposed into three circuit components: ${\cal C}_{{\rm \nu},\,x}\equiv\mathcal{C}_{Z_{2}^{f},x}\mathcal{C}_{\nu,\,\mathrm{shuffle},\,x}^{\mathrm{bosonized}}\mathcal{C}_{\nu^{f},\,x}^{\mathrm{bosonized}}$. Faces with dimmer colors and primed number labels indicate fermionic modes decoupled or waiting to be decoupled from the rest of the emergent fermions. These primed (and dimly colored) faces play a role similar to the role of blue $B$ faces in Fig.~\ref{fig:IsingTQFTrenormalizationgoal}. (a) The circuit component $\mathcal{C}_{\nu^{f},\,x}^{\mathrm{bosonized}}$. The bosonization of a fermionic quantum circuit composed of quasi-adiabatic circuits (constructed in Sec.~\ref{subsec:QLEfortopological superconductor}) for $\nu$ layers of lattice $p_{x}+ip_{y}$ topological superconductors to renormalize them horizontally. After the circuit component $\mathcal{C}_{\nu^{f},\,x}^{\mathrm{bosonized}}$, all the emergent fermionic modes on the primed faces are empty and therefore disentangled from the rest of the emergent fermionic system. The circuit is quasi-local in the sense of Fig.~\ref{fig:strictlocalityandquasilocality}.  (b) The circuit component $\mathcal{C}_{\nu,\,\mathrm{shuffle},\,x}$. We perform a series of $\left\{ {\rm SWAP}_{i,\,j}^{f}\right\} ^{\mathrm{bosonized}}$ gates to shuffle the emergent fermionic degrees of freedom. This circuit component is independent of $\nu$. (d) The circuit component $\mathcal{C}_{Z_{2}^{f},x}$. This strictly-local circuit is the same as the circuit in Fig.~\ref{fig:MERA-ToricCode-hori} and Fig.~\ref{fig:MERA-Z2f-hori}. The disentangled spins are drawn as unfilled circles. \label{fig:16foldwayhorizontalrenormalizationdetail}}
\end{figure*}

Under the zero-flux constraint, the interpolating spin Hamiltonian for the $\nu$-th Kitaev's sixteenfold way chiral spin liquid is then given by
\begin{align}
H_{\nu^{f},\,x}^{\mathrm{bosonized}}(\lambda) \equiv \left\{ H_{\nu^{f},\,x}(\lambda)\right\} ^{\mathrm{bosonized}}.\label{eq:bosonizedp+ipadiabaticHamiltonian-1}
\end{align}
The spin operator $\left\{ H_{\nu^{f},\,x}(\lambda)\right\} ^{\mathrm{bosonized}}$ is the bosonization of the fermionic Hermitian operator $H_{\nu^{f},\,x}(\lambda)$ with respect to the square lattice shown in Fig.~\ref{fig:16foldwayhorizontalrenormalizationdetail}(a).

Similar to the adiabatic evolution for lattice Ising TQFT in Sec.~\ref{subsec:MERAQLE-for-Ising}, due to the bosonization duality, the gap of the interpolating spin Hamiltonian is guaranteed by the existence of the gap in the fermionic Hamiltonian $H_{\nu^{f},\,x}(\lambda)$. As a consequence, if we were to implement the adiabatic evolution given by $H_{\nu^{f},\,x}^{\mathrm{bosonized}}(\lambda)$, we would have emptied the emergent fermionic modes on all the primed faces, which corresponds to setting emergent fermion parity operators on these faces to $W_{f}=1$. 

Having derived the gapped Hamiltonian path, we can use Eqs.\,(\ref{eq:quasiadiabaticevolution},\,\ref{eq:quasi-adabatic-continuation-operator}) to write down the quasi-adiabatic circuits on both the fermionic side and the spin side. The quasi-adiabatic circuit for the $i$-th layer of superconducting fermions ${\cal C}_{i,\,p_{x}+ip_{y},\,x}^{{\rm }}$ is generated by the quasi-adiabatic continuation operator $\mathcal{D}_{i,\,p_{x}+ip_{y},\,x}(\lambda)$ obtained by replacing $H(\lambda)\rightarrow H_{i,\,p_{x}+ip_{y},\,x}(\lambda)$ in Eq.~(\ref{eq:quasi-adabatic-continuation-operator}). Similarly, the quasi-adiabatic continuation operator on the spin side $\mathcal{D}_{\nu^{f},\,x}^{\mathrm{bosonized}}(\lambda)$ is defined by replacing $H(\lambda)\rightarrow H_{\nu^{f},\,x}^{\mathrm{bosonized}}(\lambda)$ in Eq.~(\ref{eq:quasi-adabatic-continuation-operator}). The corresponding quasi-adiabatic circuit, defined as the time evolution under $\mathcal{D}_{\nu^{f},\,x}^{\mathrm{bosonized}}(\lambda)$ for $\lambda$ from $0$ to $1$, is $\mathcal{C}_{\nu^{f},\,x}^{\mathrm{bosonized}}$. 
Since the bosonizations of the interpolating fermionic Hamiltonian for each of the 16 layers commute with each other and with the zero-flux constraint, 
following an argument similar to the one from Eq.~(\ref{eq:bosonizedp+ipadiabaticHamiltonian}) to Eq.~(\ref{eq:bosoniziedp+ipadiabaticcircuit}) in Sec.~\ref{subsec:MERAQLE-for-Ising}, we can simplify the expression for $\mathcal{C}_{\nu^{f},\,x}^{\mathrm{bosonized}}$ by writing it in terms of the fermionic quasi-adiabatic circuits ${\cal C}_{i,\,p_{x}+ip_{y},\,x}^{{\rm }}$ for the different fermion layers labeled by $i$:
\[
\mathcal{C}_{\nu^{f},\,x}^{\mathrm{bosonized}}=\prod_{i=1}^{\nu}\left\{ {\cal C}_{i,\,p_{x}+ip_{y},\,x}^{{\rm }}\right\} ^{\mathrm{bosonized}}, \label{eq:cnufx}
\]
where the ordering of the product does not matter since all the factors commute. Because we flattened sixteen layers of fermions into a single square lattice, the bosonized unitary operator $\left\{ {\cal C}_{i,\,p_{x}+ip_{y},\,x}^{{\rm }}\right\} ^{\mathrm{bosonized}}$ involves Wilson lines that cross other colors (i.e.\ layers) to connect neighboring faces of the same color (i.e.\ layer). The circuit $\mathcal{C}_{\nu^{f},\,x}^{\mathrm{bosonized}}$ is schematically illustrated in Fig.~\ref{fig:16foldwayhorizontalrenormalizationdetail}(a). From Eq.\ (\ref{eq:cnufx}), we see that $\mathcal{C}_{\nu^{f},\,x}^{\mathrm{bosonized}}$ is quasi-local in the sense of Fig.~\ref{fig:strictlocalityandquasilocality}. As in Sec.\ \ref{subsec:MERAQLE-for-Ising}, notice that the zero-flux constraint commutes with all spin Hamiltonians obtained by bosonizing fermionic Hamiltonians. This means that the spin Hamiltonians we have derived and the resulting renormalization circuits are independent of the presence of the constraint. 
Topologically, after the application of the quasi-adiabatic circuit, the spin system without the constraint is still in the same Kitaev's sixteenfold way spin liquid phase but with additional emergent fermions in the vacuum state.

\begin{figure*}[t]
\begin{centering}
\includegraphics[scale=0.70]{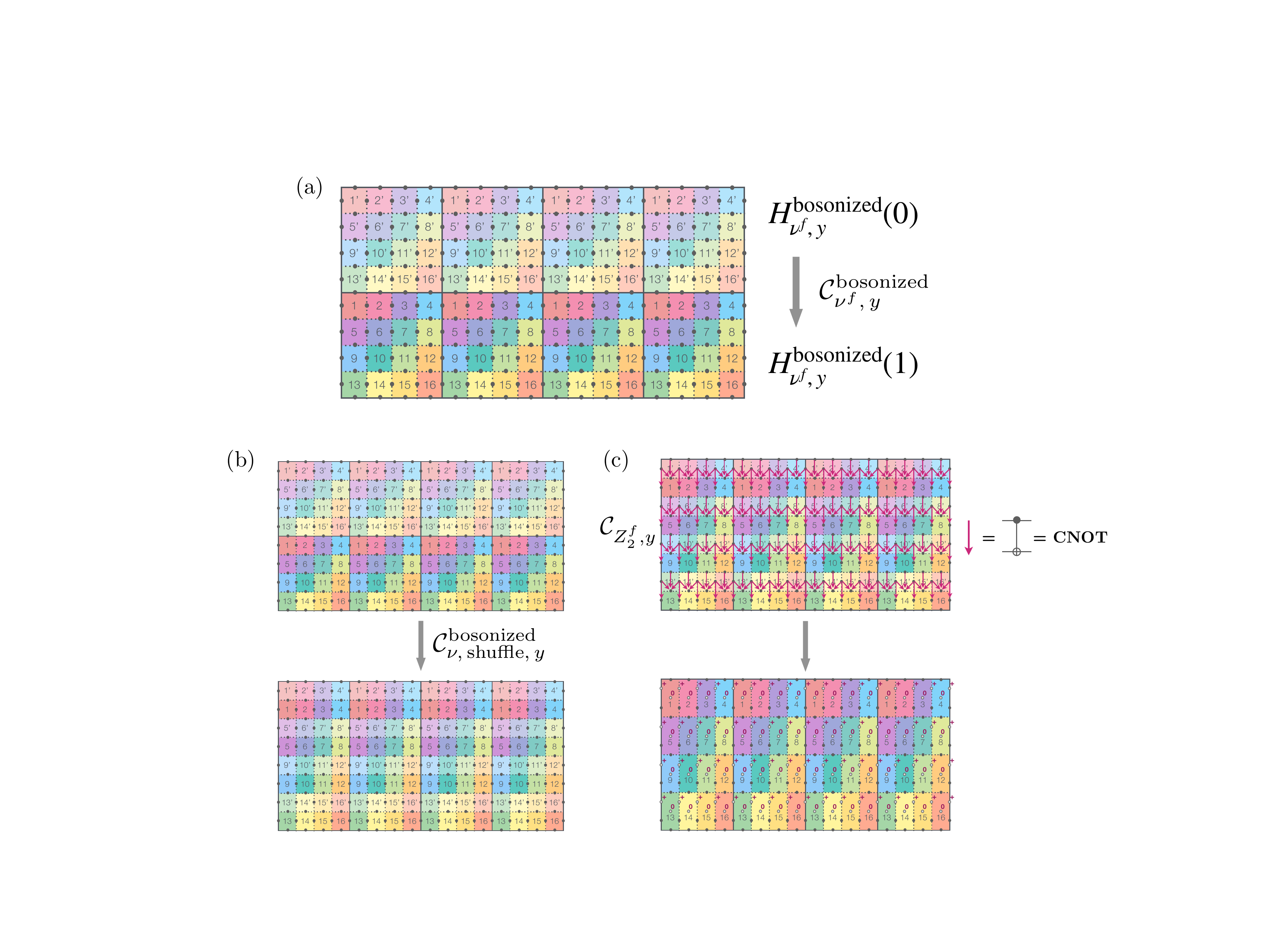}
\par\end{centering}
\centering{}\caption{The vertical entanglement renormalization subcircuit ${\cal C}_{\nu,\,y}$ for the $\nu$-th Kitaev's sixteenfold way chiral spin liquid can be decomposed into three circuit components: ${\cal C}_{{\rm \nu},\,y}\equiv\mathcal{C}_{Z_{2}^{f},y}\mathcal{C}_{\nu,\,\mathrm{shuffle},\,y}^{\mathrm{bosonized}}\mathcal{C}_{\nu^{f},\,y}^{\mathrm{bosonized}}$. Faces with dimmer colors and primed number labels indicate fermionic modes decoupled or waiting to be decoupled from the rest of the emergent fermions. These primed (and dimly colored) faces play a role similar to the role of blue $B$ faces in  Fig.~\ref{fig:IsingTQFTrenormalizationgoal}. (a) The circuit component $\mathcal{C}_{\nu^{f},\,y}^{\mathrm{bosonized}}$. The bosonization of a fermionic quantum circuit  composed of quasi-adiabatic circuits (constructed in Sec.~\ref{subsec:QLEfortopological superconductor}) for $\nu$ layers of lattice $p_{x}+ip_{y}$ topological superconductors  to renormalize them vertically. After the circuit component $\mathcal{C}_{\nu^{f},\,y}^{\mathrm{bosonized}}$, all the emergent fermionic modes on the primed faces are empty and therefore disentangled from the rest of the emergent fermionic system. The circuit is quasi-local in the sense of Fig.~\ref{fig:strictlocalityandquasilocality}. (b) The circuit component $\mathcal{C}_{\nu,\,\mathrm{shuffle},\,y}$. We perform a series of $\left\{ {\rm SWAP}_{i,\,j}^{f}\right\} ^{\mathrm{bosonized}}$ gates to shuffle the emergent fermionic degrees of freedom. This circuit component is independent of $\nu$. (d) The circuit component $\mathcal{C}_{Z_{2}^{f},y}$. This strictly-local circuit is the same as the circuit in Fig.~\ref{fig:MERA-ToricCode-verti} and Fig.~\ref{fig:MERA-Z2f-vert}. The disentangled spins are drawn as unfilled circles. \label{fig:16foldwayverticalrenormalizationdetail}}
\end{figure*}

The interpolating spin Hamiltonian $H_{\nu^{f},\,y}^{\mathrm{bosonized}}(\lambda)$ and the quasi-adiabatic circuit $\mathcal{C}_{\nu^{f},\,y}^{\mathrm{bosonized}}$ in the vertical direction can be constructed in the same way. The circuit $\mathcal{C}_{\nu^{f},\,y}^{\mathrm{bosonized}}$ is schematically shown in Fig.~\ref{fig:16foldwayverticalrenormalizationdetail}(a).

We are almost ready to apply $\mathcal{C}_{Z_{2}^{f},x}$ or $\mathcal{C}_{Z_{2}^{f},y}$ to disentangle half of the spins. However, looking more closely at the entanglement structure that $\mathcal{C}_{Z_{2}^{f},x}$ or $\mathcal{C}_{Z_{2}^{f},y}$ can renormalize---and by following the corresponding stabilizer transformations in Fig.~\ref{fig:MERA-Z2f-hori-calulation}(a,c,d) and Fig.~\ref{fig:MERA-Z2f-vert-calculation}(a,c,d) as well as the spin disentangling arguments in Sec.~\ref{subsec:MERAQLE-for-pure} for the pure $Z_{2}^{f}$ lattice gauge theory and Sec.~\ref{subsec:MERAQLE-for-Ising} for the lattice Ising TQFT---we realize that we need to first prepare an entanglement pattern that has an empty emergent fermionic modes on every other face horizontally or vertically. Making the primed sites empty by applying $\mathcal{C}_{\nu^{f},\,x}^{\mathrm{bosonized}}$ and $\mathcal{C}_{\nu^{f},\,y}^{\mathrm{bosonized}}$ does not produce such an entanglement structure. The solution is to shuffle the emergent fermionic modes after performing the quasi-adiabatic circuits. As shown in Fig.~\ref{fig:16foldwayhorizontalrenormalizationdetail}(b) (Fig.~\ref{fig:16foldwayverticalrenormalizationdetail}(b)), we want to shuffle the modes such that we get an alternating pattern of unprime-prime columns (rows) for horizontal (vertical) entanglement renormalization. In addition, we require that primed faces immediately to the right of (above) to unprimed faces should be in the same emergent layers for horizontal (vertical) entanglement renormalization. By doing so, we arrive at the desired emergent empty mode structure on every other horizontal or vertical face.

To realize the shuffling operation $\mathcal{C}_{\nu,\,\mathrm{shuffle},\,x}^{\mathrm{bosonized}}$, we first note that, if we want to swap two fermionic labeled by $i$ and $j$, we can use the following fermionic swap gate \cite{Verstraete2009,Jiang2018,Babbush2018}:

\[
{\rm SWAP}_{i,\,j}^{f}=1+c_{i}^{\dagger}c_{j}+c_{j}^{\dagger}c_{i}-c_{i}^{\dagger}c_{i}-c_{j}^{\dagger}c_{j}.
\]
The fermionic swap gate is both unitary and Hermitian. It satisfies the property that 
\begin{align}
{\rm SWAP}_{i,\,j}^{f}\,c_{i}\,{\rm SWAP}_{i,\,j}^{f} & =c_{j},\nonumber \\
{\rm SWAP}_{i,\,j}^{f}\,c_{j}\,{\rm SWAP}_{i,\,j}^{f} & =c_{i}.
\end{align}
With this gate applied multiple times for different pairs of fermion sites, we can achieve an arbitrary permutation of the fermionic degrees of freedom. However, the fermions in our case are emergent, so we have to bosonize the swap gates $\left\{ {\rm SWAP}_{i,\,j}^{f}\right\} ^{\mathrm{bosonized}}$. It can be shown that the spin operator $\left\{ {\rm SWAP}_{i,\,j}^{f}\right\} ^{\mathrm{bosonized}}$ is both unitary and hermitian regardless of the presence of the zero-flux constraint $F_v=1$.

We now use the $\left\{ {\rm SWAP}_{i,\,j}^{f}\right\} ^{\mathrm{bosonized}}$ gates constructed above to shuffle the fermionic degrees of freedom. As we have already mentioned above, the horizontal $\mathcal{C}_{\nu,\,\mathrm{shuffle},\,x}^{\mathrm{bosonized}}$ and vertical $\mathcal{C}_{\nu,\,\mathrm{shuffle},\,y}^{\mathrm{bosonized}}$ shuffling operations  for the emergent fermions are shown in Fig.~\ref{fig:16foldwayhorizontalrenormalizationdetail}(b) and Fig.~\ref{fig:16foldwayverticalrenormalizationdetail}(b), respectively. The shuffling operation puts next to each other the primed and unprimed faces belonging to the same fermionic layer $i$ and the same enlarged unit cell $\mathbf{r}$.

Now we can apply $\mathcal{C}_{Z_{2}^{f},x}$ and $\mathcal{C}_{Z_{2}^{f},y}$. We again use the stabilizer transformation shown in Fig.~\ref{fig:MERA-Z2f-hori-calulation}(a,c,d) and Fig.~\ref{fig:MERA-Z2f-vert-calculation}(a,c,d). From the $W_{f}=1$ condition on the primed faces and the zero-flux condition on every vertex, we see that half of the spins disentangle, as shown by the unfilled circles in the bottom plots of Fig.~\ref{fig:16foldwayhorizontalrenormalizationdetail}(c) and Fig.~\ref{fig:16foldwayverticalrenormalizationdetail}(c). This achieves our goal indicated in the bottom plots of Fig.~\ref{fig:16foldwayrenormalizationgoal}. Roughly speaking, the empty fermionic modes on the primed faces are shifted to the qubits labeled by unfilled circles and red state labels $0$ since the fermion parity operators $W_{f}$ are transformed into single-qubit Pauli-$Z$ operators on these qubits under $\mathcal{C}_{Z_{2}^{f},x}$ and $\mathcal{C}_{Z_{2}^{f},y}$. In addition, the transformation of the zero-flux condition $F_v=1$, with $NE(v)$ being a primed face, together with the single-qubit Pauli-$Z$ operators, leads to single-qubit Pauli-$X$ operators on the qubits at the bottom of primed faces.

However, we here encounter a problem similar to the one in the previous subsection. It is not obvious whether the remaining entangled spins on the new lattice are still in the $\nu$-th Kitaev's sixteenfold way chiral spin liquid after $\mathcal{C}_{Z_{2}^{f},x}$ or $\mathcal{C}_{Z_{2}^{f},y}$. Note that we can make edge orientation assignments, like the ones shown in Fig.~\ref{fig:newbranchingsquare}, for the new elongated square lattice for the purpose of dualizing the spin theory to a theory of fermions coupled to a $Z_2^f$ gauge field. Even though the dual gauge flux is zero everywhere (as can be seen from the transformation of stabilizer generators), the dual fermions might not correspond to $\nu$ layers of lattice $p_{x}+i p_{y}$ topological superconductors and $16-\nu$ layers of trivial insulators. Once again, this issue is addressed in Appendix~\ref{sec:appendix-compatibility}. There, we show that, with the alternating pattern of primed faces being in the emergent empty modes, under the circuit $\mathcal{C}_{Z_{2}^{f},x}$ and $\mathcal{C}_{Z_{2}^{f},y}$, the primed faces effectively disappear, while the unprimed faces become larger by consuming the original space occupied by the primed faces. This is also reflected in Fig.~\ref{fig:16foldwayhorizontalrenormalizationdetail}(c) and Fig.~\ref{fig:16foldwayverticalrenormalizationdetail}(c). Since, before $\mathcal{C}_{Z_{2}^{f},x}$ and $\mathcal{C}_{Z_{2}^{f},y}$, the unprimed faces are in the $\nu$-th Kitaev's sixteenfold way chiral spin liquid following $\mathcal{C}_{\nu^{f},\,x}^{\mathrm{bosonized}}$ and $\mathcal{C}_{\nu^{f},\,y}^{\mathrm{bosonized}}$, and since the shuffling operations $\mathcal{C}_{\nu,\,\mathrm{shuffle},\,x}^{\mathrm{bosonized}}$ and $\mathcal{C}_{\nu,\,\mathrm{shuffle},\,y}^{\mathrm{bosonized}}$ respect this structure, we can conclude that the final state is indeed the $\nu$-th Kitaev's sixteenfold way chiral spin liquid. Hence, the subcircuit for a single step of horizontal entanglement renormalization of the $\nu$-th Kitaev's sixteenfold way chiral spin liquid is ${\cal C}_{{\rm \nu},\,x}\equiv\mathcal{C}_{Z_{2}^{f},x}\mathcal{C}_{\nu,\,\mathrm{shuffle},\,x}^{\mathrm{bosonized}}\mathcal{C}_{\nu^{f},\,x}^{\mathrm{bosonized}}$. Similarly, the vertical entanglement renormalization subcircuit is ${\cal C}_{{\rm \nu},\,y}\equiv\mathcal{C}_{Z_{2}^{f},y}\mathcal{C}_{\nu,\,\mathrm{shuffle},\,y}^{\mathrm{bosonized}}\mathcal{C}_{\nu^{f},\,y}^{\mathrm{bosonized}}$. 

Finally, we can write down the entire scale-invariant entanglement renormalization circuit. It consists of successive applications of the same quantum subcircuits ${\cal C}_{{\rm \nu},\,x}$ and ${\cal C}_{{\rm \nu},\,y}$ but at different length scales. If we use an additional superscript $s\in\mathbb{N}$ to label the scale at which those subcircuits operate, the full circuit is
\[
{\cal C}_{\nu}=\prod_{s\in\mathbb{N}}\left({\cal C}_{{\rm {\rm \nu}},\,y}^{s}\,{\cal C}_{{\rm \nu},\,x}^{s}\right).
\]
The $\nu$-th Kitaev's sixteenfold way chiral spin liquid is a fixed-point wavefunction of this circuit. The application of a single layer of the MERAQLE circuit $\left({\cal C}_{{\rm \nu},\,x}^{s}\,{\cal C}_{{\rm {\rm \nu}},\,y}^{s}\right)$ disentangles $3/4$ of the original spins and leaves the remaining $1/4$ of the spins in the same $\nu$-th Kitaev's sixteenfold way chiral spin liquid but on a lattice with the unit cell length twice the size of the original one. The circuit has the structure displayed in Fig.~\ref{fig:layoutofMERAQLE}: the quasi-local evolution circuit component $\mathcal{C}_{\mathrm{ql},\,x}^{s}$ is $\mathcal{C}_{\nu^{f},\,x}^{\mathrm{bosonized}}$; the auxiliary circuit component $\mathcal{C}_{\mathrm{aux},\,x}^{s}$ is the shuffling circuit $\mathcal{C}_{\nu,\,\mathrm{shuffle},\,x}^{\mathrm{bosonized}}$ consisting of strictly local gates $\left\{ {\rm SWAP}_{i,\,j}^{f}\right\} ^{\mathrm{bosonized}}$; the circuit component $\mathcal{C}_{Z_{2}^{f},x}^{s}$ is the strictly-local subcircuit $\mathcal{C}_{Z_{2}^{f},x}$ for the pure $Z_{2}^{f}$ lattice gauge theory. We can see the corresponding structure for $\mathcal{C}_{\mathrm{ql},\,y}^{s}$, $\mathcal{C}_{\mathrm{aux},\,y}^{s}$, and $\mathcal{C}_{Z_{2}^{f},y}^{s}$ in the vertical direction as well. Here, we finally see the full power of the MERAQLE framework for chiral spin liquids with nonzero finite correlation lengths. Due to its quasi-local evolution circuit components, we circumvent the correlation-length-based no-go argument for conventional MERA circuits.


\section{Conclusions \label{sec:conclusion}}

In this paper, we solved the problem of finding scale-invariant entanglement renormalization circuits for interacting chiral topological states. We presented a new type of quantum circuit called MERAQLE to renormalize the entanglement structure of several chiral spin liquids belonging to Kitaev's sixteenfold way classification. Even though the chiral topological states we considered have nonzero finite correlation lengths and thus cannot be fixed-point wavefunctions of conventional scale-invariant MERA circuits, by inserting continuous time evolutions generated by time-dependent quasi-local Hamiltonians into the skeletons of conventional MERA circuits, we are able to overcome this issue. The reason is that quasi-local evolution can generate correlations between distant sites, and the conventional wisdom of correlation length reduction $\ell'=\ell/b,\,b>1$ for each layer of entanglement renormalization with strictly-local discrete gates no longer applies here. In other words, coarse-graining operations involving quasi-local evolutions are capable of preserving correlation lengths. It is interesting to see that the distinction between a quantum circuit based on strictly-local gates and one also equipped with quasi-local evolutions can be profound. Our analysis demonstrates that the locality constraint of a quantum circuit can be subtle, and requiring a circuit to respect strict locality or quasi-locality could have a significant difference.

We can also rephrase our result in the quantum computing language. Recall that we can define a quantum complexity class for a set of tasks that can be accomplished with access to a set of quantum operations and a certain amount of computational resources \cite{Nielsen2000, Watrous2008}. The tasks we care about here are preparing large-scale entangled quantum states \cite{Aaronson2016, Rosenthal2021, Irani2022, Metger2023} by incorporating fresh ancillary qubits under the condition that the states must be fixed-point wavefunctions throughout certain scale-invariant procedures. Therefore, in this terminology, if we define locality with respect to the (coarse-grained) lattice of each layer, then we conclude that our result separates the complexity class with access to strictly local quantum gates and quasi-local Hamiltonians from the complexity class with access to strictly local quantum gates only, provided that every layer of the scale-invariant procedures has a constant depth. This is because preparing sixteenfold way chiral spin liquids can not be done with constant depth in each layer solely using the latter set of quantum operations.

Our result also offers a new perspective on chiral topological order. Chiral topologically ordered states are interesting since they have different topological properties as compared to non-chiral topologically ordered states. For example, chiral topologically ordered states can have framing anomalies \cite{Wang2010} and gapless boundary modes \cite{Fradkin2013}. Also, because of these features, when compared to non-chiral topologically ordered states, we so far do not have many analytical tools to study the entanglement structures of chiral topologically ordered states on lattices \cite{Hasik2022}. Our analytical work of constructing exact state preparation quantum circuits adds an important cornerstone to the study of quantum many-body behavior of chiral topological order. It is worth mentioning a related demonstration of the existence of locally-commuting parent Hamiltonians for chiral $U(1)$ symmetry-protected topological states (no anyons) using infinite-dimensional on-site Hilbert spaces \cite{DeMarco2021}, which evades the no-go theorem in Ref.~\cite{Kapustin2019}. In comparison with their result, our MERAQLE circuits give fixed-point wavefunctions for intrinsic chiral topological states with a finite-dimensional Hilbert space on each site.

In this paper, we focused on constructing quantum circuits for states within Kitaev's sixteenfold way classification on square lattices, which is in contrast with other circuit constructions involving also mid-circuit measurements and feedback \cite{Tantivasadakarn2022}. 
While we focused on the case of a square lattice, we expect our approach to be immediately generalizable to other lattices, including the honeycomb lattice \cite{Kitaev2006,Yao2011,Nakai2012,Chulliparambil2020,Zhang2020,Natori2020}.
It is an open question whether we can use the MERAQLE framework to produce a chiral topological state outside the sixteenfold way classification. With an eye towards topological quantum computing, it would be useful to have entanglement renormalization circuits for chiral models with the ability to perform braiding-based universal quantum computation, such as many Read-Rezayi fractional quantum Hall states \cite{Read1999, Bonesteel2005, Hormozi2009}. On the practical side, it is also interesting to develop implementations of our scheme in the Noisy Intermediate-Scale Quantum (NISQ) era \cite{Preskill2018} for preparing chiral topological states in synthetic quantum matter, where non-chiral $Z_2$ spin liquid 
states have already been prepared 
using superconducting qubit quantum processors \cite{Satzinger2021} and Rydberg atom arrays \cite{Semeghini2021, Bluvstein2022}. We leave these 
questions to our future work.

\begin{acknowledgments}
We would like to thank David Aasen, Maissam Barkeshli, Yu-An Chen, Mohammad Hafezi, Masaki Oshikawa, Brian Swingle, and Seth Whitsitt for helpful discussions.
SKC and AVG~were supported in part by the DoE ASCR Quantum Testbed Pathfinder program (award No.\ DE-SC0019040), ARO MURI, AFOSR MURI, DoE ASCR Accelerated Research in Quantum Computing program (award No.\ DE-SC0020312), AFOSR, DoE QSA, NSF QLCI (award No.\ OMA-2120757), NSF PFCQC program, and DARPA SAVaNT ADVENT.
SKC also acknowledges the support from the Studying Abroad Scholarship by Ministry of Education in Taiwan as well as the KITP Graduate Fellowship, which was supported in part by the National Science Foundation under Grant No. NSF PHY-1748958.
GZ acknowledges support
by the U.S.~Department of Energy, Office of Science,
National Quantum Information Science Research Centers,
Co-design Center for Quantum Advantage (C2QA) under
contract number DE-SC0012704. 
\end{acknowledgments}

\appendix

\section{Details of quasi-adiabatic evolution}
\label{sec:appendix-quasiadiabatic}
In this appendix, we offer a detailed introduction to quasi-adiabatic evolution.

Traditionally, if there is a gapped path between the initial Hamiltonian and the final Hamiltonian, the adiabatic theorem dictates that, starting from the ground state of the initial Hamiltonian, we can reach the ground state of the final Hamiltonian with an arbitrarily small error provided that the adiabatic evolution is slow enough. However, for practical purposes, it is convenient to be able to implement the adiabatic process faster without introducing too much error.

The idea of quasi-adiabatic evolution solves this problem. For an adiabatic gapped path $H(\lambda)$ parameterized by $\lambda$, we can use the quasi-adiabatic continuation operator to take the ground state of the initial Hamiltonian $H(\lambda=0)$ to the ground state of the final Hamiltonian $H(\lambda=1)$ exactly in finite time. We define the quasi-adiabatic continuation operator as \cite{Osborne2007,Hastings2010,Hastings2010b}
\begin{align}
 \mathcal{D}(\lambda)=-i\int_{-\infty}^{\infty}dt\,&F(E_\mathrm{gap}t)\cdot\nonumber\\
&\exp\left(iH(\lambda )t\right)\partial_{\lambda}H(\lambda)\exp\left(-iH(\lambda)t\right).   
\end{align}
The parameter $E_\mathrm{gap}$ is chosen to be the smallest energy gap of the Hamiltonian $H(\lambda) $ along the adiabatic gapped path. The function $F(t)$ satisfies the following condition: its Fourier transform $\tilde{F}(\omega)$ is an odd function and decays as $\tilde{F}(\omega)=-1/\omega$ when $\left|\omega\right|\geq1$. Our continuous Fourier transform and inverse Fourier transform conventions are $\tilde{F}(\omega)=\int_{-\infty}^{\infty}dt\,F(t)e^{i\omega t}$, and $F(t)=\frac{1}{2\pi}\int_{-\infty}^{\infty}d\omega\,\tilde{F}(\omega)e^{-i\omega t}$. The function $F(t)$ is constructed as follows \cite{Hastings2010,Ingham1934}. Given a monotonically decreasing positive function $\epsilon(y)$ with a convergent integral $\int_{1}^{\infty}\epsilon(y)/y\,dy$ (for example, $\epsilon(y)=\frac{1}{\log\left((2+y)^{2}\right)}$ or $\epsilon(y)=\frac{1}{\log\left(\log\left((2+y)^{4}\right)\right)}$), we construct the function $F$:
\begin{equation}
F(t)=\frac{i}{2}\int du\,\left(\delta(u)-g_{\epsilon,\left\{ \rho_{n}\right\} }(u)\right){\rm sign}(t-u),\label{eq:functionFdefinition}
\end{equation}
with the function $g_{\epsilon,\left\{ \rho_{n}\right\} }(u)$ defined as \cite{Ingham1934}
\[
g_{\epsilon,\left\{ \rho_{n}\right\} }(u)=\frac{1}{\mathcal{N}}\prod_{n=1}^{\infty}\frac{\sin\rho_{n}u}{\rho_{n}u},
\]
where the sequence of parameters, $\{\rho_{n}\}_{n=1}^{\infty}$, is carefully chosen such that:
\begin{enumerate}
\item Each term is positive: $\rho_{n}>0$.
\item The sequence is monotonically decreasing: $\rho_{n}\geq\rho_{n+1}$.
\item The term $\rho_{n}$ satisfies $\rho_{n}\geq e\,\epsilon(n)/n$ for all $n\geq n_{0}$ with $n_{0}$ a positive integer and $e$ being Euler's number.
\item The series $\sum_{n=1}^{\infty}\rho_{n}$ converges with  $\sum_{n=1}^{\infty}\rho_{n}\leq1$. 
\end{enumerate}
Note that $g(u)$ is a continuous even function. The parameter $\mathcal{N}$ is set such that the Fourier transform of $g_{\epsilon,\left\{ \rho_{n}\right\} }$ satisfies $\tilde{g}_{\epsilon,\left\{ \rho_{n}\right\} }(\omega=0)=1$.

With all the requirements mentioned above being satisfied, it can be shown that $g_{\epsilon,\left\{ \rho_{n}\right\} }$ decays as \cite{Ingham1934}
\begin{equation}
g_{\epsilon,\left\{ \rho_{n}\right\} }(y)=\mathcal{O}\left(e^{-\left|y\right|\epsilon(\left|y\right|)}\right).\label{eq:gdecayequation}
\end{equation}
In addition, the Fourier transform $\tilde{g}_{\epsilon,\,\left\{ \rho_{n}\right\} }(\omega)$ of this function is identically zero for $\left|\omega\right|\geq1$. 

It is easy  to see from Eq.~(\ref{eq:gdecayequation}) that, for all $0<\alpha<1$, we can find a positive constant $C_{\alpha}$ such that $\left|g_{\epsilon,\left\{ \rho_{n}\right\} }(y)\right|\leq C_{\alpha}\exp\left(-\left|y\right|^{\alpha}\right)$. We say that the function $g_{\epsilon,\left\{ \rho_{n}\right\} }$ decays subexponentially \cite{Dziubaski1998}. 
We should always keep in mind that the function $F$, the function $g_{\epsilon,\left\{ \rho_{n}\right\} }$, and the resulting quasi-adiabatic continuation operator are $\epsilon$- and $\left\{ \rho_{n}\right\}$-dependent. However, we will drop the symbols $\epsilon$ and $\left\{ \rho_{n}\right\} $ from the subscript of $g_{\epsilon,\left\{ \rho_{n}\right\} }$, assuming that we have chosen some function $\epsilon$ and
a specific set of parameters $\left\{ \rho_{n}\right\}$ satisfying the requirements mentioned above. 
Using the fact that the function $g$ decays subexponentially, it is not hard to show, from the definition of $F$, that the function $F$ also decays subexponentially \cite{Hastings2010}.

Since the function $F$ decays subexponentially, it also decays superpolynomially, which means it decays faster than any polynomial function. We say that a Hamiltonian $H$ consists of superpolynomially decaying interactions if we can write it as $H=\sum_{\mathbf{r}}\sum_{R} H_{\mathbf{r},\,R}$, where (i) $H_{\mathbf{r},\,R}$ is an operator supported on sites within the disk of radius $R\in \mathbb{N}$ centered at position $\mathbf{r}$ and (ii) for any function decaying polynomially in $R$, the operator norm $\left\Vert H_{\mathbf{r},\,R} \right\Vert$ is bounded by some constant times this function.
In Ref.~\cite{Hastings2010}, it is shown that, if the Hamiltonian $H(\lambda)$ consists of superpolynomially decaying interactions, the quasi-adiabatic continuation operator $D(\lambda)$ also consists of superpolynomially decaying interactions. 
If the Hamiltonian is finite-range or consists of  subexponentially decaying interactions, $D(\lambda)$ is composed of subexponential interactions, i.e., $\left\Vert D_{\mathbf{r},\,R}(\lambda) \right\Vert \leq C_\alpha \exp({-R^{\alpha}})$ for any $\alpha<1$ and some $\alpha$-dependent constant $C_\alpha>0$ \cite{Osborne2007,Hastings2010}. 

It can be shown that $\partial_{\lambda}\ket{\psi_{0}(\lambda)}=iD(\lambda)\ket{\psi_{0}(\lambda)}$, where $\ket{\psi_0(\lambda)}$ is the ground state of $H(\lambda)$. Therefore, we can transfer the ground state of $H(\lambda=0)$ to the ground state of $H(\lambda=1)$ via  time evolution under $D(\lambda)$ for finite time $\lambda\in[0,1]$.

\section{Emergent fermions after a single step of entanglement renormalization }

\label{sec:appendix-compatibility}

In Sec.~\ref{subsec:MERAQLE-for-Ising}, when we discussed the entanglement renormalization of the lattice Ising TQFT, we encountered the following question: for the spins that remain entangled after the application of the subcircuit  $\mathcal{C}_{\mathrm{Ising},\,x}$ ($\mathcal{C}_{\mathrm{Ising},\,y}$), are the dual fermions defined on the new coarse-grained square lattice [see e.g.~Fig.~\ref{fig:newbranchingsquare}] still in the lattice $p_{x}+ip_{y}$ superconducting state with the right parameters? In addition, in Sec.~\ref{subsec:MERAQLE-for-16foldway}, when we discussed the entanglement renormalization of the $\nu$-th Kitaev's sixteenfold way chiral spin liquid, we also encountered a similar question: after the application of  $\mathcal{C}_{\nu,\,x}$ ($\mathcal{C}_{\nu,\,y}$), are the remaining entangled spins in the $\nu$-th Kitaev's sixteenfold way chiral spin liquid defined on the new coarse-grained square lattice? In other words, are the dual fermions in a quantum state with $\nu$ layers of topological superconductors and $16-\nu$ layers of trivial insulators?
In this appendix, we will answer these questions affirmatively by considering a very generic setup that will apply to all the cases just  mentioned.

\begin{figure}[t]
\begin{centering}
\includegraphics[width=\columnwidth]{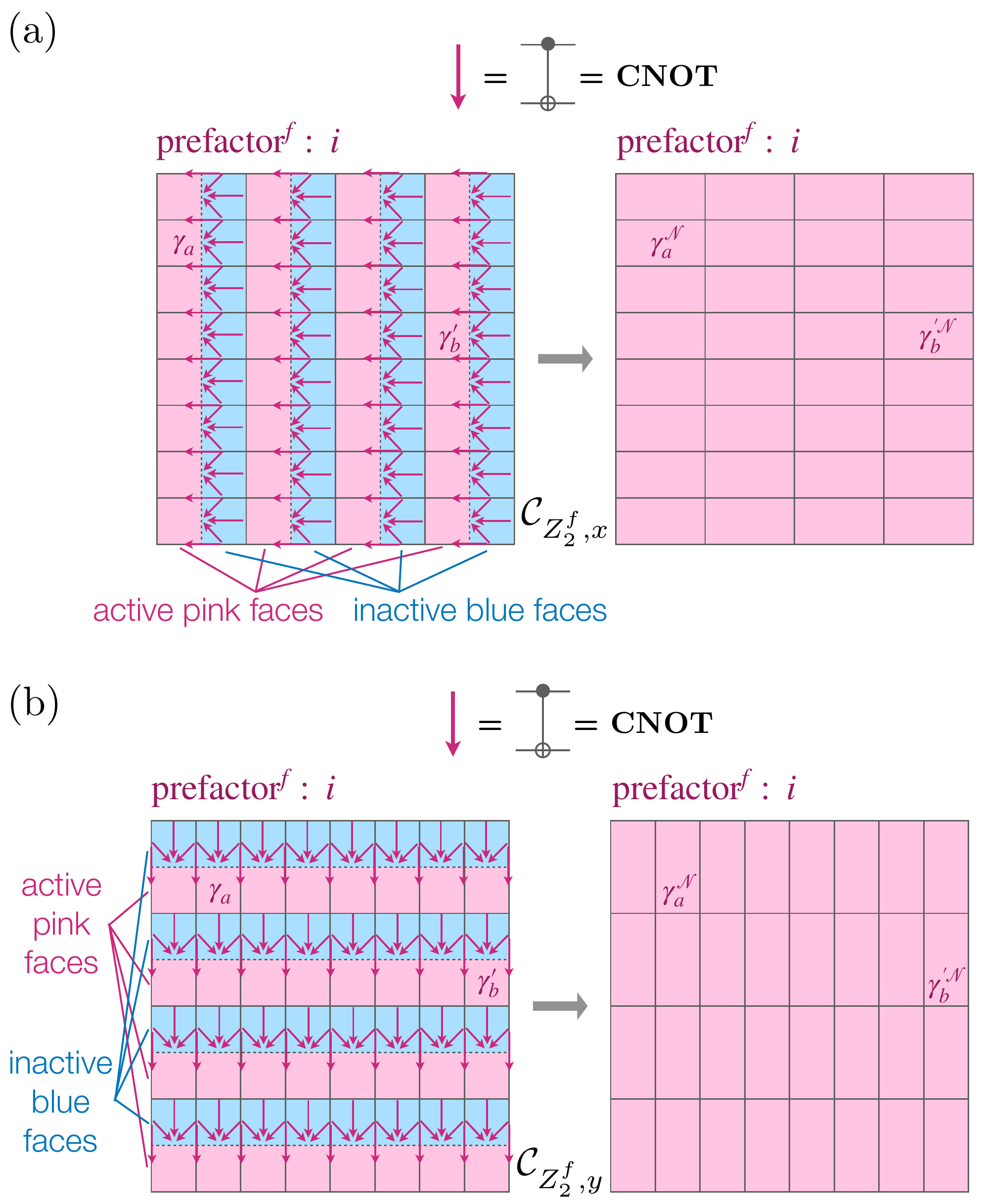}
\par\end{centering}
\centering{}\caption{We have an alternating pattern of active pink faces and inactive blue faces corresponding to the sublattice structure of the emergent (equivalently dual) fermions for the case of (a) horizontal entanglement renormalization and (b) vertical entanglement renormalization. The dual fermionic modes on the inactive blue faces are empty. We want to know how the bosonized generic quadratic fermionic term on the pink active faces, such as $\left\{i\gamma_{a}\gamma_{b}'\right\}^{\mathrm{bosonized}}$ from face $a$ to face $b$, transforms under conjugation by (a) ${\cal C}_{Z_{2}^{f},x}$ for horizontal entanglement renormalization and (b) ${\cal C}_{Z_{2}^{f},y}$ for vertical entanglement renormalization. The figure shows that, under ${\cal C}_{Z_{2}^{f},x}$ and ${\cal C}_{Z_{2}^{f},y}$, the bosonized original quadratic fermionic term $\left\{i\gamma_{a}\gamma_{b}'\right\}^{\mathrm{bosonized}}$ maps onto the corresponding bosonized quadratic fermionic term $\left\{i\gamma_{a}^\mathcal{N}{\gamma_{b}'}^\mathcal{N}\right\}^{\mathrm{bosonized}}$ defined on the new coarse-grained lattice as if the original fermion operators were living on the new elongated faces. The superscript $\mathcal{N}$ is used to label the fermion operators defined on the faces of the new lattice. The detailed calculations behind this result are shown in Fig.~\ref{fig:hoppinghori} for ${\cal C}_{Z_{2}^{f},x}$ and Fig.~\ref{fig:hoppingvert} for ${\cal C}_{Z_{2}^{f},y}$. The $\mathrm{prefactor}^{f}$ denotes the constant prefactor that must be included in front of the product of the fermionic operators shown in the figure in order to make the quadratic fermionic term under consideration. \label{fig:hoppingrenorm}}
\end{figure}

Recall that, before the circuit components $\mathcal{C}_{Z_{2}^{f},x}$ or $\mathcal{C}_{Z_{2}^{f},y}$ are applied,  we have a quantum state whose emergent (equivalently dual via the bosonization map) fermionic modes are empty on every other horizontal or vertical face. As for the fermionic modes on remaining active faces, in the lattice Ising TQFT case, they together form a lattice $p_{x}+ip_{y}$ superconducting state, whereas, in the case of the $\nu$-th Kitaev's sixteenfold way chiral spin liquid, they behave as a flattening of $\nu$ layers of topological superconductors and $16-\nu$ layers of trivial insulators. Our goal is to show that the circuit components $\mathcal{C}_{Z_{2}^{f},x}$ and $\mathcal{C}_{Z_{2}^{f},y}$ remove the empty fermionic modes and keep the active fermionic modes on the faces of the new coarse-grained lattice. This would realize our goal of having the lattice Ising TQFT state and the $\nu$-th Kitaev's sixteenfold way chiral spin liquid as fixed-point wavefunctions of the renormalization circuit. Therefore, in this appendix, to prove this functionality of $\mathcal{C}_{Z_{2}^{f},x}$ and $\mathcal{C}_{Z_{2}^{f},y}$, we consider a generalized setup of an alternating pattern of inactive blue faces, associated with empty fermionic modes, and active pink faces, associated with fermionic modes in the ground state of a quadratic fermionic Hamiltonian composed of hoppings and pairings that are not necessarily translation-invariant. For the case of horizontal entanglement renormalization, the setup is shown on the left of Fig.~\ref{fig:hoppingrenorm}(a). 
Here, the inactive (frozen in the empty fermionic state) blue faces correspond to the $B$ faces in Fig.~\ref{fig:Isinghorizontalrenormalizaitondetail}(b)  and the primed faces in Fig.~\ref{fig:16foldwayhorizontalrenormalizationdetail}(c) before the application of $\mathcal{C}_{Z_{2}^{f},x}$. A similar setup for the case of vertical entanglement renormalization is shown on the left of Fig.~\ref{fig:hoppingrenorm}(b). Here, the inactive (frozen in the empty fermionic state) blue faces correspond to the $B$ faces in Fig.~\ref{fig:Isingverticalrenormalizaitondetail}(b) and the primed faces in Fig.~\ref{fig:16foldwayverticalrenormalizationdetail}(c) before the application of $\mathcal{C}_{Z_{2}^{f},y}$.

We want to prove now that, after the circuit component $\mathcal{C}_{Z_{2}^{f},x}$ ($\mathcal{C}_{Z_{2}^{f},y}$), the new dual fermions are in the state of the old dual fermions associated with the original active pink faces. The emergent (equivalently dual) fermionic mode on an elongated face of the new lattice will be just the original emergent fermionic mode on the active pink face of the old lattice enclosed by that elongated face. Effectively, we make the inactive blue faces disappear, while the active pink faces become larger by consuming the original area occupied by the primed faces. This intuitive explanation of our objective here is reflected in the coloring of the faces of the new lattice in Fig.~\ref{fig:hoppingrenorm}. Our strategy for showing this is to work with the transformation of the parent Hamiltonian of the whole spin system under conjugation by  $\mathcal{C}_{Z_{2}^{f},x}$ ($\mathcal{C}_{Z_{2}^{f},y}$). Specifically, we want to show that the bosonized quadratic fermionic terms on the active pink faces under the zero-flux condition are mapped to the same bosonized quadratic fermionic terms on the new lattice under the corresponding new zero-flux condition as if the associated fermionic modes were just living on bigger elongated faces. Figure~\ref{fig:hoppingrenorm} demonstrates a typical situation. Initially, we have a bosonized hopping term $\left\{i\gamma_a \gamma _b^{\prime}\right\}^{\mathrm{bosonized}}$ for the emergent fermions on distant active pink faces $a$ and $b$. We want to prove that, under conjugation by  $\mathcal{C}_{Z_{2}^{f},x}$ ($\mathcal{C}_{Z_{2}^{f},y}$), this operator turns into the bosonized hopping term $\left\{i\gamma^{\mathcal{N}}_a \gamma^{\prime\mathcal{N}} _b\right\}^{\mathrm{new\,bosonized}}$. The notation $\gamma_a^{\mathcal{N}}$ or $\gamma_a^{\prime\mathcal{N}}$ indicates a Majorana fermion operator on the face of the new lattice that encloses the old active pink face $a$ (the superscript $\mathcal{N}$ stands for ``new"). The bracket notation  $\left\{\cdot \right\}^{\mathrm{new\,bosonized}}$ denotes the bosonization procedure with respect to the edge orientation assignments on the new elongated square lattice, as shown in Fig.~\ref{fig:newfermionstospins} for the horizontally coarse-grained lattice. Instead of directly working with the generic operator $\left\{i\gamma_a \gamma _b^{\prime}\right\}^{\mathrm{bosonized}}$ with arbitrary faces $a$ and $b$, we can break such an operator into a product of the generators of the algebra coming from the bosonization of the parity-conserving fermionic algebra acting only on the active pink faces. These generators are the shortest bosonized horizontal Majorana hopping terms,  the shortest bosonized vertical Majorana hopping terms, and the bosonized fermion parity operators \footnote{Recall that ``the bosonized fermion parity operator'' is the same as ``the emergent fermion parity operator.'' The former term emphasizes the fact that the operator is constructed using the bosonization technique, 
whereas the latter term emphasizes the fact that the operator measures the parity of the emergent fermionic mode on a face.}, all of which in the fermionic picture act on the fermionic modes associated only with the active pink faces. We will compute the transformation of these generators. 

The computation will be long but straightforward. We will treat the cases of horizontal entanglement renormalization (Sec.\ \ref{sec:horizontalappendix}) and vertical entanglement renormalization (Sec.\ \ref{sec:verticalappendix}) separately. The zero-flux condition $F_\nu=1$ on the original square lattice will be assumed. Following the notation in the main text, throughout our computations below, we will sometimes still use the letter $B$ or the prime symbols to label the inactive blue faces. We will also use the letter $A$ to label the active pink faces.

\subsection{Horizontal Entanglement Renormalization \label{sec:horizontalappendix}}

In this subsection, we study the action of $\mathcal{C}_{Z_{2}^{f},x}$, which is part of horizontal entanglement renormalization.

Under the zero-flux condition $F_v=1$, let $\ket{\Psi}$ be the ground state of a bosonized fermionic Hamiltonian that freezes the emergent fermionic modes on inactive blue faces in the empty state: 
\begin{equation}
H_{AB,\,x}^{\mathrm{bosonized}} =\left\{ H_{AB,\,x}^{f}\right\} ^{\mathrm{bosonized}}
\end{equation}
 with 
\[
H_{AB,\,x}^{f}=-\sum_{\mathcal{Z}_{A}}h_{\mathcal{Z}_{A}}-\sum_{\mathcal{I}_{B}}\left(-i\gamma_{\mathcal{I}_{B}}\gamma_{\mathcal{I}_{B}}^{\prime}\right).
\]

The notation $\mathcal{Z}_{A}$ is used to label different quadratic terms $h_{\mathcal{Z}_{A}}$ associated with emergent fermionic modes on the active pink faces, 
and $\mathcal{\mathcal{I}_B}$ is used to label emergent fermionic modes on the inactive blue faces.
Note that the ground state $\ket{\Psi}$ is in the sector $F_{v}=1,\, \forall v$ and has $\left\{ -i\gamma_{\mathcal{I}_{B}}\gamma'_{\mathcal{I}_{B}}\right\} ^{\mathrm{bosonized}}=Z_{N(\mathcal{I}_{B})}Z_{E(\mathcal{I}_{B})}Z_{S(\mathcal{I}_{B})}Z_{W(\mathcal{I}_{B})}=1$. The notation $N(f)$ denotes the adjacent qubit north of face $f$, and $E(f)$, $S(f)$, and $W(f)$ denote the adjacent qubits east of, south of, and west of face $f$, respectively. The fermionic Hamiltonian component $-\sum_{\mathcal{Z}_{A}}h_{\mathcal{Z}_{A}}$ need not to be translationally invariant under shifting the labels of the sites associated with the active pink faces, as is the case of the $\nu$-th Kitaev's sixteenfold way chiral spin liquid.

\begin{figure}
\begin{centering}
\includegraphics[width=\columnwidth]{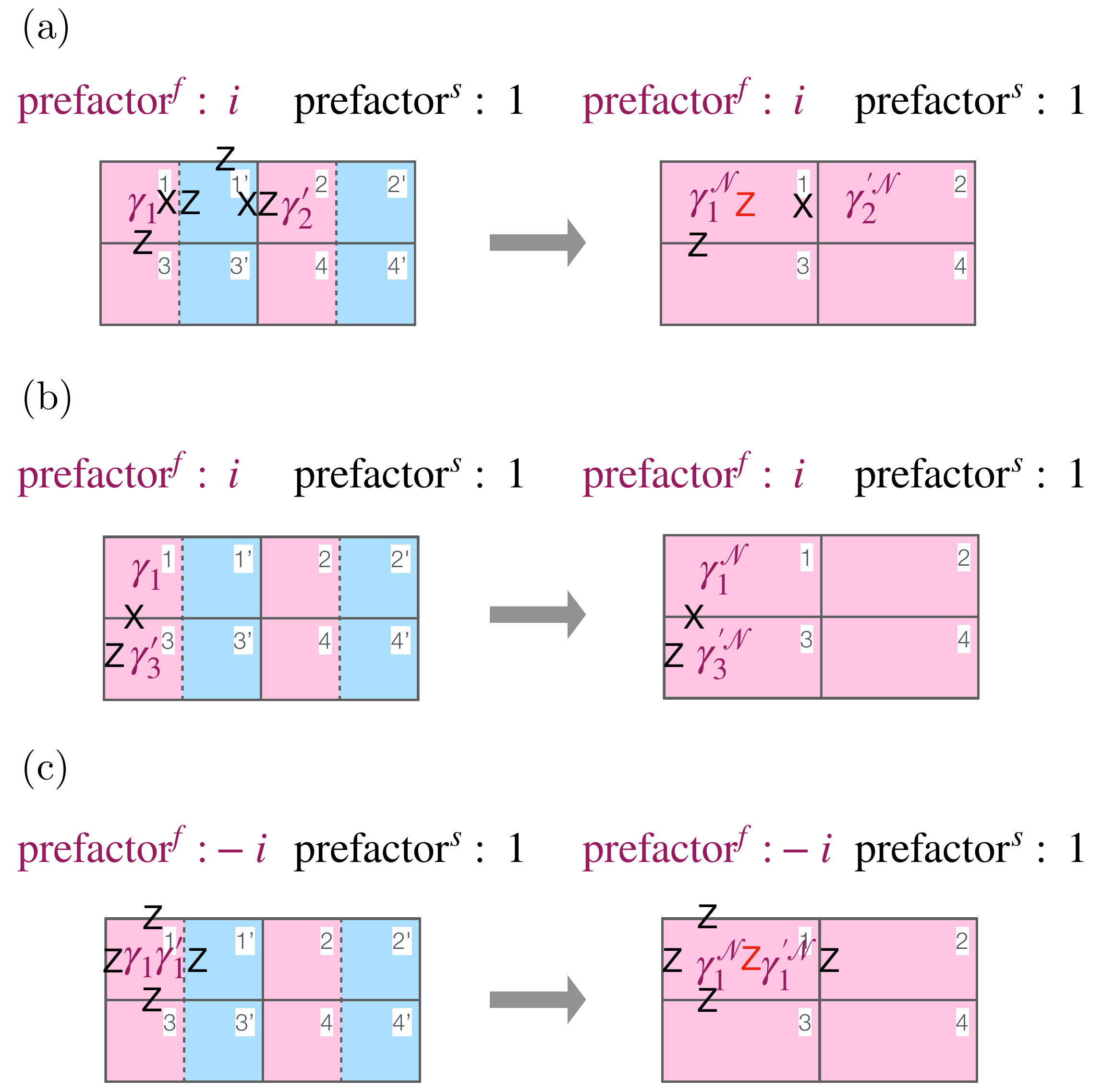}
\par\end{centering}
\centering{}\caption{The subfigures show how the bosonized generators of the fermionic algebra associated with the active pink faces change under conjugation by $\mathcal{C}_{Z_{2}^{f},x}$. Inactive blue faces are labeled by numbers with primes, while active pink faces are labeled by numbers without primes. (a) The change of the shortest bosonized horizontal Majorana hopping term $\left\{i\gamma_{1}\gamma_{2}'\right\}^{\mathrm{bosonized}}$ under $\mathcal{C}_{Z_{2}^{f},x}$. (b) The change of the shortest bosonized vertical Majorana hopping term $\left\{i\gamma_{1}\gamma_{3}'\right\}^{\mathrm{bosonized}}$ under $\mathcal{C}_{Z_{2}^{f},x}$. (c) The change of the bosonized fermion parity operator $\left\{-i\gamma_{1}\gamma_{1}'\right\}^{\mathrm{bosonized}}$ under $\mathcal{C}_{Z_{2}^{f}\,x}$. The notation $\mathrm{prefactor}^{f}$ denotes the constant prefactor that must be included in front of the product of the fermion operators shown in the figure in order to make the quadratic fermionic term under consideration, and the ordering of the fermion operators is specified above. 
The notation $\mathrm{prefactor}^{s}$ denotes the constant prefactor that must be included in front of the product of the spin operators shown in the figure. 
When an $X$ operator and a $Z$ operator both act on a qubit, the $Z$ operator acts first. The red single-qubit Pauli-$Z$ operator sitting at the center of the new face $1$ is acting on a disentangled qubit. This operator takes eigenvalue one in the transformed quantum state $\mathcal{C}_{Z_{2}^{f},x}\ket{\Psi}$. Therefore, up to this operator taking eigenvalue one, the transformed generators match the bosonized fermion operators on the new elongated faces containing the original fermion operators being transformed.  \label{fig:hoppinghori}}
\end{figure}

Now we investigate $\mathcal{C}_{Z_{2}^{f},x}\ket{\Psi}$. Our objective is to show that $\mathcal{C}_{Z_{2}^{f},x}\ket{\Psi}$ is dual (via bosonization) to the ground state of the same fermionic Hamiltonian $-\sum_{\mathcal{Z}_{A}}h_{\mathcal{Z}_{A}}$, but now defined on the faces of the new coarse-grained lattice. Notice that $\mathcal{C}_{Z_{2}^{f},x}\ket{\Psi}$ will be the ground state of the unitarily transformed Hamiltonian $\mathcal{C}_{Z_{2}^{f},x}\,H_{AB,\,x}^{\mathrm{bosonized}}\,\mathcal{C}_{Z_{2}^{f},x}^{\dagger}$. Therefore, in order to study $\mathcal{C}_{Z_{2}^{f},x}\ket{\Psi}$, we will study this new Hamiltonian, $\mathcal{C}_{Z_{2}^{f},x}\,H_{AB,\,x}^{\mathrm{bosonized}}\,\mathcal{C}_{Z_{2}^{f},x}^{\dagger}$. We first study the transformation of the component involving active pink faces, $\mathcal{C}_{Z_{2}^{f},x}\left\{ h_{\mathcal{Z}_{A}}\right\} ^{\mathrm{bosonized}}\mathcal{C}_{Z_{2}^{f},x}^{\dagger}$. As mentioned above, a generic quadratic  term involving distant active pink faces, such as $\left\{i\gamma_a \gamma _b^{\prime}\right\}^{\mathrm{bosonized}}$ on the left-hand side of Fig.~\ref{fig:hoppingrenorm}(a), can be written as a product of the shortest bosonized Majorana hopping operators and the bosonized fermion parity operators that, in the fermonic picture, are only supported on the fermionic sites on the active pink faces. Therefore, it is sufficient to study how these generators transform under conjugation by $\mathcal{C}_{Z_{2}^{f},x}$. For the shortest bosonized horizontal Majorana hopping operator between active pink faces, we take as an example the shortest bosonization of the operator $i\gamma_{1}\gamma'_{2}$ depicted on the left-hand side of Fig.~\ref{fig:hoppinghori}(a):
\begin{align}
   &\left\{ i\gamma_{1}\gamma'_{2}\right\} ^{\mathrm{bosonized}}  \nonumber\\
 = & i\left\{ \gamma_{1}\gamma'_{1'}\gamma'_{1'}\gamma{}_{1'}\gamma_{1'}\gamma'_{2}\right\} ^{\mathrm{bosonized}}\nonumber \\
 = & -\left\{ i\gamma_{1}\gamma'_{1'}\right\} ^{\mathrm{bosonized}}\left\{ -i\gamma{}_{1'}\gamma'_{1'}\right\} ^{\mathrm{bosonized}}\left\{ i\gamma_{1'}\gamma'_{2}\right\} ^{\mathrm{bosonized}}\nonumber \\
 = & -\left(X_{E(1)}Z_{S(1)}\right)\left(Z_{N(1')}Z_{E(1')}Z_{S(1')}Z_{W(1')}\right)\cdot\nonumber\\
   & \quad \quad \left(X_{E(1')}Z_{S(1')}\right)\nonumber \\
 = & \left(Z_{S(1)}Z_{N(1')}Z_{E(1')}Z_{W(1')}\right)\left(X_{E(1)}X_{E(1')}\right).
\end{align}
We used $\gamma_{f}^{2}=\gamma_{f}^{\prime2}=1$. The corresponding bosonized result is illustrated  on the left-hand side of Fig.~\ref{fig:hoppinghori}(a). Note that our way of denoting qubits is redundant: e.g., $E(1)=W(1')$. We now conjugate $\left\{ i\gamma_{1}\gamma'_{2}\right\} ^{\mathrm{bosonized}}$ by $\mathcal{C}_{Z_{2}^{f},x}$:
\begin{align}
& \mathcal{C}_{Z_{2}^{f},x}\left\{ i\gamma_{1}\gamma'_{2}\right\} ^{\mathrm{bosonized}}\mathcal{C}_{Z_{2}^{f},x}^{\dagger}  \nonumber\\
= & \mathcal{C}_{Z_{2}^{f},x}\left(Z_{S(1)}Z_{N(1')}Z_{E(1')}Z_{W(1')}\right)\mathcal{C}_{Z_{2}^{f},x}^{\dagger}\cdot\nonumber \\
 & \quad \mathcal{C}_{Z_{2}^{f},x}\left(X_{E(1)}X_{E(1')}\right)\mathcal{C}_{Z_{2}^{f},x}^{\dagger}\nonumber \\
= & Z_{S(1)}Z_{W(1')}X_{E(1')}\nonumber \\
= & Z_{W(1')}\left(X_{E(1')}Z_{S(1)}\right)\nonumber \\
= & Z_{W(1')}\left\{ i\gamma_{1}^{\mathcal{N}}\gamma^{\prime}{}_{2}^{\mathcal{N}}\right\} ^{\mathrm{new\,bosonized}}.\label{eq:hori_hopping_horirenorm}
\end{align}
The bracket with a superscript $\left\{\cdot \right\} ^{\mathrm{new\,bosonized}}$ denotes bosonization on the new horizontally elongated square lattice, and the superscript $\mathcal{N}$ for $\gamma_{f}^{\mathcal{N}}$ denotes an emergent Majorana fermion operator on face $f$ of the new lattice. We used labels on the old active pink face to indicate the new faces containing them since we expect there will be a correspondence between the new faces and the old active pink faces. Note that, throughout the calculation, for the subscripts of all the $X$ and $Z$ operators, the qubit labeling $N(f)$, $E(f)$, $S(f)$, and $W(f)$ is associated with face $f$ of the old lattice. It is only for the new Majorana operators $\gamma_{f}^{\mathcal{N}}$ and $\gamma_{f}^{\prime\mathcal{N}}$ do we use the face labeling on the new lattice. We will also use this convention throughout the later calculations. The result of the above computation is shown on the right-hand side of Fig.~\ref{fig:hoppinghori}(a).

With the same labeling of the lattice faces, we now consider the shortest bosonized vertical Majorana hopping operator $\left\{ i\gamma_{1}\gamma'_{3}\right\} ^{\mathrm{bosonized}}$. We can easily see that
\begin{align}
& \mathcal{C}_{Z_{2}^{f},x}\left\{ i\gamma_{1}\gamma'_{3}\right\} ^{\mathrm{bosonized}}\mathcal{C}_{Z_{2}^{f},x}^{\dagger}  \nonumber \\
= & \,\mathcal{C}_{Z_{2}^{f},x}\left(X_{S(1)}Z_{W(3)}\right)\mathcal{C}_{Z_{2}^{f},x}^{\dagger} \nonumber \\
= & X_{S(1)}Z_{W(3)} \nonumber \\
= & \left\{ i\gamma_{1}^{\mathcal{N}}\gamma{}_{3}^{\mathcal{\prime N}}\right\} ^{\mathrm{new\,bosonized}}.\label{eq:vert_hopping_horirenorm}
\end{align}
This is shown in Fig.~\ref{fig:hoppinghori}(b). 

For the bosonized fermion parity operator on face $1$, we have
\begin{align}
& \mathcal{C}_{Z_{2}^{f},x}\left\{ -i\gamma_{1}\gamma'_{1}\right\} ^{\mathrm{bosonized}}\mathcal{C}_{Z_{2}^{f},x}^{\dagger}  \nonumber \\
= & \, \mathcal{C}_{Z_{2}^{f},x}Z_{N(1)}Z_{E(1)}Z_{S(1)}Z_{W(1)}\mathcal{C}_{Z_{2}^{f},x}^{\dagger}\nonumber \\
= & Z_{N(1)}Z_{E(1)}Z_{S(1)}Z_{W(1)}Z_{E(1')}\nonumber \\
= & Z_{E(1)}\left(Z_{N(1)}Z_{S(1)}Z_{W(1)}Z_{E(1')}\right)\nonumber \\
= & Z_{W(1')}\left\{ -i\gamma_{1}^{\mathcal{N}}\gamma^{\prime}{}_{1}^{\mathcal{N}}\right\} ^{\mathrm{new\,bosonized}}.\label{eq:fparity_horirenorm}
\end{align}
This is shown in Fig.~\ref{fig:hoppinghori}(c). We can see that the old bosonized hopping and fermion-parity operators are mapped to the corresponding new bosonized hopping and fermion-parity operators on the new lattice up to some Pauli-$Z$ operators. Therefore, a generic bosonized quadratic term $\left\{ h_{\mathcal{Z}_{A}}\right\} ^{\mathrm{bosonized}}$ will be mapped to $\left\{ h_{\mathcal{Z}_{A}}^{\mathcal{N}}\right\} ^{\mathrm{new\,bosonized}}$ up to a product of Pauli-$Z$ operators acting on qubits residing on the left edges of inactive blue faces. We use the notation $h_{\mathcal{Z}_{A}}^{\mathcal{N}}$ to mean the same as $h_{\mathcal{Z}_{A}}$ but with all the Majorana operators replaced according to $\gamma_f\rightarrow \gamma_f^{\cal{N}}$, $\gamma_f^{\prime}\rightarrow \gamma_f^{\prime\cal{N}}$.

We will now analyze the transformation of the operator $\left\{ -i\gamma_{\mathcal{I}_{B}}\gamma'_{\mathcal{I}_{B}}\right\} ^{\mathrm{bosonized}}$ on the inactive blue faces:
\begin{align}
& \mathcal{C}_{Z_{2}^{f},x}\left\{ -i\gamma_{\mathcal{I}_{B}}\gamma'_{\mathcal{I}_{B}}\right\} ^{\mathrm{bosonized}}\mathcal{C}^{\dagger}_{Z_{2}^{f},x}  \nonumber \\
= & \mathcal{C}_{Z_{2}^{f},x}Z_{N(\mathcal{I}_{B})}Z_{E(\mathcal{I}_{B})}Z_{S(\mathcal{I}_{B})}Z_{W(\mathcal{I}_{B})}\mathcal{C}_{Z_{2}^{f},x}^{\dagger} \nonumber \\
= & Z_{W(\mathcal{I}_{B})}.
\label{eq:transformation_of_parity_hori}
\end{align}
This is nothing but the computation done in Fig.~\ref{fig:MERA-Z2f-hori-calulation}(a).

Combining the above results, we finally obtain the transformation of the full Hamiltonian:
\begin{align}
& \mathcal{C}_{Z_{2}^{f},x}\,H_{AB,\,x}^{\mathrm{bosonized}}\,\mathcal{C}_{Z_{2}^{f},x}^{\dagger}  \nonumber \\
= & -\sum_{\mathcal{Z}_{A}}\left(\prod_{\mathcal{I}_{B}\in \mathcal{S}_B[h_{\mathcal{Z}_{A}}]}Z_{W(\mathcal{I}_{B})}\right) \cdot \left\{ h_{\mathcal{Z}_{A}}^{\mathcal{N}}\right\} ^{\mathrm{new\,bosonized}}\nonumber\\
& \quad \quad-\sum_{\mathcal{I}_{B}}Z_{W(\mathcal{I}_{B})},
\label{eq:horirenormalizedHamiltonian}
\end{align}
where 
\begin{equation}
\mathcal{S}_B[h_{\mathcal{Z}_{A}}]\equiv
\{
\mathcal{I}_{B}|Z_{W(\mathcal{I}_{B})}
\in 
\mathcal{C}_{Z_{2}^{f},x}
\,
\left\{h_{\mathcal{Z}_{A}}\right\}^{\mathrm{bosonized}}
\,
\mathcal{C}_{Z_{2}^{f},x}^{\dagger}
\}.
\end{equation}
The
operator product $\left(\prod_{\mathcal{I}_{B}\in \mathcal{S}_B[h_{\mathcal{Z}_{A}}]}Z_{W(\mathcal{I}_{B})}\right)$ in front of $\left\{ h_{\mathcal{Z}_{A}}^{\mathcal{N}}\right\} ^{\mathrm{new\,bosonized}}$ is simply due to the single-qubit Pauli-$Z$ operators resulting from the transformation of the bosonized horizontal hoppings and fermion parity operators in Eq.~(\ref{eq:hori_hopping_horirenorm}) and Eq.~(\ref{eq:fparity_horirenorm}).

We are now prepared to state some properties of the state $\mathcal{C}_{Z_{2}^{f},x}\ket{\Psi}$. Due to the emptiness of the old inactive blue faces in the ground  state $\ket{\Psi}$, we have $Z_{N(\mathcal{I}_{B})}Z_{E(\mathcal{I}_{B})}Z_{S(\mathcal{I}_{B})}Z_{W(\mathcal{I}_{B})}=1$. From Eq.~(\ref{eq:transformation_of_parity_hori}), we learn that  $Z_{W(\mathcal{I}_{B})}=1$ in the transformed ground state $\mathcal{C}_{Z_{2}^{f},x}\ket{\Psi}$.
In addition, the transformation of the zero-flux condition $F_v=1$ computed in Fig.~\ref{fig:MERA-Z2f-hori-calulation}(c,d) gives $F^{{\left[{\mathrm{Fig.\,}\ref{fig:MERA-Z2f-hori-calulation}(\mathrm{c})}\right]}_{\mathrm{RHS}}}_{v^\mathcal{N}}=1$ and $F^{{\left[{\mathrm{Fig.\,}\ref{fig:MERA-Z2f-hori-calulation}(\mathrm{d})}\right]}_{\mathrm{RHS}}}_{v^\mathcal{N}}=1$ for $\mathcal{C}_{Z_{2}^{f},x}\ket{\Psi}$. The notation $v^{\mathcal{N}}$ denotes a vertex on the new lattice. The shorthand notation $F^{{\left[{\mathrm{Fig.\,}\ref{fig:MERA-Z2f-hori-calulation}(\mathrm{c})}\right]}_{\mathrm{RHS}}}_{v^\mathcal{N}}$ denotes the operator shown on the right-hand side of Fig.~\ref{fig:MERA-Z2f-hori-calulation}(c), whereas the notation $F^{{\left[{\mathrm{Fig.\,}\ref{fig:MERA-Z2f-hori-calulation}(\mathrm{d})}\right]}_{\mathrm{RHS}}}_{v^\mathcal{N}}$ denotes the operator shown on the right-hand side of Fig.~\ref{fig:MERA-Z2f-hori-calulation}(d). Therefore, with $Z_{W(\mathcal{I}_B)}=1$ and $F^{{\left[{\mathrm{Fig.\,}\ref{fig:MERA-Z2f-hori-calulation}(\mathrm{c})}\right]}_{\mathrm{RHS}}}_{v^\mathcal{N}}=1$, we obtain $X_{S(\mathcal{I}_B)}=1$ for $\mathcal{C}_{Z_{2}^{f},x}\ket{\Psi}$. That is, $W(\mathcal{I}_{B})$ and $S(\mathcal{I}_{B})$ are disentangled 
qubits in states $\ket{0}$ and $\ket{+}$, respectively. Additionally, from $Z_{W(\mathcal{I}_B)}=1$, $X_{S(\mathcal{I}_B)}=1$, and $F^{{\left[{\mathrm{Fig.\,}\ref{fig:MERA-Z2f-hori-calulation}(\mathrm{d})}\right]}_{\mathrm{RHS}}}_{v^\mathcal{N}}=1$, we see that, for $\mathcal{C}_{Z_{2}^{f},x}\ket{\Psi}$, we have $F_{v^{\mathcal{N}}}^{\mathcal{N}}=1$, where $F_{v^{\mathcal{N}}}^{\mathcal{N}}$  is the flux measuring operator associated with the vertex $v^\mathcal{N}$ defined on the new coarse-grained lattice shown as the black 6-qubit operator in Fig.~\ref{fig:MERA-Z2f-hori}(c). That is, the state $\mathcal{C}_{Z_{2}^{f},x}\ket{\Psi}$ satisfies the new zero-flux condition on the new coarse-grained lattice. The new zero-flux condition justifies the definition of the new bosonization mapping $\left\{\cdot \right\} ^{\mathrm{new\,bosonized}}$ for the new lattice.

Putting together the above properties of $\mathcal{C}_{Z_{2}^{f},x}\ket{\Psi}$, we arrive at the desired result that $\mathcal{C}_{Z_{2}^{f},x}\ket{\Psi}$ is
the ground state of the Hamiltonian
\begin{equation}
H_{A,\,x}^{\mathcal{N}\,\mathrm{new\,bosonized}}-\sum_{\mathcal{I}_{B}}Z_{W(\mathcal{I}_{B})}-\sum_{\mathcal{I}_{B}}X_{S(\mathcal{I}_{B})}
\label{eq:horirenormalizedHamiloniantarget}
\end{equation}
with
\begin{align}
    H_{A,\,x}^{\mathcal{N}\,\mathrm{new\,bosonized}}&\equiv\left\{ H_{A,\,x}^{\mathcal{N}\,f}\right\} ^{\mathrm{new\,bosonized}}\nonumber\\
    H_{A,\,x}^{\mathcal{N}\,f}&\equiv-\sum_{\mathcal{Z}_{A}}h_{\mathcal{Z}_{A}}^{\mathcal{N}}
\end{align}
under the new zero-flux condition $F_{v^{\mathcal{N}}}^{\mathcal{N}}=1$.
The operator product $\left(\prod_{\mathcal{I}_{B}\in \mathcal{S}_B[h_{\mathcal{Z}_{A}}]}Z_{W(\mathcal{I}_{B})}\right)$ in Eq.~(\ref{eq:horirenormalizedHamiltonian}) is not shown here since it is equal to one for $\mathcal{C}_{Z_{2}^{f},x}\ket{\Psi}$, so $\mathcal{C}_{Z_{2}^{f},x}\ket{\Psi}$ is the ground state of Eq.\ (\ref{eq:horirenormalizedHamiloniantarget}) both with and without $\left(\prod_{\mathcal{I}_{B}\in \mathcal{S}_B[h_{\mathcal{Z}_{A}}]}Z_{W(\mathcal{I}_{B})}\right)$. The Hamiltonian in Eq.~(\ref{eq:horirenormalizedHamiloniantarget}) has the simple interpretation as the bosonized original Hamiltonian $-\sum_{\mathcal{Z}_{A}}h_{\mathcal{Z}_{A}}$ on the active pink faces but now defined on the new faces and with the new zero-flux condition. The terms $-\sum_{\mathcal{I}_{B}}Z_{W(\mathcal{I}_{B})}$ and $-\sum_{\mathcal{I}_{B}}X_{S(\mathcal{I}_{B})}$
describe the disentangled
qubits.

\subsection{Vertical Entanglement Renormalization \label{sec:verticalappendix}}

In this subsection, we present a similar argument for vertical entanglement renormalization.

Under the zero-flux condition $F_v=1$, let $\ket{\Psi}$ be the ground state of the following parent Hamiltonian with a vertically alternating pattern of active pink faces and inactive blue faces: 
\begin{equation}
H_{AB,\,y}^{\mathrm{bosonized}} =\left\{ H_{AB,\,y}^{f}\right\} ^{\mathrm{bosonized}},
\end{equation}
where
\[
H_{AB,\,y}^{f}=-\sum_{\mathcal{Z}_{A}}h_{\mathcal{Z}_{A}}-\sum_{\mathcal{I}_{B}}\left(-i\gamma_{\mathcal{I}_{B}}\gamma_{\mathcal{I}_{B}}^{\prime}\right).
\]

We can borrow all the arguments from our discussion of horizontal entanglement renormalization above and show that the transformed quantum state $\mathcal{C}_{Z_{2}^{f},y}\ket{\Psi}$ is the ground state of 
\begin{equation}
H_{A,\,y}^{\mathcal{N}\,\mathrm{new\,bosonized}}-\sum_{\mathcal{I}_{B}}X_{W(\mathcal{I}_{B})}-\sum_{\mathcal{I}_{B}}Z_{S(\mathcal{I}_{B})}
\label{eq:vertrenormalizedHamiloniantarget}
\end{equation}
with
\begin{align}
    H_{A,\,y}^{\mathcal{N}\,\mathrm{new\,bosonized}}&\equiv\left\{ H_{A,\,y}^{\mathcal{N}\,f}\right\} ^{\mathrm{new\,bosonized}}\nonumber\\
    H_{A,\,y}^{\mathcal{N}\,f}&\equiv-\sum_{\mathcal{Z}_{A}}h_{\mathcal{Z}_{A}}^{\mathcal{N}}.
\end{align}
The bracket with a superscript $\left\{\cdot \right\} ^{\mathrm{new\,bosonized}}$ denotes bosonization on the new vertically elongated square lattice under the new zero-flux condition $F_{v^{\mathcal{N}}}^{\mathcal{N}}=1$, where the operator $F_{v^{\mathcal{N}}}^{\mathcal{N}}$ is the flux measuring operator associated with the vertex $v^\mathcal{N}$ defined on the new vertically coarse-grained lattice shown as the black 6-qubit operator in Fig.~\ref{fig:MERA-Z2f-vert}(c). The qubits $W(\mathcal{I}_{B})$ and $Z(\mathcal{I}_{B})$ are disentangled qubits. Through the computations in Fig.~\ref{fig:MERA-Z2f-vert-calculation}(a,c,d), we find that $X_{W(\mathcal{I}_{B})}=1$, $Z_{S(\mathcal{I}_{B})}=1$, and $F_{v^{\mathcal{N}}}^{\mathcal{N}}=1$ for $\mathcal{C}_{Z_{2}^{f},y}\ket{\Psi}$. The quadratic fermionic  term $h_{\mathcal{Z}_{A}}^{\mathcal{N}}$ is the same as $h_{\mathcal{Z}_{A}}$ but with all the Majorana operators replaced according to $\gamma_f\rightarrow \gamma_f^{\cal{N}}$, $\gamma_f^{\prime}\rightarrow \gamma_f^{\prime\cal{N}}$. Here the superscript $\mathcal{N}$ for Majorana fermion operators labels the fermionic operators defined on the faces of the vertically coarse-grained lattice, and we used the label $f$ on the old active pink face to denote the new face containing it. The Hamiltonian in Eq.~(\ref{eq:vertrenormalizedHamiloniantarget}) has a simple interpretation as the bosonized original Hamiltonian $-\sum_{\mathcal{Z}_{A}}h_{\mathcal{Z}_{A}}$ on the old active pink faces but now defined on the new vertically elongated faces with the new zero-flux condition. The terms $-\sum_{\mathcal{I}_{B}}X_{W(\mathcal{I}_{B})}$ and $-\sum_{\mathcal{I}_{B}}Z_{S(\mathcal{I}_{B})}$ describe the disentangled qubits.

\begin{figure}
\begin{centering}
\includegraphics[width=\columnwidth]{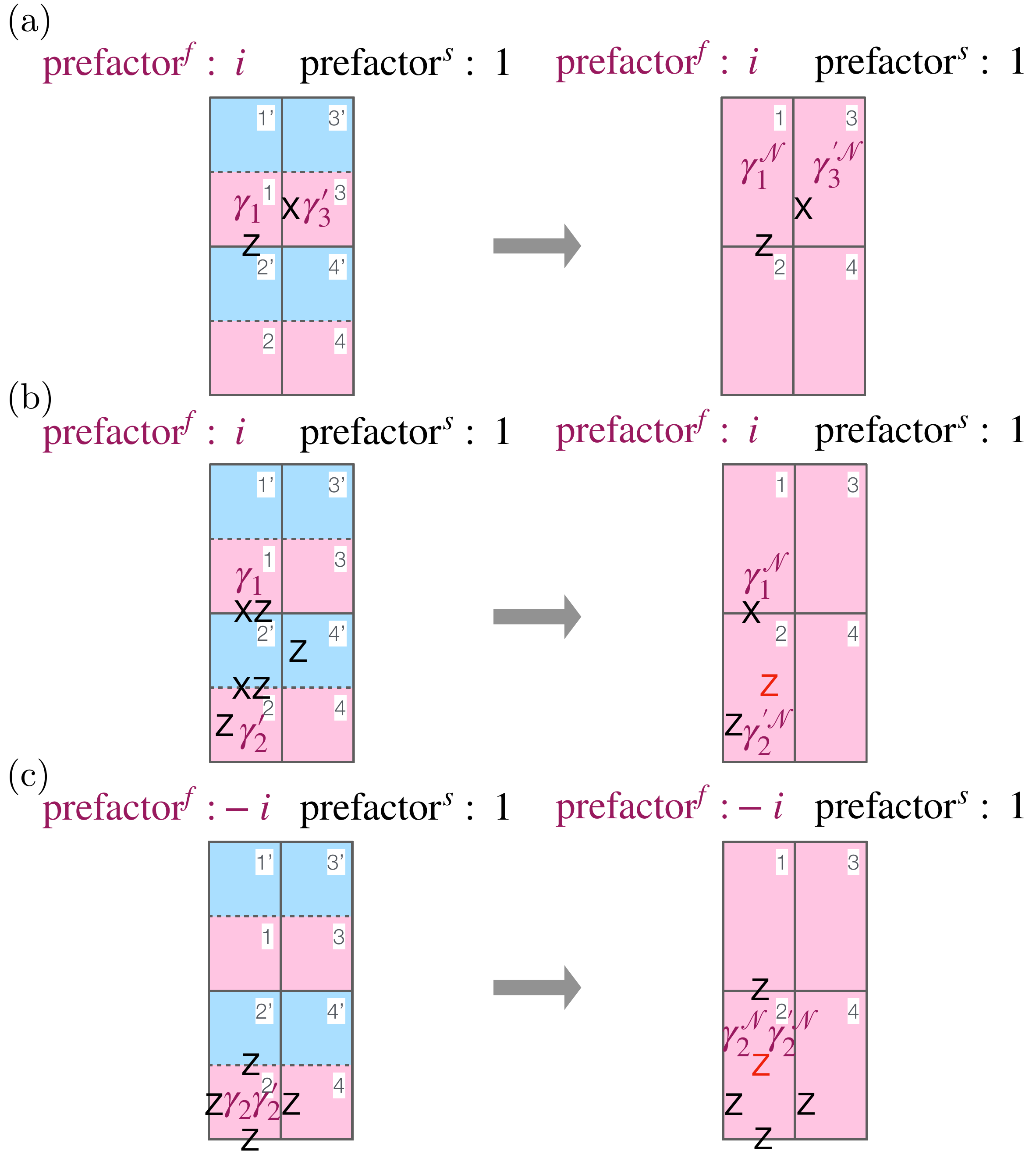}
\par\end{centering}
\centering{}\caption{The subfigures show how the bosonized generators of the fermionic algebra associated with the active pink faces change under conjugation by $\mathcal{C}_{Z_{2}^{f},y}$. Inactive blue faces are labeled by numbers with primes, while active pink faces are labeled by numbers without primes. (a) The change of the shortest bosonized horizontal Majorana hopping term $\left\{i\gamma_{1}\gamma_{3}'\right\}^{\mathrm{bosonized}}$ under $\mathcal{C}_{Z_{2}^{f},y}$. (b) The change of the shortest bosonized vertical Majorana hopping term $\left\{i\gamma_{1}\gamma_{2}'\right\}^{\mathrm{bosonized}}$ under $\mathcal{C}_{Z_{2}^{f},y}$. (c) The change of the bosonized fermion parity operator $\left\{-i\gamma_{2}\gamma_{2}'\right\}^{\mathrm{bosonized}}$ under $\mathcal{C}_{Z_{2}^{f}\,y}$. The notation $\mathrm{prefactor}^{f}$ denotes the constant prefactor that must be included in front of the product of the fermion operators shown in the figure in order to make the quadratic fermionic term under consideration, and the ordering of the fermion operators is specified above. The notation $\mathrm{prefactor}^{s}$ denotes the constant prefactor that must be included in front of the product of the spin operators shown in the figure. When an $X$ operator and a $Z$ operator both act on a qubit, the $Z$ operator acts first. The red single-qubit Pauli-$Z$ operator sitting at the center of the new face $2$ is acting on a disentangled qubit. This operator takes eigenvalue one in the transformed quantum state $\mathcal{C}_{Z_{2}^{f},y}\ket{\Psi}$. Therefore, up to this operator taking eigenvalue one, the transformed generators match the bosonized fermion operators on the new elongated faces containing the original fermion operators being transformed.  \label{fig:hoppingvert}}
\end{figure}

Just as in the case of horizontal entanglement renormalization, we need to analyze the transformed bosonized quadratic fermionic term $\mathcal{C}_{Z_{2}^{f},x}\left\{ h_{\mathcal{Z}_{A}}\right\} ^{\mathrm{bosonized}}\mathcal{C}_{Z_{2}^{f},x}^{\dagger}$.
The transformation of a typical quadratic term like $\left\{i\gamma_a \gamma _b^{\prime}\right\}^{\mathrm{bosonized}}$ on the left-hand side of Fig.~\ref{fig:hoppingrenorm}(b) can be computed from the transformations of the generators of the bosonized fermionic algebra supported on the active pink faces, as shown in Fig.~\ref{fig:hoppingvert}.
For the shortest bosonized horizontal Majorana hopping operator between active pink faces like $\left\{ i\gamma_{1}\gamma_{3}'\right\} ^{\mathrm{bosonized}}$ on the left-hand side of Fig.~\ref{fig:hoppingvert}(a), its conjugation by $\mathcal{C}_{Z_{2}^{f},y}$ is
\begin{align}
& \mathcal{C}_{Z_{2}^{f},y}\left\{ i\gamma_{1}\gamma_{3}'\right\} ^{\mathrm{bosonized}}\mathcal{C}_{Z_{2}^{f},y}^{\dagger}  \nonumber \\
= & \, \mathcal{C}_{Z_{2}^{f},y}\left(X_{E(1)}Z_{S(1)}\right)\mathcal{C}_{Z_{2}^{f},y}^{\dagger} \nonumber \\
= & X_{E(1)}Z_{S(1)} \nonumber \\
= & \left\{ i\gamma_{1}^{\mathcal{N}}\gamma_{3}^{\mathcal{\prime N}}\right\} ^{\mathrm{new\,bosonized}}.
\end{align}
The qubit labeling subscripts of all the $X$ and $Z$ operators are associated with the faces of the old lattice. It is only for the new Majorana operators $\gamma_{f}^{\mathcal{N}}$ and $\gamma_{f}^{\prime\mathcal{N}}$ do we use the face labeling on the new vertically coarse-grained lattice. The result of the computation is shown on the right-hand side of Fig.~\ref{fig:hoppingvert}(a). 

For the shortest bosonized vertical Majorana hopping operator, like $\left\{ i\gamma_{1}\gamma_{2}'\right\} ^{\mathrm{bosonized}}$ on the left-hand side of Fig.~\ref{fig:hoppingvert}(b), we first express it in terms of spin operators:
\begin{align}
& \left\{ i\gamma_{1}\gamma'_{2}\right\} ^{\mathrm{bosonized}}  \nonumber \\
= & -\left\{ i\gamma_{1}\gamma'_{1'}\right\} ^{\mathrm{bosonized}}\left\{ -i\gamma{}_{1'}\gamma'_{1'}\right\} ^{\mathrm{bosonized}}\left\{ i\gamma_{1'}\gamma'_{2}\right\} ^{\mathrm{bosonized}}\nonumber \\
= & -\left(X_{S(1)}Z_{W(2')}\right)\left(Z_{N(2')}Z_{E(2')}Z_{S(2')}Z_{W(2')}\right)\cdot \nonumber\\
  & \quad \quad \left(X_{S(2')}Z_{W(2)}\right)\nonumber \\
= & \left(Z_{N(2')}Z_{E(2')}Z_{S(2')}Z_{W(2)}\right)\left(X_{S(1)}X_{S(2')}\right).
\end{align}
The conjugation of $\left\{ i\gamma_{1}\gamma'_{2}\right\} ^{\mathrm{bosonized}}$ under $\mathcal{C}_{Z_{2}^{f},y}$ is therefore
\begin{align}
& \mathcal{C}_{Z_{2}^{f},y}\left\{ i\gamma_{1}\gamma'_{2}\right\} ^{\mathrm{bosonized}}\mathcal{C}_{Z_{2}^{f},y}^{\dagger}  \nonumber \\
= &\, \mathcal{C}_{Z_{2}^{f},y}\left(Z_{N(2')}Z_{E(2')}Z_{S(2')}Z_{W(2)}\right)\mathcal{C}_{Z_{2}^{f},y}^{\dagger}\cdot \nonumber \\
& \quad \mathcal{C}_{Z_{2}^{f},y}\left(X_{S(1)}X_{S(2')}\right)\mathcal{C}_{Z_{2}^{f},y}^{\dagger}\nonumber \\
= & Z_{S(2')}Z_{W(2)}X_{S(1)}\nonumber \\
= & Z_{S(1')}\left\{ i\gamma_{1}^{\mathcal{N}}\gamma^{\prime}{}_{2}^{\mathcal{N}}\right\} ^{\mathrm{new\,bosonized}}.\label{eq:vert_hopping_vertrenorm}
\end{align}
The result of the computation is shown on the right-hand side of Fig.~\ref{fig:hoppingvert}(b). 

For bosonized fermion-parity operators on active pink faces like the one on the left-hand side of Fig.~\ref{fig:hoppingvert}(c), we have its conjugation under $\mathcal{C}_{Z_{2}^{f},y}$ given by
\begin{align}
& \mathcal{C}_{Z_{2}^{f},y}\left\{ -i\gamma_{2}\gamma'_{2}\right\} ^{\mathrm{bosonized}}\mathcal{C}_{Z_{2}^{f},y}^{\dagger}  \nonumber \\
= & \, \mathcal{C}_{Z_{2}^{f},x}Z_{N(2)}Z_{E(2)}Z_{S(2)}Z_{W(2)}\mathcal{C}_{Z_{2}^{f},y}^{\dagger}\nonumber \\
= & Z_{N(2)}Z_{E(2)}Z_{S(2)}Z_{W(2)}Z_{N(2')}\nonumber \\
= & Z_{N(2)}\left(Z_{N(2')}Z_{E(2)}Z_{S(2)}Z_{W(2)}\right)\nonumber \\
= & Z_{S(1')}\left\{ -i\gamma_{2}^{\mathcal{N}}\gamma^{\prime}{}_{2}^{\mathcal{N}}\right\} ^{\mathrm{new\,bosonized}}.\label{eq:fparity_vertrenorm}
\end{align}
These calculations for the generators show that a generic bosonized quadratic fermionic term $\left\{ h_{\mathcal{Z}_{A}}\right\} ^{\mathrm{bosonized}}$ can be mapped to $\left\{ h_{\mathcal{Z}_{A}}^{\mathcal{N}}\right\} ^{\mathrm{new\,bosonized}}$ up to a product of Pauli-$Z$ operators acting on qubits that live on bottom edges of inactive blue faces. 

As in the case of horizontal entanglement renormalization, we can put these results together to argue that  $\mathcal{C}_{Z_{2}^{f},y}\ket{\Psi}$ is the ground state of the Hamiltonian in Eq.~(\ref{eq:vertrenormalizedHamiloniantarget}).

\bibliography{library}

\end{document}